\documentclass{revtex4-2}
\usepackage[utf8]{inputenc}
\usepackage{amsbsy}
\usepackage{amsopn}
\usepackage{amstext}
\usepackage{amsmath,amsthm,amsfonts,amssymb}
\usepackage[mathcal]{eucal}
\usepackage{mathrsfs}
\usepackage{braket}
\usepackage[english]{babel}
\usepackage{color}
\usepackage{esint}
\usepackage{graphicx}
\usepackage{subcaption}
\usepackage{float}
\usepackage{units}
\usepackage{cancel}
\usepackage{textcomp}
\usepackage{orcidlink}

\DeclareGraphicsExtensions{.png,PNG,.pdf,.PDF}
\usepackage{hyperref}
\usepackage{slashed}
\newcommand{\ie}{\begin{equation}}
\newcommand{\fe}{\end{equation}}
\newcommand{\se}{\begin{eqnarray}}
\newcommand{\ff}{\end{eqnarray}}
\begin{document}

\title{Charged Dirac fermions with anomalous magnetic moment in the presence of the chiral magnetic effect and of a noncommutative phase space}
\author{R. R. S. Oliveira\,\orcidlink{0000-0002-6346-0720}}
\email{rubensrso@fisica.ufc.br}
\affiliation{Departamento de F\'isica, Universidade Federal da Para\'iba, Caixa Postal 5008, 58051-900, Jo\~ao Pessoa, PB, Brazil}


\date{\today}

\begin{abstract}
In this paper, we analyze the relativistic energy spectrum (or relativistic Landau levels) for charged Dirac fermions with anomalous magnetic moment (AMM) in the presence of the chiral magnetic effect (CME) and of a noncommutative (NC) phase space, where we work with the $(3+1)$-dimensional Dirac equation in cylindrical coordinates. Using a similarity transformation, we obtain four coupled first-order differential equations. Subsequently, obtain four non-homogeneous second-order differential equations. To solve these equations exactly and analytically, we use a change of variable, the asymptotic behavior, and the Frobenius method. Consequently, we obtain the relativistic spectrum for the electron/positron, where we note that this spectrum is quantized in terms of the radial quantum number $n$ and the angular quantum number $m_j$, and explicitly depends on the helicity $h$ (describes the projection of spin in the direction of linear momentum), position and momentum NC parameters $\theta$ and $\eta$ (describes the NC phase space), cyclotron frequency $\omega_c$ (an angular frequency that depends on the electric charge $e$, mass $m$, and external magnetic field $B$, i.e., $\omega_c=eB/m$), anomalous magnetic energy $E_m$ (an energy generated through the interaction of the AMM with the magnetic field), $z$-momentum $k_z$ (linear momentum along the $z$-axis), and on the fermion and chiral chemical potential $\mu$ and $\mu_5$ (describes the CME). However, through $\theta$, $\eta$, and $m$, we define two types of ``NC angular frequencies'', given by $\omega_\theta=4/m\theta$ and $\omega_\eta=\eta/m$ (our spectrum depends on three angular frequencies). Comparing our spectrum with other papers, we verified that it generalizes several particular cases found in the literature. Besides, we also graphically analyze the behavior of the spectrum as a function of $B$, $\mu$, $\mu_5$, $k_z$, $\theta$, and $\eta$ for three different values of $n$ and $m_j$.
\end{abstract}
\maketitle

\section{Introduction}

In Relativistic Quantum Mechanics (RQM), the so-called Dirac fermions are spin-1/2 massive particles (e.g., electrons, protons, neutrons, muons, taus, and quarks) governed/described by the Dirac equation (DE), which is a relativistic wave equation (of first order in both time and space) derived by British physicist Paul Dirac in 1928 \cite{P1,P2,Greiner,Bjorken,Grandy,Thomson,Griffiths,Halzen}. In literature, these particles are also often so-called Dirac particles or even matter particles (e.g., electrons, protons, and neutrons, or simply electrons and quarks). So, in addition to the DE naturally explaining the spin, helicity, and chirality of spin-1/2 particles, it is predicted that for each one of these particles, there are their respective antiparticles (e.g., positrons, antiprotons, antineutrons, etc) \cite{Greiner,Bjorken,Grandy,Griffiths,Thomson,Halzen}. Already in Particle Physics (PP), the Dirac fermions together with the bosons (integer spin particles) form the two fundamental classes of particles that make up the well-known Standard Model (SM) \cite{Greiner,Bjorken,Grandy,Griffiths,Thomson,Halzen}. The DE has many applications (somewhat incalculable), for example, it can be used to model/study the Dirac oscillator \cite{Moshinsky,Franco,Oliveira1,Oliveira2,Cunha}, Aharonov-Bohm and Aharonov-Casher effects \cite{Oliveira1,Oliveira2,Cunha,Aharonov,Hagen}, Aharonov-Bohm quantum rings \cite{Oliveira3,Oliveira4}, cold atoms \cite{Boada}, trapped ions \cite{Lamata}, quantum Hall effect \cite{Schakel,Miransky,Oliveira5}, quantum computing \cite{Huerta}, quantum phase transitions \cite{Bermudez1}, mesoscopic superposition states \cite{Bermudez2}, neutrinos \cite{Oliveiraneutrinos}, Dirac materials (graphene, fullerenes, metamaterials, semimetals, carbon nanotubes, and topological insulators) \cite{Novoselov,Gonzalez,McCann,Ahrens,Armitage,Hasan,Qi}, wormhole \cite{Guvendi}, and so on.

Besides, the DE in curved spacetimes or black holes also has many applications, such as in the cosmic string spacetime \cite{Oliveira1,Oliveira4,Oliveira5,Oliveiraneutrinos,Oliveira6,Bakke,O}, Gödel-type spacetime \cite{Oliveira7,Ahmed}, global monopole spacetime \cite{Bragança}, Schwarzschild black hole \cite{Cho}, Kerr black hole \cite{Dolan}, BTZ black hole \cite{Hendi}, Reissner-Nordstrom black hole \cite{Finster}, regular Bardeen black hole \cite{Badawi}, and so on. On the other hand, in quantum electrodynamics (QED), the so-called anomalous magnetic moment (AMM) is a dimensionless physical quantity that arises from loop Feynman diagrams for fermions interacting with strong external electromagnetic fields \cite{Greiner,Bjorken,Grandy,Griffiths,Halzen,Thomson,Broderick,Bautista,Behncke,Dyck}, i.e., it is a type of additional magnetic contribution to the total magnetic dipole moment (MDM) of the fermion \cite{Bubnov}. In this context, the DE (or Dirac Lagrangian) for charged fermions with AMM in external magnetic fields is also very relevant for physics since it allows the investigation of several interesting problems, for example, the quantum Hall effect \cite{Oliveira5}, axial-vector condensate \cite{Bubnov}, thermodynamics of neutrons \cite{Ferrer}, synchrotron radiation \cite{Frolov}, neutral dense quark matter \cite{Kawaguchi}, compact stars \cite{Paulucci}, electron-positron vacuum \cite{Rodionov}, electroweak matter \cite{Dvornikov}, De Sitter geometry \cite{Struckmeier}, and so on. It is also important to mention that the first papers to calculate the energy spectrum of neutral and charged fermions with AMM were Refs. \cite{Ternov,Connell}.

The chiral magnetic effect (CME) is a quantum phenomenon in which an electromagnetic current is induced (or generated) along the direction of an external magnetic field when there is an imbalance between right-handed and left-handed chiral fermions ($\mu_R\neq \mu_L$), i.e., due to electromagnetic axial anomaly \cite{Fukushima,Fuk} (see also Refs. \cite{Kharzeev,Warringa} for a pictorial representation of the CME). In other words, if a magnetic field is applied to a system with an asymmetry between the number of right-handed and left-handed fermions ($n_R\neq n_L$), an electric current is induced along the magnetic field \cite{Fukushima,Fuk}. Consequently, this effect breaks/violates parity (P) and charge-parity (CP) symmetry. Already in Ref. \cite{Li}, the CME is a macroscopic manifestation of the quantum anomaly in relativistic field theory of chiral fermions (i.e., massless spin 1/2-particles with a definite projection of spin on momentum). Also, according to \cite{Li}, the CME was observed experimentally in ZrTe$_5$ from the effective transmutation of a Dirac semimetal into a Weyl semimetal. So, defining the right-handed and left-handed chemical potential as $\mu_R=\mu+\mu_5$ and $\mu_L=\mu-\mu_5$, where $\mu$ denotes the quark chemical potential, we can define another chemical potential, so-called chiral chemical potential, denoted by $\mu_5$, with $\mu_5\equiv(\mu_R-\mu_L)/2\neq 0$, i.e., this new potential (in which it describes the CME) couples to the difference between the number of right-handed and left-handed fermions \cite{Fukushima,Fuk}. In particular, the induced current of the CME is given by $\vec{J}_{CME}=\frac{e^2\mu_5}{2\pi^2}\vec{B}$, where $e$ and $\vec{B}$ are the electric charge and the external magnetic field, while the Lagrangian (density) of the CME is given by $\mathcal{L}_{CME}=\mathcal{L}_{int}+\mathcal{L}_{\mu_5}$, being $\mathcal{L}_{int}$ the interaction Lagrangian or the Dirac Lagragian with minimal coupling (or covariant derivative) and $\mathcal{L}_{\mu_5}=\mu_5\bar{\psi}\gamma^0\gamma^5\psi$ is the ``chiral Lagrangian’’, where $\gamma^0$ and $\gamma^5$ are the gamma matrices (there are others), and $\psi$ and $\bar{\psi}$ are the adjoint and Dirac spinors \cite{Fukushima,Fuk,Greiner,Yamamoto}.

Besides, the chiral chemical potential (has no sign problem), which is a more static quantity, is the energy necessary to put a right-handed quark on its Fermi surface or to remove a left-handed quark from its Fermi surface, that is, to change a left-handed fermion in a right-handed fermion requires removing a particle from the left-handed Fermi surface and adding it to the right-handed Fermi surface \cite{Fukushima,Fuk} (in this reference, was computed the magnitude of the induced current as a function of magnetic field, chirality, temperature, and baryon chemical potential as well as the relativistic Landau levels for the CME via DE). So, since it was introduced into the literature, the CME has been studied/applied in different physical contexts/problems, such as in the Early Universe plasma \cite{Dvornikov2}, hydrodynamical \cite{Isachenkov}, nuclear collisions \cite{Kharzeev2,Li2}, lattice gauge theory \cite{Buividovich}, fluid-gravity model \cite{Kalaydzhyan}, protoneutron stars \cite{Sigl}, Nambu–Jona–Lasinio (NJL) model \cite{Fukushima2,Ghosh1,Ghosh2,Farias,Pasqualotto}, Wigner functions \cite{Sheng}, and so on. It is interesting to mention that the chiral chemical potential $\mu_5$ is also the basis for modeling another remarkable physical effect, the so-called Chiral Vortical Effect (CVE) \cite{Kharzeev2}. On the other hand, besides the CME, magnetic fields can influence many quantum chromodynamics (QCD) processes, for example, dynamical chiral symmetry breaking \cite{Gusynin}, quark/chiral condensate \cite{Shushpanov,Cohen}, quark–hadron/chiral phase transition \cite{Fraga,Agasian}, color superconductor \cite{Alford,Ferrer2,Fukushima3}, anomalous axion interactions \cite{Metlitski}, $\pi^0$-domain walls \cite{Son}, and so on.

In recent years, charged fermions with AMM have been studied in the presence of the CME \cite{Chaudhuri}. In this work, the authors (based on Refs. \cite{Connell,Fukushima,Sheng}) incorporated the AMM of quarks in the framework of the Polyakov loop extended Nambu–Jona–Lasinio (PNJL) model to study hot and dense magnetised matter with chiral imbalance, i.e., under the influence of the CME. For this purpose, the relativistic energy spectrum (or relativistic eigenenergy) of the Dirac equation was obtained in the presence of a constant background magnetic field and chiral chemical potential, along with the AMM of the fermion. However, although there is a marginal enhancement in the inverse magnetic catalysis (IMC) behaviour of the quark condensate due to the combined effects of AMM and chiral chemical potential, the authors found that the overall behaviour of the Polyakov loop and the chiral charge density is dominated by the chiral chemical potential. Consequently, the authors showed that the AMM effects in the presence of the CME remain insignificant even after consideration of thermo-magnetically modified moments.

The concept of noncommutative (NC) spaces, or NC spacetimes, first emerged in 1947 through two papers by the American physicist Hartland Snyder on quantized spacetimes (that is, a spacetime where position no longer commutes with position or with time) \cite{Snyder1,Snyder2}. For Snyder, although the spacetime of special relativity (i.e., Minkowski spacetime) is a continuum, this assumption is not technically/obligatorily required by Lorentz invariance. Therefore, Snyder proposed/defined a model of a discrete Lorentz-invariant spacetime inspired by the commutation relations of quantum mechanics. According to Refs. \cite{Moffat,Majid,Szabo1,Addazi}, a NC spacetime could be considered a possible scenario for the short-range (Planck length) behavior of some physical theories, such as quantum gravity (Planck-scale gravity). Thus, NC spacetimes could provide a natural background for a possible regularization of a quantum field theory (QFT) for gravity \cite{Ho,Bertolami1}. Posteriorly, a more general concept was introduced into the literature, the so-called NC phase space (now momentum no longer commutes with momentum), which obeys a formalism more rigorous, that is, the NC geometry (NCG) \cite{Seiberg,Gomis,Douglas,Szabo,Djemai}. In essence, the NC phase space must satisfy the following commutation relations: $[x^\mu_{NC}, x^\nu_{NC}]=i\theta^{\mu\nu}$ and $[p^\mu_{NC}, p^\nu_{NC}]=i\eta^{\mu\nu}$ ($\mu,\nu=0,1,2,3$), where $x^\mu_{NC}$ and $p^\mu_{NC}$ are the NC position and momentum four-vectors (position and momentum are NC variables), and $\theta^{\mu\nu}$ and $\eta^{\mu\nu}$ are antisymmetric constant tensors with dimensions of $(length)^2$ and $(momentum)^2$, respectively (however, for $\eta^{\mu\nu}=0$, we have only the NC spacetime) \cite{Moffat,Bertolami1}. From a phenomenological point of view, supposed signatures of NC were investigated in the decay of kaons and vector bosons \cite{Melic,Buric}, photon-neutrino interaction \cite{Schupp}, vacuum birefringence \cite{Abel}, and quantum optics \cite{Pikovski}.

Besides, NC (phase) spaces have already been applied in several areas of physics, such as in string theory \cite{Seiberg}, QCD \cite{Carlson}, QED \cite{Riad}, black holes \cite{Nicolini}, quantum cosmology \cite{Bastos1}, and standard model \cite{Calmet}, as well as in several problems involving nonrelativistic quantum mechanics (NRQM) and RQM, such as in the gravitational quantum well \cite{Berto}, quantum Hall effect (or Landau system) \cite{Dayi,Gangopadhyay}, 3D and 2D quantum harmonic oscillator \cite{Gamboa,Nascimento,Santos}, Fisher information and Shannon entropies \cite{Nascimento}, hydrogen and muonic hydrogen atomic spectra \cite{Chaichian,Haghighat}, Landau analog levels for electric and magnetic dipoles \cite{Ribeiro,Yu}, modified Coulomb plus inverse-square potential \cite{Maireche}, Klein-Gordon oscillator \cite{Hassanabadi1,Maluf}, Duffin-Kemmer-Petiau oscillator \cite{Yang}, and so on. Now, considering especially the DE in an NC (phase) space, we can investigate problems involving the quantum Hall effect with AMM in a conical Gödel-type spacetime and in a rotating frame in the spinning cosmic string spacetime \cite{Oliveira5,Oliveira8}, thermodynamic properties \cite{Oliveira9}, Dirac oscillator with electric dipole moment \cite{Oliveira10,Oliveira11}, hydrogen atom spectrum and the Lamb shift \cite{Adorno,Alavi}, 
crossed electric and magnetic fields \cite{Nath}, minimal length \cite{Boumali}, 
zitterbewegung (ZBW) phenomenon \cite{Abyaneh}, Aharonov-Bohm, Aharonov-Casher, He–Mckellar–Wilkens and Anandan effects \cite{Mirza,Jing1,Jing2,Ma1}, persistent charged current \cite{Ma2}, Spavieri effect \cite{Ma3}, modified Eckart plus Hulthen potentials \cite{Maireche2}, graphene \cite{Bastos}, and so on. Recently, Ref. \cite{Qolizadeh} studied the thermodynamics of the NC Dirac oscillator with the Mie-Type Potential, while Ref. \cite{Halder} studied the gauge-invariant description of a Dirac electron moving in a NC gauge field background.

The present paper has as its goal to analyze the relativistic energy spectrum (or relativistic Landau levels) for charged Dirac fermions (electron/positron) with AMM in the presence of the CME and of a NC phase space. Therefore, we will work with the DE in cylindrical coordinates (i.e., ``cylindrical DE'') in the $(3+1)$-dimensional Minkowski spacetime (or four-dimensional flat spacetime). In other words, we will obtain the NC version of the spectrum from Ref. \cite{Chaudhuri}, that is, to investigate the effects/influence of NC on the usual spectrum. However, unlike \cite{Chaudhuri}, here, we will use a similarity transformation (given by a unitary operator/matrix $U$) to convert the curvilinear/cylindrical gamma matrices into the Cartesian/usual gamma matrices \cite{Oliveira5,Oliveira2,Schluter,Villalba} (in fact, this is necessary since we want to solve exactly and analytically the DE). Consequently, we will generalize not only the spectrum of Ref. \cite{Chaudhuri} but also of other references, that is, excluding the NC as well as other parameters in the spectrum (and also adjusting/redefining the quantum numbers), we will be able to obtain several particular cases from the literature. Besides, for the sake of avoiding the problem of unitarity (i.e., violation of causality), here, we will only work on the case of the space-like NC (or simply space NC), where $\theta^{0i}=\eta^{0i}=0$ ($i=1,2,3$) \cite{Szabo,Gomis} (in general this is what is done in NRQM and RQM \cite{Berto,Dayi,Gangopadhyay,Gamboa,Nascimento,
Santos,Chaichian,Haghighat,Ribeiro,Yu,Maireche,Hassanabadi1,Maluf,Yang,Oliveira5,Oliveira8,Oliveira9,Oliveira10,Oliveira11,Adorno,Alavi,Nath,Boumali,Abyaneh,Mirza,Jing1,Jing2,Ma1,Ma2,Ma3,Maireche2,Bastos,Qolizadeh,Halder}), and for simplicity, we will only consider the NC along the $z$-axis (as it will also be for the external magnetic field), that is, the NC vectors associated with position and momentum will be given by $\vec{\theta}=(0,0,\theta)=\theta\vec{e}_z$ ($\theta=\theta_3$) and $\vec{\eta}=(0,0,\eta)=\eta\vec{e}_z$ ($\eta=\eta_3$) \cite{Maluf,Nascimento,Santos,Yang}, where $\theta\geq 0$ and $\eta\geq 0$ are the position and momentum NC parameters, respectively. 

The structure of this paper is organized as follows. In Sect. \ref{sec2}, we introduce the noncommutative Dirac equation of the system (or better, for the electron and positron), where we use a similarity transformation (to simplify the problem), and we obtain four coupled first-order differential equations. Subsequently, we obtain four non-homogeneous second-order differential equations, in which each equation depends on two of the four components of the spinor. In Sect. \ref{sec3}, we solve exactly and analytically these four differential equations through a change of variable, the asymptotic behavior, and the Frobenius method (power-series method). Consequently, we obtain the relativistic energy spectrum (or relativistic Landau levels) for the electron/positron with AMM in the presence of the CME and of a NC phase space, where we discuss in detail several characteristics of this spectrum. Comparing our spectrum with other works/papers, we verified that it generalizes several particular cases found in the literature. In Sect. \ref{sec4}, we graphically analyze (via 2D graphs) the behavior of the spectrum (of the electron/positron) as a function of the magnetic field $B$, fermion and chiral chemical potential $\mu$ and $\mu_5$, $z$-momentum $k_z$, and the position and momentum NC parameters $\theta$ and $\eta$ for three different values of the radial quantum number $n$ and the angular quantum number $m_j$. In Sect. \ref{sec5}, we present our conclusions. Some more detailed calculations on how the energy spectrum was obtained are provided in the Appendix \ref{sec6}. Here, we use the natural units system $(\hslash=c=G=1)$ and the spacetime (or metric) with signature $(+,-,-,-)$.

\section{The noncommutative Dirac equation for a charged fermion with anomalous magnetic moment in the presence of the chiral magnetic effect \label{sec2}}

The Dirac Lagrangian (density) for spin-1/2 charged fermions with AMM in the presence of the CME is given by \cite{Chaudhuri,Greiner,Bjorken,Grandy,Oliveira5,Oliveira8,Sheng}
\begin{equation}\label{dirac1}
\mathcal{L}=\mathcal{L}_{CME}=\bar{\psi}\left[i\gamma^\mu \left(\partial_\mu+iq A_\mu^{ext}\right)-m+\mu \gamma^0+\mu_5\gamma^0 \gamma^5+\frac{1}{2}\mu_m\sigma^{\mu\nu}F_{\mu\nu}^{ext}\right]\psi, \ \ (\mu,\nu=0,1,2,3),
\end{equation}
where $\gamma^{\mu}=g^{\mu\nu}\gamma_\nu=(\gamma^0,\gamma^i)=(\gamma^0,\vec{\gamma})$ are the gamma matrices, in which must satisfy the anticommutation relation of the Clifford algebra, given by $\{\gamma^\mu,\gamma^\nu\}=2g^{\mu\nu}I_4$ (or $\{\gamma_\mu,\gamma_\nu\}=2g_{\mu\nu}I_4$), being $g^{\mu\nu}=g_{\mu\nu}$=diag$(+1,-1,-1,-1)$ the Minkowski metric tensor (or simply metric) and $I_4$ the $4\times 4$ identity matrix (i.e., $I_4$=diag$(1,1,1,1)$), $\partial_\mu=\partial/\partial x^\mu=(\partial_0,\partial_i)=(\partial_t,\vec{\nabla})$ are the partial derivatives (gradient four-vector or four-dimensional gradient operator), being $x^\mu=g^{\mu\nu}x_\nu=(x^0,x^i)=(t,\vec{x})$ the position four-vector (or four-dimensional position operator), $\sigma^{\mu\nu}=(i/2)[\gamma^\mu,\gamma^\nu]=i\gamma^\mu\gamma^\nu$ ($\mu\neq \nu$) is an antisymmetric tensor formed by the gamma matrices, $F_{\mu\nu}=\partial_\mu A_\nu^{ext}-\partial_\nu A_\mu^{ext}$ is the electromagnetic field tensor (or field strength tensor), being $A_\mu^{ext}=(A_0,A_i)=(0,-\vec{A})$ the external electromagnetic field (electromagnetic potential four-vector or simply potential four-vector), $\gamma^5=\gamma_5=i\gamma^0\gamma^1\gamma^2\gamma^3$ is the five gamma matrix (or chirality matrix/operator), $\mu$ and $\mu_5$ are the fermion and chiral chemical potential (or fermion-number and chiral-charge chemical potential \cite{Sheng}), $\mu_m=a\mu_B$ \cite{Grandy,Connell,Oliveira5,Oliveira8} is the (anomalous) magnetic dipole moment (or simply magnetic dipole) of the fermion, being $a$ the AMM and $\mu_B$ the Bohr magneton (in \cite{Greiner,Bjorken} $\kappa$ is used instead of $a$ to symbolize the AMM), $m=m_{rest}=m_0>0$ and $q=\pm e$ (with $e>0$) are the rest mass and electric charge (negative or positive) of the fermion (or antifermion), and $\bar{\psi}=\bar{\psi}(\vec{x},t)$ and $\psi=\psi(\vec{x},t)$ are the four-component adjoint and Dirac spinors, respectively. However, instead of $\partial_\mu$ we also could use the momentum four-vector (or four-dimensional momentum operator), given by $p_\mu=i\partial_\mu=(p_0,p_i)=(i\partial_t,-\vec{p})$. In particular, in the case of the electron ($q_{e^-}=-e<0$), $a$ is given by $a_{e^-}=0.0011596521884$, while its antiparticle, that is, the positron ($q_{e^+}=+e>0$), has a very close value, given by $a_{e^+}=0.0011596521879$ (i.e., one of the most accurately verified predictions in the history of physics) \cite{Dyck}. Already in the case of the mass, it is the same for both (i.e., $m_{e^-}=m_{e^+}=m$). So, in the standard/Dirac representation, the gamma matrices are written as \cite{P1,P2,Chaudhuri,Greiner,Bjorken,Grandy,Thomson,Griffiths,Halzen}
\begin{eqnarray}\label{gammamatrices}
&& \gamma^0=\left(
\begin{array}{cc}
 I_2 & 0 \\
 0 & -I_2 \\
\end{array}
\right)=\left(
\begin{array}{cccc}
 1 & 0 & 0 & 0 \\
 0 & 1 & 0 & 0 \\
 0 & 0 & -1 & 0 \\
 0 & 0 & 0 & -1
\end{array}
\right),
\ \
\gamma^5=\left(\begin{array}{cc}
0 & I_2 \\
I_2 & 0 
\end{array}\right)=\left(
\begin{array}{cccc}
 0 & 0 & 1 & 0 \\
 0 & 0 & 0 & 1 \\
 1 & 0 & 0 & 0 \\
 0 & 1 & 0 & 0
\end{array}
\right),
\nonumber\\
&& \gamma^i=\left(\begin{array}{cc}
0 & \sigma^i \\
-\sigma^i & 0 
\end{array}\right)=\begin{cases}
\gamma^1=\left(
\begin{array}{cccc}
 \ 0 & \ 0 & 0 & 1 \\
 \ 0 & \ 0 & 1 & 0 \\
 \ 0 & -1 & 0 & 0 \\
 -1 & \ 0 & 0 & 0
\end{array}
\right), \\
\gamma^2=\left(
\begin{array}{cccc}
 \ 0 & 0 & 0 & -i \\
 \ 0 & 0 & i & \ 0 \\
 \ 0 & i & 0 & \ 0 \\
 -i & 0 & 0 & \ 0
\end{array}
\right), \\
\gamma^3=\left(
\begin{array}{cccc}
 \ 0 & 0 & 1 & \ 0 \\
 \ 0 & 0 & 0 & -1 \\
 -1 & 0 & 0 & \ 0 \\
 \ 0 & 1 & 0 & \ 0
\end{array}
\right),
\end{cases}
\end{eqnarray}
where $\sigma^1$, $\sigma^2$, and $\sigma^3$ are the $2\times 2$ Pauli spin matrices, i.e., $\vec{\sigma}=(\sigma^1,\sigma^2,\sigma^3)=(\sigma_x,\sigma_y,\sigma_z)$, and $\gamma^0$ and $\gamma^i$ (or $\vec{\gamma}$=off-diag$(\vec{\sigma},-\vec{\sigma})$) can be defined in terms of the standard/usual Dirac matrices as $\gamma^0=\gamma_0\equiv\beta$, and $\gamma^i\equiv\beta \alpha^i$ or $\vec{\gamma}\equiv\beta\vec{\alpha}$ (with $\gamma^1=\beta\alpha^1=\beta \alpha_x$, $\gamma^2=\beta\alpha^2=\beta \alpha_y$, and $\gamma^3=\beta\alpha^3=\beta \alpha_z$), being $\vec{\alpha}=(\alpha^1,\alpha^2,\alpha^3)=(-\alpha_1,-\alpha_2,-\alpha_3)=(\alpha_x,\alpha_y,\alpha_z)$, in which also must satisfy the anticommutation relation of the Clifford algebra, given by $\{\alpha^i,\alpha^j\}=2\delta^{ij}I_4$.

However, knowing that the Euler-Lagrange equation is given by \cite{Greiner}\label{EulerLagrangeequation}
\begin{equation}
\partial_\mu\frac{\partial\mathcal{L}}{\partial (\partial_\mu\bar{\psi})}-\frac{\partial\mathcal{L}}{\partial\bar{\psi}}=0,
\end{equation}
we then obtain the following tensorial/covariant DE for a negatively charged fermion (electron) with AMM in the presence of the CME \cite{Chaudhuri,Greiner,Bjorken,Grandy,Thomson}
\begin{equation}\label{dirac2}
\left[i\gamma^\mu \left(\partial_\mu-ieA_\mu^{ext}\right)-m+\mu \gamma^0+\mu_5\gamma^0 \gamma^5+\frac{1}{2}\mu_m\sigma^{\mu\nu}F_{\mu\nu}^{ext}\right]\psi=0.
\end{equation}

Before proceeding, it is interesting to obtain and comment a little on the DE for the antifermion, that is, for the positron (which is the antiparticle of the electron, i.e., a ``positive electron''). In particular, this can be easily done using Refs. \cite{Greiner,Thomson,Halzen}. However, before this is done, it is convenient to rewrite Eq. \eqref{dirac2} as follows
\begin{equation}\label{D1}
\left[\gamma^\mu\left(\partial_\mu-ieA_\mu^{ext}\right)+im-i\mu \gamma^0-i\mu_5\gamma^0\gamma^5+\frac{1}{2}\mu_m\gamma^\mu\gamma^\nu F_{\mu\nu}^{ext}\right]\psi=0,
\end{equation}
where we use the fact that $\sigma^{\mu\nu}=\frac{i}{2}[\gamma^\mu,\gamma^\nu]=\frac{i}{2}[\gamma^\mu\gamma^\nu-\gamma^\nu\gamma^\mu]=\frac{i}{2}[\gamma^\mu\gamma^\nu+\gamma^\mu\gamma^\nu]=i\gamma^\mu\gamma^\nu$. So, according to Refs. \cite{Greiner,Thomson,Halzen} (and especially \cite{Thomson}), taking the complex
conjugate of Eq. \eqref{D1}, we have
\begin{equation}\label{D2}
\left[(\gamma^\mu)^\star\left(\partial_\mu+ieA_\mu^{ext}\right)-im+i\mu \gamma^0+i\mu_5\gamma^0\gamma^5+\frac{1}{2}\mu_m(\gamma^\mu)^\star(\gamma^\nu)^\star F_{\mu\nu}^{ext}\right]\psi^\star=0.
\end{equation}

Now, multiplying on the left the above equation by $-i\gamma^2$ \cite{Thomson}, we have
\begin{equation}\label{D3}
\left[-i\gamma^2(\gamma^\mu)^\star\left(\partial_\mu+ieA_\mu^{ext}\right)-m\gamma^2+\mu \gamma^2\gamma^0+\mu_5\gamma^2\gamma^0\gamma^5-\frac{i}{2}\mu_m\gamma^2(\gamma^\mu)^\star(\gamma^\nu)^\star F_{\mu\nu}^{ext}\right]\psi^\star=0.
\end{equation}

Knowing that in the Dirac representation the gamma matrices satisfy: $(\gamma^0)^\star=\gamma^0$, $(\gamma^1)^\star=\gamma^1$, $(\gamma^2)^\star=-\gamma^2$, and $(\gamma^3)^\star=\gamma^3$, with $\gamma^2\gamma^\mu=-\gamma^\mu\gamma^2$ (for $\mu\neq 2$) and $\gamma^\mu\gamma^5=-\gamma^5\gamma^\mu$ (for $\mu=0,1,2,3$) \cite{Thomson}, and using $D_\mu=\partial_\mu+ieA_\mu^{ext}$, implies that
\begin{align}\label{impliesthat}
-i\gamma^2(\gamma^\mu)^\star D_\mu&=-i\gamma^2(\gamma^0)^\star D_0-i\gamma^2(\gamma^1)^\star D_1-i\gamma^2(\gamma^2)^\star D_2-i\gamma^2(\gamma^3)^\star D_3,
\nonumber\\
&=i\gamma^0\gamma^2 D_0+i\gamma^1\gamma^2 D_1+i\gamma^2\gamma^2 D_2+i\gamma^3\gamma^2 D_3,
\nonumber\\
&=i\gamma^\mu \gamma^2 D_\mu=i\gamma^\mu D_\mu \gamma^2,
\nonumber\\
\mu\gamma^2\gamma^0 &=-\mu\gamma^0\gamma^2,
\nonumber\\
\mu_5\gamma^2\gamma^0\gamma^5 &=-\mu_5\gamma^0\gamma^2\gamma^5=\mu_5\gamma^0\gamma^5\gamma^2,
\nonumber\\
-i\gamma^2(\gamma^\mu)^\star(\gamma^\nu)^\star F_{\mu\nu}^{ext} &=i\gamma^\mu\gamma^2(\gamma^\nu)^\star F_{\mu\nu}^{ext}=-i\gamma^\mu \gamma^\nu\gamma^2 F_{\mu\nu}^{ext}=i\gamma^\mu \gamma^\nu F_{\mu\nu}^{ext}\gamma^2.
\end{align}

Consequently, Eq. \eqref{D3} becomes
\begin{equation}\label{D4}
\left[i\gamma^\mu\left(\partial_\mu+ieA_\mu^{ext}\right)-m-\mu\gamma^0+\mu_5\gamma^0\gamma^5+\frac{i}{2}\mu_m\gamma^\mu\gamma^\nu F_{\mu\nu}^{ext}\right]\gamma^2\psi^\star=0,
\end{equation}
or yet
\begin{equation}\label{D5}
\left[\gamma^\mu\left(\partial_\mu+ieA_\mu^{ext}\right)+im+i\mu\gamma^0-i\mu_5\gamma^0\gamma^5+\frac{1}{2}\mu_m\gamma^\mu\gamma^\nu F_{\mu\nu}^{ext}\right](i\gamma^2\psi^\star)=0.
\end{equation}

So, organizing the equation above (and using $\sigma^{\mu\nu}$), we obtain the following DE for the positron (with AMM in the presence of the CME)
\begin{equation}\label{D6}
\left[i\gamma^\mu\left(\partial_\mu+ieA_\mu^{ext}\right)-m-\mu\gamma^0+\mu_5\gamma^0\gamma^5+\frac{1}{2}\mu_m\sigma^{\mu\nu}F_{\mu\nu}^{ext}\right]\psi'=0,
\end{equation}
where we define $\psi'=\psi_C\equiv i\gamma^2\psi^\star=C\psi$, which can be interpreted as the antiparticle wavefunction, or better, the positron spinor ($\psi'=\psi_{e^+}$), or simply the Dirac spinor of the positron (while $\psi=\psi_{e^-}$ is the Dirac spinor of the electron, i.e., the antiparticle spinor explicitly depends on the complex conjugate of the particle spinor), and $C$ is the charge-conjugation operator (which transforms a particle spinor into the corresponding antiparticle spinor, or yet, is a symmetry operation that replaces a particle by its antiparticle). In particular, the (positive energy) antiparticle spinor is defined/expressed in terms of the negative energy particle solutions (or negative energy particle spinor) of the DE \eqref{dirac2} \cite{Thomson}. Besides, the last expression in \eqref{impliesthat} can also be seen as: $\frac{1}{2}\mu_m\bar{\psi}\sigma^{\mu\nu}F^{ext}_{\mu\nu}\psi\xrightarrow{C}\frac{1}{2}\mu_m\bar{\psi}'(-\sigma^{\mu\nu})(-F^{ext}_{\mu\nu})\psi'=\frac{1}{2}\mu_m\bar{\psi}'\sigma^{\mu\nu}F^{ext}_{\mu\nu}\psi'$, i.e., the Lagragian of the AMM is invariant under charge conjugation \cite{Greiner,Thomson}. In this way, comparing \eqref{D6} with \eqref{dirac2}, we see that the DE for $\psi'$ differs from the DE for $\psi$ just by the sign of the charge and the fermion chemical potential. Therefore, the electric charge and the chemical potential of the positron are opposite to that of the electron ($q_{e^+}=-q_{e^-}=+e$ and $\mu_{e^+}=-\mu_{e^-}=-\mu$ \cite{Thomas}), i.e., the electron has negative charge and positive chemical potential, while the positron has positive charge and negative chemical potential, respectively. However, since our focus here is exclusively on the energy spectrum of the fermion/antifermion (electron/positron), we can write \eqref{D6} and \eqref{dirac2} into a single equation, such as (for convenience, we will still use $\psi$)
\begin{equation}\label{D7}
\left[i\gamma^\mu\left(\partial_\mu+iqA_\mu^{ext}\right)-m+\mu\gamma^0+\mu_5\gamma^0\gamma^5+\frac{1}{2}\mu_m\sigma^{\mu\nu}F_{\mu\nu}^{ext}\right]\psi=0,
\end{equation}
where $q=q_{e^+}=+e$ is for the positron, and $q=q_{e^-}=-e$ is for the electron, respectively.
It is important to highlight that we prefer to continue using $\psi$ instead of $\psi'$ for a good reason that we will see in the next section, that is, the positive energies of DE \eqref{D7} will be the electron energies, while the negative energies will be used to build the positive energies of the positron, i.e., $E_{positron}=-E_{electron \ with \ negative \ energy}^{q_{e^-}\to q_{e^+}}>0$ \cite{Greiner} (in other words, the positron spinor should be a solution of \eqref{D6} with positive energy! \cite{Greiner}). A consequence of this is that $\mu$ will have the opposite sign to the electron in the spectrum. Therefore, here, we do not need to use $\pm\mu$ or $\mu \gtrless 0$ in \eqref{D7} (i.e., the minus sign in front of the electron spectrum with negative energy (and $-e\to+e$) already plays this role).

So, knowing that \cite{Greiner,Bjorken,Grandy,Thomson}: $i\gamma^\mu (\partial_\mu+iqA_\mu^{ext})=i\gamma^0\partial_0+i\gamma^i (\partial_i+iqA_i)=i\gamma^0\partial_t+\gamma^i (p_i-qA_i)=i\gamma^0\partial_t-\gamma^i(p^i-qA^i)=i\gamma^0\partial_t-\vec{\gamma}\cdot(\vec{p}-q\vec{A})$, and $\frac{1}{2}\mu_m\sigma^{\mu\nu}F_{\mu\nu}^{ext}=i\mu_m\gamma^\mu\gamma^\nu F_{\mu\nu}^{ext}=\mu_m (i\vec{\alpha}\cdot\vec{E}-\vec{S}\cdot\vec{B})=-\mu_m \vec{\Sigma}\cdot\vec{B}=-2\mu_m\vec{S}\cdot\vec{B}$, we obtain
\begin{equation}\label{dirac3}
\left[i\gamma^0\partial_t-\vec{\gamma}\cdot(\vec{p}-q\vec{A})-m+\mu\gamma^0+\mu_5\gamma^0\gamma^5-2\mu_m \vec{S}\cdot\vec{B}\right]\psi=0,
\end{equation}
or yet
\begin{equation}\label{dirac4}
\left[i\gamma^0\partial_t-\vec{\gamma}\cdot\left(\vec{p}-\frac{q}{2}\vec{B}\times\vec{x}\right)-m+\mu\gamma^0+\mu_5\gamma^0\gamma^5-2\mu_m \vec{S}\cdot\vec{B}\right]\psi=0,
\end{equation}
where $\vec{E}$ and $\vec{B}$ are the electric and magnetic field vectors (here $\vec{E}=0$), $\vec{S}=\frac{1}{2}\vec{\Sigma}$ is the $4\times 4$ spin operator/vector, being $\vec{\Sigma}$ a matrix (``$4\times 4$ Pauli matrices'') given by $\vec{\Sigma}=(\Sigma^1,\Sigma^2,\Sigma^3)$=diag$(\vec{\sigma},\vec{\sigma})$, or in index notation, as $\Sigma^i=\frac{i}{2}\varepsilon^{ijk}\gamma_j \gamma_k$ ($i,j,k=1,2,3$) \cite{Oliveira10}, and we use the fact that the vector potential is given by $\vec{A}=\frac{1}{2}\vec{B}\times\vec{x}$ (or $A_i=\frac{1}{2}\varepsilon_{ijk}B^j x^k$).

Now, let us obtain the noncommutative Dirac equation (NCDE). So, in a NC phase space \cite{Oliveira5,Oliveira8,Oliveira10}, Eq. \eqref{dirac4} becomes 
\begin{equation}\label{dirac5}
\left[i\gamma^0\partial_t-\vec{\gamma}\cdot\left(\vec{p}_{NC}-\frac{q}{2}\vec{B}\times\vec{x}_{NC}\right)-m+\mu\gamma^0+\mu_5\gamma^0 \gamma^5-2\mu_m \vec{S}\cdot\vec{B}\right]\star\psi=0,
\end{equation}
where $x_i^{NC}$ and $p_i^{NC}$ are the NC position and momentum operators (in index notation), in which they must satisfy the commutation relations $[x_i^{NC},x_j^{NC}]=i\theta_{ij}$ and $[p_i^{NC},p_j^{NC}]=i\theta_{ij}$ \cite{Maluf,Abyaneh,Nascimento,Santos,Yang}, and the symbol $\star$ is the so-called star product or Moyal product, and now the product of two arbitrary functions $F(\vec{x})$ and $G(\vec{x})$ is given by
\begin{equation}
F(\vec{x},\theta)\star G(\vec{x},\theta)=F(\vec{x})e^{(i/2)(\overleftarrow{\partial}_{x_i}\theta_{ij}\overrightarrow{\partial}_{x_j})}G(\vec{x}).
\end{equation}

So, knowing that the NC operators $x_{i}^{NC}$ and  $p^{NC}_{i}$ are written as follows (in index and vector notation) \cite{Maluf,Abyaneh,Nascimento,Santos,Yang}
\begin{eqnarray}\label{NCoperators}
&& x_i^{NC}=x_i-\frac{1}{2}\theta_{ij}p^j=x_i-\frac{1}{2}\varepsilon_{ijk}\theta^k p^j, \ \ \longleftrightarrow \ \ \vec{x}_{NC}=\vec{x}+\frac{\vec{\theta}\times\vec{p}}{2},
\nonumber\\
&& p^{NC}_{i}=p_i+\frac{1}{2}\eta_{ij}x^{j}=p_i+\frac{1}{2}\varepsilon_{ijk}\eta^{k}x^{j}, \ \ \longleftrightarrow \ \ \vec{p}_{NC}=\vec{p}-\frac{\vec{\eta}\times\vec{x}}{2}.
\end{eqnarray}

Therefore, using the above NC operators in \eqref{dirac5}, we get the following NCDE (with $\star\psi\to\psi_{NC}$)
\begin{equation}\label{dirac6.0}
\left[i\gamma^0\partial_t-\vec{\gamma}\cdot\left(\vec{p}-\frac{\vec{\eta}\times\vec{x}}{2}-\frac{q}{2}\vec{B}\times\left(\vec{x}+\frac{\vec{\theta}\times\vec{p}}{2}\right)\right)-m+\mu\gamma^0+\mu_5\gamma^0 \gamma^5-2\mu_m\vec{S}\cdot\vec{B}\right]\psi_{NC}=0,
\end{equation}

On the other hand, knowing that in cylindrical coordinates $(r,\phi,z)$ the operators $\vec{p}$ and $\vec{x}$, NC vectors $\vec{\eta}$ and $\vec{\theta}$, and the magnetic field $\vec{B}$ are written as follows
\begin{eqnarray}
&& \vec{p}=-i\vec{\nabla}=(p_r,p_\phi,p_z)=-i\left(\vec{e}_r\partial_r+\frac{\vec{e}_\phi}{r}\partial_\phi+\vec{e}_z\partial_z\right), \ \ \vec{x}=(r,0,z)=r\vec{e}_r+z\vec{e}_z,
\nonumber\\
&& \vec{\eta}=(0,0,\eta)=\eta\vec{e}_z, \ \ \vec{\theta}=(0,0,\theta)=\theta\vec{e}_z, \ \ \vec{B}=(0,0,B)=B\vec{e}_z,
\end{eqnarray}
with $0\leq r<\infty, 0\leq\phi\leq 2\pi$, and $-\infty<z<+\infty$, and using the vector triple product identity given by $\vec{B}\times(\vec{\theta}\times\vec{p})=\vec{\theta}(\vec{B}\cdot\vec{p})-\vec{p}(\vec{B}\cdot\vec{\theta})$, we obtain from \eqref{dirac6.0} the following equation
\begin{equation}\label{dirac6}
\left[i\gamma^0\partial_t+i\tau\left(\gamma^r\partial_r+\frac{\gamma^\phi}{r}\partial_\phi\right)+i\gamma^z\partial_z+\frac{q\lambda Br}{2}\gamma^\phi-m+\mu\gamma^0+\mu_5\gamma^0\gamma^5-\mu_m\Sigma^z B\right]\psi_{NC}=0,
\end{equation}
where $\gamma^r\equiv \vec{\gamma}\cdot\vec{e}_r=\gamma^1 \cos{\phi}+\gamma^2\sin{\phi}$, $\gamma^\phi\equiv \vec{\gamma}\cdot\vec{e}_\phi=-\gamma^1 \sin{\phi}+\gamma^2\cos{\phi}$, $\gamma^z\equiv \vec{\gamma}\cdot\vec{e}_z=\gamma^3$, and $\Sigma^z\equiv \vec{\Sigma}\cdot\vec{e}_z=\Sigma^3=\Sigma_3$ (since $\Sigma^3=i\gamma^1\gamma^2$=diag$(\sigma^3,\sigma^3)$) are the curvilinear gamma matrices (in the case of $\gamma^r$ and $\gamma^\phi$, it is as if they were ``rotated counterclockwise'' by an angle $\phi$ in the polar plane) \cite{Oliveira5,Oliveira2,Schluter,Villalba}, $\psi_{NC}=\psi_{NC}(t,r,\phi,z)$ is the NC spinor, and $\tau$ and $\lambda$ are dimensionless real parameters, defined as $\tau=\tau_q\equiv 1+\frac{qB\theta}{4}$ and $\lambda=\lambda_q\equiv 1+\frac{\eta}{qB}$ \cite{Oliveira8}. In particular, here, we consider that $\tau\neq 0$ and $\lambda\neq 0$, or better, $\tau>0$ and $\lambda>0$ \cite{Oliveira8,Oliveira9}; otherwise, i.e., if $\tau$ and $\lambda$ could be zero, then it would ``break'' the quantum dynamics of the system, i.e., it would not be possible to reach the bound states (or quantized spectrum). Of course, the parameters $\tau$ and $\lambda$ can also be negative \cite{Oliveira5} (i.e., the important thing is to always have $\tau\lambda>0$). But for simplicity and without loss of generality, here, we will consider them positive. Besides, the product $\tau\lambda$ for $q=-e$ and $\eta\to\bar{\theta}$ appears in the nonrelativistic spectrum (39) of Ref. \cite{Gangopadhyay} (we will see later that it will also appear in our relativistic spectrum).

So, we see that it is difficult to proceed (because of $\cos{\phi}$ and $\sin{\phi}$) without a simplification of Eq. \eqref{dirac6}. However, this obstacle can be perfectly removed through a similarity transformation (``clockwise rotation'') in the equation, where the unitary operator/matrix of such a transformation is given by $U=U(\phi)=e^{-i\phi\Sigma^3/2}$ ($U^\dagger U=I_4$) \cite{Oliveira5,Oliveira2,Schluter,Villalba}. In particular, this unitary operator aims to transform (``rotate'') the curvilinear gamma matrices (difficult to work with) into usual/Cartesian gamma matrices (easy to work with), that is,
\begin{equation}
U^{-1}\gamma^r U=\gamma^1, \ \ U^{-1}\gamma^\phi U=\gamma^2,
\end{equation}
where $U$ commutes with $\gamma^0$ ($U\gamma^0=\gamma^0U$), $\Sigma^3$ ($U\Sigma^3=\Sigma^3U$), $\gamma^3$ ($U\gamma^3=\gamma^3U$), and $\gamma^5$ ($U\gamma^5=\gamma^5U$).

In this way, Eq. \eqref{dirac6} becomes
\begin{equation}\label{dirac7}
\left[i\gamma^0\partial_t+i\tau\gamma^1\left(\partial_r+\frac{1}{2r}\right)+\gamma^2\left(\frac{i\tau}{r}\partial_\phi+\frac{q\lambda Br}{2}\right)+i\gamma^3\partial_z-m+\mu\gamma^0+\mu_5\gamma^0 \gamma^5-\mu_m\Sigma^3 B\right]\psi_{NC}=0,
\end{equation}
or in terms of the Dirac Hamiltonian $H_D=H_{q}=H_{\pm e}$, such as
\begin{equation}\label{dirac8}
i\partial_t\psi_{NC}=H_D\psi_{NC}, \ \ H_D=\left[-i\tau\gamma^0\gamma^1\left(\partial_r+\frac{1}{2r}\right)-\gamma^0\gamma^2\left(\frac{i\tau}{r}\partial_\phi+\frac{q\lambda Br}{2}\right)-i\gamma^0\gamma^3\partial_z+m\gamma^0-\mu-\mu_5 \gamma^5+\mu_m\gamma^0\Sigma^3 B\right],
\end{equation}
being $\varphi_{NC}$ a new spinor (``rotated spinor''), defined as $\varphi_{NC}\equiv U^{-1}\psi_{NC}=e^{i\phi\Sigma^3/2}\psi_{NC}$. In particular, $\varphi_{NC}$ and $\psi_{NC}$ must satisfy the following periodicity conditions: $\varphi_{NC}(\phi\pm 2\pi)=-\varphi_{NC}(\phi)$ and $\psi_{NC}(\phi\pm 2\pi)=\psi_{NC}(\phi)$ \cite{Oliveira5,Oliveira2,Schluter,Villalba}. Explicitly, Eq. \eqref{dirac6} can be written in the following matrix form (with $\Sigma^3$=diag$(\sigma^3,\sigma^3)$=diag$(+1,-1,+1,-1)$)
\begin{equation}\label{dirac9}
\left(
\begin{array}{cccc}
 -m-\mu_m B+i\partial_t+\mu & 0 & i\partial_z+\mu_5 & \chi^1_- \\
 0 & -m+\mu_m B+i\partial_t+\mu  & \chi^1_+ & -i\partial_z+\mu_5 \\
 -i\partial_z-\mu_5 & \chi^2_- & -m-\mu_m B-i\partial_t-\mu  & 0 \\
\chi^2_+ & i\partial_z-\mu_5 & 0 & -m+\mu_m B-i\partial_t-\mu
\end{array}
\right)\varphi_{NC}=0,
\end{equation}
where we define
\begin{equation}
\chi^1_\pm \equiv i\tau\left[\left(\partial_r+\frac{1}{2r}\right)\pm\left(\frac{i}{r}\partial_\phi+\frac{\sigma\lambda m\omega_cr}{2\tau}\right)\right], \ \ \chi^2_\pm \equiv -i\tau\left[\left(\partial_r+\frac{1}{2r}\right)\pm\left(\frac{i}{r}\partial_\phi+\frac{\sigma\lambda m\omega_cr}{2\tau}\right)\right],
\end{equation}
and we use $q=\sigma e$, being $\sigma=\pm 1$ a ``charge parameter'' ($\sigma=\sigma_{e^+}=+1$ for the positron, and $\sigma=\sigma_{e^-}=-1$ for the electron), where $\omega_c=eB/m\geq 0$ ($\vec{\omega}_c=\omega_c\vec{e}_z$) is the well-known/famous cyclotron frequency, which is the angular velocity or circular motion of the fermion in the plane (due to the interaction of its charge with the magnetic field).

Now, assuming that our system is a stationary system (i.e., we are seeking stationary bound states), we can then define an ansatz for the spinor $\varphi_{NC}$ in the following form \cite{Oliveira5,Oliveira2,Schluter,Villalba,O}
\begin{equation}\label{spinor}
\varphi_{NC}(r,\phi,z,t)=\frac{1}{\sqrt{2\pi}}e^{-iEt}e^{im_j\phi}e^{ik_z z}\left(
           \begin{array}{c}
            f_1(r) \\
            f_2(r) \\
            f_3(r) \\
            f_4(r) \\
           \end{array}
         \right),
\end{equation}
where $f_{1,2,3,4}(r)$ are radial functions (or spatial/radial components of the spinor), $E$ is the relativistic total energy (with $E\gtrless 0$ or $E=E^\pm=\pm\vert E^\pm\vert$) or the (two) relativistic energy eigenvalues (of the electron/positron), $k_z=const.$ ($-\infty<k_z<\infty$) is the intensity/projection (or $z$-component) of the (angular) wavevector (or $k$-vector) along the $z$-direction ($k_z=\vec{k}\cdot\vec{e}_z$), i.e., $k_z$ can be called (angular) wavenumber, $z$-wavevector or even $z$-momentum (in SI units, we have $k_z=p_z/\hslash$ and $\vec{k}=\vec{p}/\hslash$), and $m_j=m_{\phi}=\pm 1/2,\pm 3/2,\pm 5/2,\ldots$ (or $m_j=\pm \vert m^\pm_j\vert$) is the total magnetic quantum number (i.e., eigenvalues of $J_z=-i\partial_\phi$, that is the total angular momentum operator along the $z$-axis, and arise as a consequence of the periodicity of $\varphi_{NC}$), or simply angular quantum number (to differentiate from the radial quantum number that we will see later). In particular, this quantum number can also be written in terms of two others (which also appear in NRQM), such as: $m_j=m_l+m_s$, being $m_l=0,\pm 1,\pm 2,\pm 3,\ldots$ and $m_s=\pm 1/2=s/2=\uparrow\downarrow$ ($s=\pm 1$ is a ``spin parameter'' or spin variable \cite{Connell}) the orbital and spin magnetic quantum numbers (i.e., eigenvalues of $L_z$ and $S_z$, where $L_z$ and $S_z$ are the orbital and intrinsic angular momentum (quantized) along the $z$-axis, being $J_z=L_z+S_z$). With this, for a given value of $m_j$ ($m_j>0$ or $m_j<0$), we have two spin states, i.e., $m_j=m_l\pm 1/2>0$ is satisfied for $m_l\geq 0$ with spin up $m_s=+1/2$, or for $m_l>0$ with spin down $m_s=-1/2$ (in this two cases, we have a positive orbital angular momentum), and $m_j=m_l\pm 1/2<0$ is satisfied for $m_l<0$ with spin up $m_s=+1/2$, or for $m_l\leq 0$ with spin down $m_s=-1/2$ (in these two cases, we have a negative orbital angular momentum). Furthermore, it is important to highlight that the spinor $\psi_{NC}$ is written as $\psi_{NC}=(\psi_1,\psi_2,\psi_3,\psi_4)^T$, being $\psi_{1,3}=\frac{1}{\sqrt{2\pi}}e^{-iEt}e^{i(m_j-1/2)\phi}e^{ik_z z}f_{1,3}(r)$ and $\psi_{2,4}=\frac{1}{\sqrt{2\pi}}e^{-iEt}e^{i(m_j+1/2)\phi}e^{ik_z z}f_{2,4}(r)$ \cite{Leary}, where we use $U=e^{-i\phi\Sigma^3/2}$=diag$(e^{-i\phi\sigma^3/2},e^{-i\phi\sigma^3/2})$=diag$(e^{-i\phi/2},e^{+i\phi/2},e^{-i\phi/2},e^{+i\phi/2})$ (for $m_j=m_l-1/2$, with $m_l\equiv l$, we obtain exactly the components of Refs. \cite{Chaudhuri,Connell}). According to Refs. \cite{Greiner,Bjorken,Grandy,Griffiths,Thomson,Halzen}, $\psi_{1,2}$ (first two components or the first two independent solutions) describes a positive energy particle (positive energy components or positive energy states/solutions) with spin up ($\uparrow$) and down ($\downarrow$), while $\psi_{3,4}$ (last two components or the last two independent solutions) describes a negative energy particle (negative energy components or negative energy states/solutions) also with spin up ($\uparrow$) and down ($\downarrow$), i.e., $\psi_{NC}=(\psi_{E>0}^{\uparrow},\psi_{E>0}^{\downarrow},\psi_{E<0}^{\uparrow},\psi_{E<0}^{\downarrow})^T$ (in other words, $\psi_{1,2}$ describes the spin-up and spin-down positive energy solutions and $\psi_{3,4}$ the spin-up and spin-down negative energy solutions \cite{Thomson}). 

According to Refs. \cite{Thomson}, the DE provides a beautiful mathematical framework for the RQM of spin-1/2 fermions in which the properties of spin and magnetic moments emerge naturally, where the presence of negative energy solutions is unavoidable. Besides, in RQM, a complete set of basis states is required to span the vector space; therefore, the negative energy solutions cannot simply be discarded as being unphysical. In this way, it is necessary to provide a physical interpretation for the negative energy solutions \cite{Thomson}. Indeed, according to Refs. \cite{Greiner,Bjorken,Grandy,Griffiths,Thomson,Halzen} (and especially \cite{Thomson}. See 4.7.3 Antiparticle spinors - page 98), these negative energy solutions are (re)interpreted (by reversing the signs of the energy and momentum, i.e., $(E,\Vec{p})\to(-E,-\Vec{p})$ or $p^\mu\to -p^\mu$) as positive energy solutions for an antiparticle (or simply antiparticle solutions), that is, a positive energy antiparticle (therefore, $\psi_{4,3}(-E,-\vec{p})$ represents the spin-up and spin-down positive energy antiparticle solutions). However, these positive energy components for the antiparticle can also be obtained from $\psi'=i\gamma^2\psi^\star$, where $\psi'_{1,2}(E,\Vec{p})=\psi_{4,3}(-E,-\Vec{p})$, or better, $\psi_{1,2}(E,\Vec{p})\xrightarrow{C}\psi'_{1,2}(E,\Vec{p})=\psi_{4,3}(-E,-\Vec{p})$ \cite{Thomson} (due to $\gamma^2$, the large/upper and small/lower components are exchanged under charge conjugation \cite{Greiner,Thomson}). Therefore, the effect of the charge-conjugation operator on the particle spinor (first two components) is to transform them, respectively, to the antiparticle spinor (where the energy that appears in the spinor is the positive physical energy of the particle/antiparticle) \cite{Thomson} (see 4.7.5 Charge conjugation - page 102). In addition, the electromagnetic four-vector current for the electron with energy $E$, momentum $\vec{p}$, and charge $-e$, is given by $J^\mu (e^-)=-2e\vert N\vert^2(E,\vec{p})$, while for the positron with the same $E$ and $\vec{p}$, but with charge $+e$, is given by $J^\mu(e^+)=-2e\vert N\vert^2(-E,-\vec{p})=+2e\vert N\vert^2(E,\vec{p})$ (or yet $J^\mu (e^-,E,\vec{p})\xrightarrow{C}J^\mu (e^+,E,\vec{p})=J^\mu (e^-,-E,-\vec{p})$), that is, the current for the positron is exactly the same as the current for the electron with $-E$, $-\vec{p}$, and $-e$ (therefore, an electron with $-E$, $-\vec{p}$, and $-e$ (and spin $-m_s$) is actually or reinterpreted as a positron with $+E$, $+\vec{p}$, and $+e$ (and spin $+m_s$) \cite{Greiner,Bjorken,Grandy,Griffiths,Thomson,Halzen}).

So, using \eqref{spinor} in \eqref{dirac9}, we can obtain four coupled first-order differential equations (i.e., each equation couples or depends on three spinor components), given by
\begin{eqnarray}
&& E_{-}f_1(r)=-A_-f_3(r)-i\left(\frac{d}{dr}-m\sigma \omega r+\frac{m_+}{r}\right)f_4(r),
\label{dirac10}\\
&& \bar{E}_{-}f_2(r)=-A_+f_4(r)-i\left(\frac{d}{dr}+m\sigma\omega r-\frac{m_-}{r}\right)f_3(r),
\label{dirac11}\\
&& E_{+}f_3(r)=-A_-f_1(r)-i\left(\frac{d}{dr}-m\sigma\omega r+\frac{m_+}{r}\right)f_2(r),
\label{dirac12}\\
&& \bar{E}_{+}f_4(r)=-A_+f_2(r)-i\left(\frac{d}{dr}+m\sigma\omega r-\frac{m_-}{r}\right)f_1(r),
\label{dirac13}
\end{eqnarray}
where we define
\begin{equation}\label{define}
E_\pm\equiv\frac{(E+\mu)\pm (m+\mu_m B)}{\tau}, \ \ \bar{E}_\pm\equiv\frac{(E+\mu)\pm (m-\mu_m B)}{\tau}, \ \ A_\pm\equiv\frac{(\mu_5 \pm k_z)}{\tau}, \ \ m_\pm\equiv\left(m_j\pm\frac{1}{2}\right),
\end{equation}
being $\omega=\omega^{\sigma}_{eff}=\omega^{\pm}_{eff}=\omega^{e^\pm}_{eff}\equiv\frac{\lambda\omega_c}{2\tau}$ a type of ``effective cyclotron frequency'' or simply ``effective frequency''.

Now, multiplying both sides of Eqs. \eqref{dirac10} and \eqref{dirac12} by $-i\left(\frac{d}{dr}+m\sigma\omega r-\frac{m_-}{r}\right)$, and, using Eqs. \eqref{dirac11} and \eqref{dirac13} \cite{Chaudhuri,Connell}, we obtain two non-homogeneous second-order differential equations (i.e., second-order differential equations coupled by the components $f_4 (r)$ and $f_2 (r)$), given as follows
\begin{eqnarray}
&& \left[\frac{d^2}{dr^2}+\frac{1}{r}\frac{d}{dr}-\frac{(m_j+\frac{1}{2})^2}{r^2}-(m\omega r)^2+2m\sigma\omega\left(m_j-\frac{1}{2}\right)+\mathbb{B}'_1\right]f_4(r)=\mathbb{B}'_2 f_2 (r),
\label{dirac14}\\
&& \left[\frac{d^2}{dr^2}+\frac{1}{r}\frac{d}{dr}-\frac{(m_j+\frac{1}{2})^2}{r^2}-(m\omega r)^2+2m\sigma\omega\left(m_j-\frac{1}{2}\right)+\mathbb{D}'_1\right]f_2(r)=\mathbb{D}'_2 f_4 (r),
\label{dirac15}
\end{eqnarray}
where we define
\begin{equation}\label{BD}
\mathbb{B}'_1\equiv (E_-\bar{E}_+ + A_+ A_-), \ \  \mathbb{B}'_2\equiv -(A_-\bar{E}_- + E_- A_+), \ \ \mathbb{D}'_1\equiv (E_+\bar{E}_- + A_+ A_-), \ \  \mathbb{D}'_2\equiv -(A_-\bar{E}_+ + E_+ A_+).
\end{equation}
 
Similarly, multiplying both sides of Eqs. \eqref{dirac11} and \eqref{dirac13} by $-i\left(\frac{d}{dr}-m\sigma\omega r+\frac{m_+}{r}\right)$, and, using Eqs. \eqref{dirac10} and \eqref{dirac12} \cite{Chaudhuri,Connell} (``unfortunately'', these references only worked with $f_4 (r)$ and $f_2 (r)$), we obtain two non-homogeneous second-order differential equations (i.e., second-order differential equations coupled by the components $f_3 (r)$ and $f_1 (r)$), given as follows
\begin{eqnarray}
&& \left[\frac{d^2}{dr^2}+\frac{1}{r}\frac{d}{dr}-\frac{(m_j-\frac{1}{2})^2}{r^2}-(m\omega r)^2+2m\sigma\omega\left(m_j+\frac{1}{2}\right)+\mathbb{C}'_1\right]f_3(r)=\mathbb{C}'_2 f_1 (r),
\label{dirac16}\\
&& \left[\frac{d^2}{dr^2}+\frac{1}{r}\frac{d}{dr}-\frac{(m_j-\frac{1}{2})^2}{r^2}-(m\omega r)^2+2m\sigma\omega\left(m_j+\frac{1}{2}\right)+\mathbb{A}'_1\right]f_1(r)=\mathbb{A}'_2 f_3 (r),
\label{dirac17}
\end{eqnarray}
where we define
\begin{equation}\label{CA}
\mathbb{C}'_1\equiv (E_+\bar{E}_- + A_+ A_-), \ \  \mathbb{C}'_2\equiv -(A_-\bar{E}_- + E_- A_+), \ \ \mathbb{A}'_1\equiv (E_-\bar{E}_+ + A_+ A_-), \ \  \mathbb{A}'_2\equiv -(A_-\bar{E}_+ + E_+ A_+).
\end{equation}

\section{The relativistic energy spectrum\label{sec3}}

To solve exactly and analytically Eqs. \eqref{dirac14}, \eqref{dirac15}, \eqref{dirac16}, and \eqref{dirac17}, let us first introduce a new variable in our problem, given by: $\rho=m\omega r^2$ (i.e., a dimensionless variable). Thus, through a change of variable, Eqs. \eqref{dirac14}, \eqref{dirac15}, \eqref{dirac16}, and \eqref{dirac17} take the following form (for convenience, we put them in a matrix form)
\begin{equation}\label{dirac18}
\left(
           \begin{array}{c}
            \left[\rho\frac{d^2}{d\rho^2}+\frac{d}{d\rho}-\frac{(m_j+\frac{1}{2})^2}{4\rho}-\frac{\rho}{4}+\frac{\sigma}{2}\left(m_j-\frac{1}{2}\right)+\mathbb{B}_1\right]f_4(\rho) \\
            \left[\rho\frac{d^2}{d\rho^2}+\frac{d}{d\rho}-\frac{(m_j+\frac{1}{2})^2}{4\rho}-\frac{\rho}{4}+\frac{\sigma}{2}\left(m_j-\frac{1}{2}\right)+\mathbb{D}_1\right]f_2(\rho) \\
            \left[\rho\frac{d^2}{d\rho^2}+\frac{d}{d\rho}-\frac{(m_j-\frac{1}{2})^2}{4\rho}-\frac{\rho}{4}+\frac{\sigma}{2}\left(m_j+\frac{1}{2}\right)+\mathbb{C}_1\right]f_3(\rho) \\
            \left[\rho\frac{d^2}{d\rho^2}+\frac{d}{d\rho}-\frac{(m_j-\frac{1}{2})^2}{4\rho}-\frac{\rho}{4}+\frac{\sigma}{2}\left(m_j+\frac{1}{2}\right)+\mathbb{A}_1\right]f_1(\rho) \\
           \end{array}
         \right)=\left(
           \begin{array}{c}
            \mathbb{B}_2 f_2 (\rho) \\
            \mathbb{D}_2 f_4 (\rho) \\
            \mathbb{C}_2 f_1 (\rho) \\
            \mathbb{A}_2 f_3 (\rho) \\
           \end{array}
         \right),
\end{equation}
where 
\begin{equation}\label{BDCA}
\mathbb{B}_{1,2}=\frac{\mathbb{B}'_{1,2}}{4m\omega}, \ \ \mathbb{D}_{1,2}=\frac{\mathbb{D}'_{1,2}}{4m\omega}, \ \ \mathbb{C}_{1,2}=\frac{\mathbb{C}'_{1,2}}{4m\omega}, \ \ \mathbb{A}_{1,2}=\frac{\mathbb{A}'_{1,2}}{4m\omega}.
\end{equation}

According to Refs. \cite{Chaudhuri,Connell}, the equations in \eqref{dirac18} can be solved by the Frobenius method (power-series method). In this way, a good choice for the functions $f_1 (\rho)$, $f_2 (\rho)$, $f_3 (\rho)$, and $f_4 (\rho)$ are given as follows
\begin{equation}\label{functions}
f_1 (\rho)=e^{-\rho/2}\rho^{\gamma}\sum_{N=0}^\infty a_N \rho^N, \ \  f_2 (\rho)=e^{-\rho/2}\rho^{\gamma'}\sum_{N=0}^\infty b_N \rho^N, \ \ f_3 (\rho)=e^{-\rho/2}\rho^{\gamma}\sum_{N=0}^\infty c_N \rho^N, \ \ f_4 (\rho)=e^{-\rho/2}\rho^{\gamma'}\sum_{N=0}^\infty d_N \rho^N,
\end{equation}
where $\gamma=\frac{\vert m_j-\frac{1}{2}\vert}{2}\geq 0$ and $\gamma'=\frac{\vert m_j+\frac{1}{2}\vert}{2}\geq 0$. For example, for $m_j=l-\frac{1}{2}$ (where it implies $\gamma'=\vert l\vert/2$), we obtain exactly the functions of Refs. \cite{Chaudhuri,Connell}. However, these references did not explain very well why functions can be written in the form/configuration given in \eqref{functions}. In particular, this can be easily explained by considering the asymptotic behavior/limit of the equations for $\rho\to 0$ and $\rho\to\infty$ (such as is done in Refs. \cite{Oliveira5,Oliveira2,Oliveiraneutrinos,Villalba,O,Bakke,Ahmed,Cunha,Ribeiro}). That is, for $\rho\to 0$, we obtain an equation whose solutions are given by $\rho^{\pm\gamma/2}$ (or $\rho^{\pm\gamma'/2}$), while for $\rho\to\infty$, we obtain an equation whose solutions are given by $e^{\pm\rho/2}$, respectively. So, in order to have regular/finite solutions (i.e., to avoid divergence difficulties at the origin and infinity), it implies that a physically acceptable solution (or normalizable solution) is given by $e^{-\rho/2}\rho^{\gamma/2}F(\rho)$ (or $e^{-\rho/2}\rho^{\gamma'/2}G(\rho)$), where $F(\rho)$ (or $G(\rho)$) is an unknown function to be determined (and they must also be a finite/regular function).

So, substituting \eqref{functions} in \eqref{dirac18}, we obtain the following recursion relations (see Appendix \ref{sec6})
\begin{equation}\label{dirac19}
\left(
           \begin{array}{c}
            \left\{\left[\mathbb{B}_1-N-\frac{1}{2}-\frac{\vert m_j+\frac{1}{2}\vert-\sigma(m_j-\frac{1}{2})}{2}\right]d_{N}-\mathbb{B}_2 b_{N}+(N+1)(N+1+\vert m_j+\frac{1}{2}\vert) d_{N+1}\right\}\\
            \left\{\left[\mathbb{D}_1-N-\frac{1}{2}-\frac{\vert m_j+\frac{1}{2}\vert-\sigma(m_j-\frac{1}{2})}{2}\right]b_{N}-\mathbb{D}_2d_{N}+(N+1)(N+1+\vert m_j+\frac{1}{2}\vert) b_{N+1}\right\}\\
           \left\{\left[\mathbb{C}_1-N-\frac{1}{2}-\frac{\vert m_j-\frac{1}{2}\vert-\sigma(m_j+\frac{1}{2})}{2}\right]c_{N}-\mathbb{C}_2 a_{N}+(N+1)(N+1+\vert m_j-\frac{1}{2}\vert)c_{N+1}\right\}\\
            \left\{\left[\mathbb{A}_1-N-\frac{1}{2}-\frac{\vert m_j-\frac{1}{2}\vert-\sigma(m_j+\frac{1}{2})}{2}\right]a_{N}-\mathbb{A}_2 c_{N}+(N+1)(N+1+\vert m_j-\frac{1}{2}\vert)a_{N+1}\right\}\\
           \end{array}
         \right)=0.
\end{equation}

According to Refs. \cite{Chaudhuri,Connell}, to obtain a well-behaved function (finite or normalizable solution), the power series must terminate at some $N=N'$, such that the recursion relations spits out $a_{N'+1}=b_{N'+1}=c_{N'+1}=d_{N'+1}=0$ (for convenience, here, we will use $n$ instead of $N'$). Therefore, applying this condition in \eqref{dirac19}, we obtain
\begin{equation}\label{dirac20}
\left(
\begin{array}{cc}
 \mathbb{B}_1-N_{2,4} & \ \ -\mathbb{B}_2 \\
 -\mathbb{D}_2 & \ \ \mathbb{D}_1-N_{2,4} \\
\end{array}
\right)\left(
           \begin{array}{c}
            d_{n} \\
            b_{n} \\
           \end{array}
         \right)=0, \ \ \left(
\begin{array}{cc}
 \mathbb{C}_1-N_{1,3} & \ \ -\mathbb{C}_2 \\
 -\mathbb{A}_2 & \ \ \mathbb{A}_1-N_{1,3} \\
\end{array}
\right)\left(
           \begin{array}{c}
            c_{n} \\
            a_{n} \\
           \end{array}
         \right)=0,
\end{equation}
where we define
\begin{equation}\label{N}
N_{2,4}\equiv\left(n+\frac{1}{2}+\frac{\vert m_j+\frac{1}{2}\vert-\sigma(m_j-\frac{1}{2})}{2}\right), \ \ N_{1,3}\equiv\left(n+\frac{1}{2}+\frac{\vert m_j-\frac{1}{2}\vert-\sigma(m_j+\frac{1}{2})}{2}\right),
\end{equation}
being $N_{2,4}$ for the functions $f_{2,4}(r)$ (or spinor components $\psi_{2,4}$), and $N_{1,3}$ for the functions $f_{1,3}(r)$ (or spinor components $\psi_{1,3}$), respectively.

According to Refs. \cite{Chaudhuri,Connell}, for the equations in \eqref{dirac20} to have non-trivial solutions ($f_{1,2,3,4}(\rho)\neq 0$), we must have
\begin{equation}\label{dirac22}
det\left(
\begin{array}{cc}
 \mathbb{B}_1-N_{2,4} & \ \ -\mathbb{B}_2 \\
 -\mathbb{D}_2 & \ \ \mathbb{D}_1-N_{2,4} \\
\end{array}
\right)=0, \ \ det\left(
\begin{array}{cc}
 \mathbb{C}_1-N_{1,3} & \ \ -\mathbb{C}_2 \\
 -\mathbb{A}_2 & \ \ \mathbb{A}_1-N_{1,3} \\
\end{array}
\right)=0,
\end{equation}
where implies that
\begin{equation}\label{dirac23}
\mathbb{B}_1 \mathbb{D}_1-\mathbb{B}_2 \mathbb{D}_2-N_{2,4}[\mathbb{B}_1+\mathbb{D}_1]+N_{2,4}^2=0, \ \ \mathbb{C}_1 \mathbb{A}_1-\mathbb{C}_2 \mathbb{A}_2-N_{1,3}[\mathbb{C}_1+\mathbb{A}_1]+N_{1,3}^2=0,
\end{equation}
or yet
\begin{equation}\label{dirac24}
\mathbb{B}'_1 \mathbb{D}'_1-\mathbb{B}'_2 \mathbb{D}'_2-4m\omega N_{2,4}[\mathbb{B}'_1+\mathbb{D}'_1]+16m^2\omega^2N_{2,4}^2=0, \ \ \mathbb{C}'_1 \mathbb{A}'_1-\mathbb{C}'_2 \mathbb{A}'_2-4mN_{1,3}[\mathbb{C}'_1+\mathbb{A}'_1]+16m^2\omega^2 N_{1,3}^2=0,
\end{equation}
where we use \eqref{BDCA}.

Therefore, using \eqref{define}, \eqref{BD}, \eqref{CA}, and \eqref{dirac24}, we obtain the following relativistic energy spectra (relativistic Landau levels) for a charged Dirac fermion with AMM in the presence of the CME and of a NC phase space (see Appendix \ref{sec6})
\begin{align}\label{E1}
E^\pm_{N_{2,4}} &=-\mu\pm\sqrt{k^2_z+m^2+\mu_5^2+(\mu_m B)^2+2m\tau\lambda\omega_c N_{2,4}+4h\sqrt{[m\mu_m B-\mu_5 k_z]^2+2m\tau\lambda\omega_c N_{2,4}[(\mu_m B)^2+\mu_5^2]}},
\\
E^\pm_{N_{1,3}} &=-\mu\pm\sqrt{k^2_z+m^2+\mu_5^2+(\mu_m B)^2+2m\tau\lambda\omega_c N_{1,3}+4h\sqrt{[m\mu_m B-\mu_5 k_z]^2+2m\tau\lambda\omega_c N_{1,3}[(\mu_m B)^2+\mu_5^2]}}.
\end{align}

On the other hand, we can write the two spectra above into a single spectrum (due to the spin of each component of the spinor) in the following form (i.e., the two different eigenvalues of the Dirac Hamiltonian)
\begin{equation}\label{E2}
E^\kappa_{N_{eff}}=-\mu+\kappa\sqrt{k^2_z+m^2+\mu_5^2+(\mu_m B)^2+2m\tau\lambda\omega_c N_{eff}+4h\sqrt{[m\mu_m B-\mu_5 k_z]^2+2m\tau\lambda\omega_c N_{eff}[(\mu_m B)^2+\mu_5^2]}},
\end{equation}
where $N_{eff}$ is an ``effective quantum number'', which is defined as follows
\begin{equation}\label{N1}
N_{eff}=N_{\sigma}\equiv\left(n+\frac{1}{2}+\frac{\Big|m_j-\frac{s}{2}\Big|-\sigma\left(m_j+\frac{s}{2}\right)}{2}\right)=\left(n+\frac{1}{2}+\frac{\vert m_j-m_s\vert-\sigma(m_j+m_s)}{2}\right)\geq 0, \ \ (m_s=s/2),
\end{equation}
or yet
\begin{equation}\label{N2}
N_{eff}\equiv\left(n+\frac{1-2\sigma m_s}{2}+\frac{\vert m_j-m_s\vert-\sigma(m_j-m_s)}{2}\right)=\left(n+\frac{1-2\sigma m_s}{2}+\frac{\vert m_l\vert-\sigma m_l}{2}\right)\geq 0, \ \ (m_j=m_l+m_s),
\end{equation}
being $s=+1$ (or $m_s=+1/2$) for the functions $f_{1,3}(\rho)$ (or spinor components $\psi_{1,3}$ with spin up), and $s=-1$ (or $m_s=-1/2$) for the functions $f_{2,4}(\rho)$ (or spinor components $\psi_{2,4}$ with spin down), $n=n_r=0,1,2,3,\ldots$ is the radial quantum number (since it arises from a radial differential equation), or Landau (level) index/number, and $\kappa$ is a real parameter (``energy parameter'') which describes/represents the positive-energy states/solutions ($\kappa_{E>0}=+1$) as well as the negative-energy states/solutions ($\kappa_{E<0}=-1$), respectively. Therefore, we can also call $N_{eff}$ the total quantum number ($N_{eff}=N_{total}$) since it depends on all the others. In particular (since $E=E^\pm=\pm \vert E^\pm\vert$), the spectrum of the particle (electron with $\sigma_{e^-}=-1$) is given by $E_{particle}=E_{electron}^{e^-}=E^+_{N_{eff}}=\vert E^+_{N_{eff}}\vert>0$, while the spectrum of the antiparticle (positron with $\sigma_{e^+}=+1$) is given by $E_{antiparticle}=E_{positron}^{e^+}=-E^-_{N_{eff}}=\vert E^-_{N_{eff}}\vert>0$, or better, $E_{particle}=E_{electron\ with\ pos.\ energy}^{e^-}>0$, and $E_{antiparticle}=-E_{electron\ with\ neg.\ energy}^{\sigma=+1}=E_{positron\ with\ pos.\ energy}^{e^+}>0$ \cite{Greiner,Bjorken,Grandy,Thomson}. Therefore, both the electron and the positron have positive energies (and they can have the same helicity $h=\pm 1/2$, since helicity is invariant under charge conjugation); however, here, with different values, i.e., we have an asymmetric spectrum where $E_{electron}^{e^-}\neq E_{positron}^{e^+}$. In fact, this happens because we have $\mu\neq 0$. For example, if $\mu=0$ (and considering the same helicity, of course), we would have a symmetrical spectrum, that is, $E_{electron}^{e^-}=E_{positron}^{e^+}$ (we will see this better soon when we analyze the spectrum for $m_j>0$ and $m_j<0$). In other words (general case), a particle with negative energy $E_p<0$, charge $q_p=q_{particle}=\pm e$ and chemical potential $\mu_p=\mu_{particle}=\pm\mu$ (i.e., $E_p=E_{particle}^{q_p,\mu_p}<0$), is actually (or (re)interpreted as) an antiparticle with positive energy $E_a>0$, charge $q_a=q_{antiparticle}=\mp e$ and chemical potential $\mu_a=\mu_{antiparticle}=\mp\mu$ (i.e., $E_a=E_{antiparticle}^{q_a,\mu_a}>0$), which is a consequence of charge conjugation (in which transform a particle with negative energy, charge $q_p$ and chemical potential $\mu_p$ in an antiparticle with positive energy and opposite charge and chemical potential ($q_a=-q_p$ and $\mu_a=-\mu_p$), where the expected value of the Hamiltonian of both satisfies: $\langle H_{antiparticle \ with \ E>0}^{q_a,\mu_a}\rangle_c=-\langle H_{particle \ with \ E<0}^{q_p,\mu_p}\rangle$, or better, $\langle H_{antiparticle \ with \ E>0}^{q_a,\mu_a}\rangle=-\langle H_{particle \ with \ E<0}^{q_p,\mu_p}\rangle_c$, with $E=\langle H\rangle=(\langle H\rangle_c)_c$ \cite{Greiner}), or even of the Feynman-Stückelberg interpretation (in which it states that negative-energy electrons going/moving/propagating backward in space and time (or for the past) are (seen/interpreted) as positive-energy positrons going/moving/propagating forward in space and time (or for the future), i.e., negative-energy solutions of the DE are (re)interpreted as positive-energy antiparticles) \cite{Greiner,Bjorken,Grandy,Griffiths,Thomson,Halzen}. In fact, if the energy of the positron were really/physically negative (or any other antiparticle), then in the pair annihilation phenomenon (or particle-antiparticle annihilation) the final energy would be zero, i.e., the gamma photon(s) created would have null/zero energy ($E_{e^-}+E_{e^+}=E_{\gamma}=0$), which makes no sense (this would violate the conservation of energy and, therefore, unphysical \cite{Greiner,Bjorken,Grandy,Griffiths,Thomson,Halzen}). Besides, it is important to mention that since $q_{e^+}=-q_{e^-}(=+e)$ and $\mu_{e^+}=-\mu_{e^-}(=-\mu)$, it implies that $q_{e^+}+q_{e^-}=q_{\gamma}=0$ and $\mu_{e^+}+\mu_{e^-}=\mu_{\gamma}=0$, i.e., the electric charge and the chemical potential of the gamma photon is null/zero (therefore, the charge conservation and the conservation of lepton number are satisfied) \cite{Thomas}. In short, if the spectrum of the electron is given by
\begin{equation}\label{spectrumoftheelectron}
E^{e^-}_{N_{eff}}=-\mu_{e^-}+\sqrt{k^2_z+m^2+\mu_5^2+(\mu_m B)^2+2m\tau\lambda\omega_c N_{eff}+4h\sqrt{[m\mu_m B-\mu_5 k_z]^2+2m\tau\lambda\omega_c N_{eff}[(\mu_m B)^2+\mu_5^2]}}>0,
\end{equation}
or simply
\begin{equation}\label{spectrumoftheelectron}
E^{e^-}_{N_{eff}}=-\mu+\sqrt{k^2_z+m^2+\mu_5^2+(\mu_m B)^2+2m\tau\lambda\omega_c N_{eff}+4h\sqrt{[m\mu_m B-\mu_5 k_z]^2+2m\tau\lambda\omega_c N_{eff}[(\mu_m B)^2+\mu_5^2]}}>0,
\end{equation}
where
\begin{equation}
N_{eff}=N_{e^-}=\left(n+\frac{1+2m_s}{2}+\frac{\vert m_l\vert+m_l}{2}\right), \ \ \tau\lambda=\tau_{e^-}\lambda_{e^-}=\left(1-\frac{eB\theta}{4}\right)\left(1-\frac{\eta}{eB}\right),
\end{equation}
then the spectrum of the positron is given by (since $E_{positron}=-E_{electron \ with \ neg. \ energy}^{\sigma=+1}=-E^{-}_{N_{eff}}$ and $\mu_{e^+}=-\mu_{e^-}$)
\begin{equation}\label{spectrumofthepositron}
E^{e^+}_{N_{eff}}=-\mu_{e^+}+\sqrt{k^2_z+m^2+\mu_5^2+(\mu_m B)^2+2m\tau\lambda\omega_c N_{eff}+4h\sqrt{[m\mu_m B-\mu_5 k_z]^2+2m\tau\lambda\omega_c N_{eff}[(\mu_m B)^2+\mu_5^2]}}>0,
\end{equation}
or simply
\begin{equation}\label{spectrumofthepositron}
E^{e^+}_{N_{eff}}=\mu+\sqrt{k^2_z+m^2+\mu_5^2+(\mu_m B)^2+2m\tau\lambda\omega_c N_{eff}+4h\sqrt{[m\mu_m B-\mu_5 k_z]^2+2m\tau\lambda\omega_c N_{eff}[(\mu_m B)^2+\mu_5^2]}}>0,
\end{equation}
where
\begin{equation}
N_{eff}=N_{e^+}=\left(n+\frac{1-2m_s}{2}+\frac{\vert m_l\vert-m_l}{2}\right),\ \ \tau\lambda=\tau_{e^+}\lambda_{e^+}=\left(1+\frac{eB\theta}{4}\right)\left(1+\frac{\eta}{eB}\right).
\end{equation}

Thus, we clearly see that, in addition to the spectrum \eqref{E2} (or spectra \eqref{spectrumoftheelectron} and \eqref{spectrumofthepositron}) being quantized in terms of the radial quantum number $n$ and the angular quantum number $m_j$ (or by $m_l$ and $m_s$), or simply by $N_{eff}$ (which labels the spectrum), it explicitly depends on the helicity $h$ (describes the fermion spin projection in the direction of momentum), position and momentum NC parameters $\theta$ and $\eta$ (in $\tau$ and $\lambda$), cyclotron frequency $\omega_c$, or NC cyclotron frequency $\omega^{NC}_c\equiv\tau\lambda\omega_c$ (this generates a ``NC cyclotron spectrum'', defined as $E^{NC}_{cyclotron}\equiv 2m\omega^{NC}_c N_{eff}$ \cite{Oliveira8}), anomalous magnetic potential energy (or simply anomalous magnetic energy), defined as $E_m=E_{AMM}\equiv \mu_m B\geq 0$ (electron and positron have practically the same value, i.e., $E^{e^-}_m \cong E^{e^+}_m$ \cite{Oliveira8,Dyck}), $z$-momentum $k_z$, and on the fermion and chiral chemical potential $\mu$ and $\mu_5$ (or fermion and chiral chemical energy since we are in natural units), respectively. Regarding the chemical potential $\mu$, it affects/modifies the spectrum of the electron and the positron differently, that is, in the case of the electron, it has the function of decreasing the energies (or shifting all energy levels to lower values), while in the case of the positron, it has the function of increasing the energies (or shifting all energy levels to higher values). Therefore, the energies of the electron are higher in the absence of $\mu$ (since it appears with a negative sign in the spectrum), while in the case of the positron, they are higher in its presence (since it appears with a positive sign in the spectrum). In particular, something very similar/analogous to this occurs with Dirac neutrinos/antineutrinos interacting with matter, where the matter potential is positive in the spectrum of the neutrino (i.e., shifting all energy levels to higher values) and negative in the spectrum of the antineutrino (i.e., shifting all energy levels to lower values) \cite{Pal,Studenikin,Akhmedov}. Now, regarding helicity $h$, it can affect/modify the spectrum ``positively'' or ``negatively'', that is, if the electron and the positron have positive helicity ($h_{e^-}=h_{e^+}=+1/2$), the spectrum will be maximal, or better, will have maximum energies (as we will see later, this will require $m_j>0$ for the electron and $m_j<0$ for the positron). In this case, the spectrum will be positively affected by the helicity if the spin is aligned parallel/positively to the linear momentum. However, if the electron and the positron have negative helicity ($h_{e^-}=h_{e^+}=-1/2$), the spectrum will be minimal, or better, will have minimum energies (as we will see later, this will require $m_j<0$ for the electron and $m_j>0$ for the positron). In this case, the spectrum will be negatively affected by helicity if the spin is aligned anti-parallel/negatively to the linear momentum.

However, knowing that $\tau$ and $\lambda$ are dimensionless quantities, we can rewrite them as: $\tau=(1+\sigma\omega_c/\omega_\theta)$ and $\lambda=(1+\sigma\omega_{\eta}/\omega_c)$, where $\omega_\theta\equiv\frac{4}{m\theta}$ and $\omega_{\eta}\equiv\frac{\eta}{m}$ are ``NC angular frequencies'' \cite{Oliveira8}, being $\omega_\theta$ the position NC frequency (decreases with increasing $\theta$), and $\omega_{\eta}$ the momentum NC frequency (increases with increasing $\eta$), respectively. Therefore, our spectrum depends on three frequencies, given by $\omega_c$, $\omega_\theta$, and $\omega_\eta$. Besides, as $\tau>0$ and $\lambda>0$, it implies that the three frequencies must satisfy some conditions depending on the sign of $\sigma$. For example, for $\sigma=-1$, we must have $\omega_c/\omega_\theta<1$, and $\omega_\eta/\omega_c<1$, while for $\sigma=+1$, we can have $\omega_c/\omega_\theta<1$ or $\omega_c/\omega_\theta>1$ (whatever), and $\omega_\eta/\omega_c<1$ or $\omega_\eta/\omega_c>1$ (whatever), respectively. That is, in the case of the electron, there are no ``resonance states'' in the system and, therefore, the cyclotron frequency cannot coincide (or ``oscillate'') with the same value of the two NC frequencies ($\omega_c\neq\omega_\theta$ and $\omega_c\neq\omega_\eta$). In that way, we see that the NC phase space affects/modifies the spectrum of the electron and positron differently. Furthermore, we also note that even in the absence of the magnetic field ($B=0$), the spectrum \eqref{E2} still remains quantized/discrete due to the presence of $\eta$ (or $\omega_\eta$), i.e., such NC parameter (or NC frequency) acts as a type of ``NC field or potential'' (in a way, that mimics a uniform magnetic field). Explicitly, this spectrum is written as follows
\begin{equation}\label{E3}
E^\kappa_{\bar{N}_{eff}}=-\mu+\kappa\sqrt{k^2_z+m^2+\mu_5^2+2m\omega_\eta \bar{N}_{eff}+4h\mu_5\sqrt{k_z^2+2m\omega_\eta\bar{N}_{eff}}}=-\mu+\kappa\sqrt{m^2+\left(\sqrt{k_z^2+2m\omega_\eta \bar{N}_{eff}}+2h\mu_5\right)^2},
\end{equation}
where $\bar{N}_{eff}$ is defined as (and does not depend on $\sigma$ since the vector potential is zero, that is, $\vec{A}=\frac{1}{2}\vec{B}\times\vec{x}=0$)
\begin{equation}\label{N3}
\bar{N}_{eff}\equiv\left(n+\frac{1}{2}+\frac{\Big|m_j-\frac{s}{2}\Big|-\left(m_j+\frac{s}{2}\right)}{2}\right)=\left(n+\frac{1}{2}+\frac{\vert m_j-m_s\vert-(m_j+m_s)}{2}\right)\geq 0.
\end{equation}

On the other hand, it is also important to analyze the spectrum of the electron \eqref{spectrumoftheelectron} and the positron \eqref{spectrumofthepositron} according to the (positive and negative) values of the quantum number $m_j$ (or of $m_l$ and $m_s$). So, analyzing $N_{eff}$ for $m_j>0$ (electron/positron with positive total angular momentum), as well as for $m_j<0$ with (electron/positron with negative total angular momentum), we obtain Table \ref{tab1}, which shows four possible settings for the spectrum (two for the electron and two for the positron), where we define $K^2\equiv k^2_z+m^2$, $\chi^2\equiv (\mu_m B)^2+\mu_5^2$, and $\Xi^2\equiv [m\mu_m B-\mu_5 k_z]^2$, being $\tau\lambda=\tau_{e^+}\lambda_{e^+}=(1+\omega_c/\omega_\theta)(1+\omega_\eta/\omega_c)>0$ and $\tau\lambda=\tau_{e^-}\lambda_{e^-}=(1-\omega_c/\omega_\theta)(1-\omega_\eta/\omega_c)>0$.
\begin{table}[h]
\centering
\begin{small}
\caption{Relativistic spectrum depends on the values of $m_j$.} \label{tab1}
\begin{tabular}{ccc}
\hline
Setting & $m_j$ & Spectrum \\
\hline
1& \ \ $m_j>0$ & \ \ \ \ $E^{e^-}_{n,m_j}=-\mu+\sqrt{K^2+\chi^2+m\tau\lambda\omega_c (2n+1+2m_j)+4h\sqrt{\Xi^2+m\tau\lambda\omega_c\chi^2(2n+1+2m_j)}}$\\
2& \ \ $m_j<0$ & \ \ \ \ $E^{e^-}_{n,m_s}=-\mu+\sqrt{K^2+\chi^2+m\tau\lambda\omega_c (2n+1+2m_s)+4h\sqrt{\Xi^2+m\tau\lambda\omega_c\chi^2(2n+1+2m_s)}}$\\
3& \ \ $m_j>0$ & \ \ \ \ $E^{e^+}_{n,m_s}=\mu+\sqrt{K^2+\chi^2+m\tau\lambda\omega_c (2n+1-2m_s)+4h\sqrt{\Xi^2+m\tau\lambda\omega_c\chi^2(2n+1-2m_s)}}$\\
4& \ \ $m_j<0$ & \ \ \ \ $E^{e^+}_{n,m_j}=\mu+\sqrt{K^2+\chi^2+m\tau\lambda\omega_c (2n+1-2m_j)+4h\sqrt{\Xi^2+m\tau\lambda\omega_c\chi^2(2n+1-2 m_j)}}$\\
\hline
\end{tabular}
\end{small}
\end{table}

So, according to the table \ref{tab1}, we see that the spectrum of the electron (settings 1 and 2) will always depend on the spin $m_s$ (as well as on $n$) regardless of the value of $m_j$; however, can or not depend on the orbital angular momentum $m_l$ depending on the value of $m_j$ (or even $m_l$), that is, for $m_j>0$ (with $m_l\geq 0$ and $m_s=+1/2$, or $m_l>0$ and $m_s=-1/2$), the spectrum will depend on $m_l$, while for $m_j<0$ (with $m_l<0$ and $m_s=+1/2$, or $m_l\leq 0$ and $m_s=-1/2$), it will not. In other words, if the total/orbital angular momentum of the electron is positive, its spectrum will be affected by $m_l$, and if it is negative, it will not. Consequently, this implies that the spectrum will be larger for $m_j>0$ ($J_z$ is aligned parallel to the magnetic field, i.e., $ \vec{B}\uparrow\uparrow\vec{J}$); therefore, we have $E^{e^-}_{n,m_j>0}>E^{e^-}_{n,m_j<0}$ (with this, the spectrum will be maximum for $h=+1/2$ and $m_j>0$ (i.e., positive helicity and spin up or down), and minimum for $h=-1/2$ and $m_j<0$ (i.e., negative helicity and also spin up or down), respectively). In particular, even for $m_l=0$, i.e., a null/zero orbital angular momentum (setting 1 for $m_s=+1/2$ and setting 2 for $m_s=-1/2$), the spectra are still not equal (in this case, the spectrum is larger for $m_s=+1/2$, where the ground state ($n=0$) will still depend on the NC parameters). Furthermore, it is important to mention that both $m_j>0$ and $m_j<0$ result in two spectra, where the spectra for each case are the same. That is, for $m_j>0$, we have $E^{e^-}_{n,m_l\geq 0, m_s=+1/2}=E^{e^-}_{n,m_l>0,m_s=-1/2}$ (this means that an electron with orbital angular momentum greater than or equal to zero and spin up, or with orbital angular momentum greater than zero and spin down, it will have the same spectrum), while for $m_j<0$, we have $E^{e^-}_{n,m_l<0, m_s=+1/2}=E^{e^-}_{n,m_l\leq 0,m_s=-1/2}$ (this means that an electron with orbital angular momentum less than zero and spin up, or with orbital angular momentum less than or equal to zero and spin down, it will have the same spectrum).

Already for the case of the positron (settings 3 and 4), something similar and opposite to the electron happens (we can consider this as a consequence of the charge conjugation operation). For example, similar to the case of the electron, the spectrum of the positron will always depend on the spin $m_s$ (as well as on $n$), regardless of the value of $ m_j$; however, it can or cannot depend on the orbital angular momentum $m_l$ depending on the value of $m_j$ (or even $m_l$). However, unlike the case of the electron, the spectrum of the positron for $m_j>0$ (with $m_l\geq 0$ and $m_s=+1/2$, or $m_l>0$ and $m_s=-1/2$), will not depend on $m_l$, while for $m_j<0$ (with $m_l<0$ and $m_s=+1/2$, or $m_l\leq 0$ and $m_s=-1/2$), it will. In other words, if the total/orbital angular momentum of the positron is positive, its spectrum will not be affected by $m_l$, and if it is negative, it will. Consequently, this implies that the spectrum of the positron will be larger for $m_j<0$ ($J_z$ is aligned antiparallel to the magnetic field, i.e., $\vec{B}\uparrow\downarrow\vec{J}$); therefore, we have $E^{e^+}_{n,m_j<0}>E^{e^+}_{n,m_j>0}$ (with this, the spectrum will be maximum for $h=+1/2$ and $m_j<0$ (i.e., positive helicity and spin up or down), and minimum for $h=-1/2$ and $m_j>0$ (i.e., negative helicity and also spin up or down), respectively). Now, similar to the electron, even for $m_l=0$, i.e., a null/zero orbital angular momentum (setting 3 for $m_s=+1/2$ and setting 4 for $m_s=-1/2$), the spectra of the positron are still not equal (in this case, the spectrum is larger for $m_s=-1/2$, where the ground state ($n=0$) will still depend on the NC parameters). Besides (still similar to the electron), both $m_j>0$ and $m_j<0$ also result in two spectra, where the spectra for each case are the same, that is, for $m_j>0$, we have $E^{e^+}_{n,m_l\geq 0, m_s=+1/2}=E^{e^+}_{n,m_l>0,m_s=-1/2}$ (this means that a positron with orbital angular momentum greater than or equal to zero and spin up, or with orbital angular momentum greater than zero and spin down, it will have the same spectrum), while for $m_j<0$, we have $E^{e^+}_{n,m_l<0, m_s=+1/2}=E^{e^+}_{n,m_l\leq 0,m_s=-1/2}$ (this means that a positron with orbital angular momentum less than zero and spin up, or with orbital angular momentum less than or equal to zero and spin down, it will have the same spectrum).

In addition, and according to the charge conjugation where $\langle L_z\rangle_c=-L_z$ and $\langle S_z\rangle_c=-S_z$ \cite{Greiner,Bjorken,Grandy,Thomson} (see the last expression in \eqref{N2}), if an electron ($\sigma=-1$) has an orbital angular momentum and a spin whose eigenvalues are given by $m_l$ ($=0,\pm 1,\pm 2,\ldots$) and by $m_s$ ($=\pm 1/2$), then the positron ($\sigma=+1$) has opposite values, that is, has an orbital angular momentum and a spin whose eigenvalues are given by $-m_l$ ($=0,\mp 1,\mp 2,\ldots$) and by $-m_s$ ($=\mp 1/2$), respectively. Besides, to avoid complex/imaginary energies for the electron/positron with negative helicity ($h=-1/2$), we must have $K^2+\chi^2+m\tau\lambda\omega_c (2n+1-2m_j)-2\sqrt{\Xi^2+m\tau\lambda\omega_c\chi^2(2n+1-2 m_j)}>0$. However, even if the result between them is zero, which also implies in real energy, i.e., $E_{e^{\pm}}=\mu$, we discard this since for $\mu=0$, results in a null/zero energy, which would violate the RQM and, therefore, would result in an unobservable quantum state or unreal physical state. That is, in RQM, the lowest energy that a relativistic massive particle/antiparticle can have is its own rest energy $E=\pm m_0\neq 0$. So, as we have already discussed (and we can clearly see in Table \ref{tab1}), here, the spectra are asymmetric since the spectra of the electron and positron have different values due to the presence of $\mu\neq 0$ (i.e., $E^{e^-}_{n,m_j>0}\neq E^{e^+}_{n,m_j<0}$ and $E^{e^-}_{n,m_j<0}\neq E^{e^+}_{n,m_j>0}$). For example, assuming that both the electron and the positron have the same helicity ($h_{e^-}=h_{e^+}=+1/2$ or $h_{e^-}=h_{e^+}=-1/2$), we have $E^{e^-}_{n,m_j>0}<E^{e^+}_{n,m_j<0}$ (settings 1 and 4) and $E^{e^-}_{n,m_j<0}< E^{e^+}_{n,m_j>0}$ (settings 2 and 3), i.e., the energies of the positron are always greater than those of the electron (we will see this in all the graphs of this paper). Now, assuming that the electron and the positron do not have the same helicity ($h_{e^-}=-h_{e^+}=+1/2$ or $h_{e^-}=-h_{e^+}=-1/2$), then the positron can or cannot have higher energies than the electron (or vice versa). For example, if the positron has $h_{e^+}=+1/2$ and the electron has $h_{e^-}=-1/2$, implies that $E^{e^-}_{n,m_j>0}<E^{e^+}_{n,m_j<0}$ and $E^{e^-}_{n,m_j<0}< E^{e^+}_{n,m_j>0}$ (i.e., the energies of the positron are greater than those of the electron); however, if the positron has $h_{e^+}=-1/2$ and the electron has $h_{e^-}=+1/2$, implies that $E^{e^-}_{n,m_j>0}>E^{e^+}_{n,m_j<0}$ and $E^{e^-}_{n,m_j<0}>E^{e^+}_{n,m_j>0}$ (i.e., the energies of the electron are greater than those of the positron).

Before concluding this section, let us compare our spectrum \eqref{E2} with other works/papers found in the literature. Thus, we verified that our spectrum generalizes several particular/specific cases found in the literature; that is, several particular cases can be obtained when we exclude some parameters/quantities and adjust (or redefine) the quantum numbers in our spectrum. For example, for $\theta=\eta$ (absence of NC phase space), with $m_s=-1/2$ (spin down) or $s=-1$ (functions $f_{2,4}(\rho)$ or spinor components $\psi_{2,4}$), $\sigma=-1$ (electron), and doing $k_z \to p_z$, $\mu_m \to a$, $n\to N'\geq 0$ and $m_l \to l\geq 0$ (where implies in $N_{eff}=N'+l$), \eqref{E2} reduces to the following spectrum (actually relativistic dispersion relation)
\begin{equation}\label{L1}
(E+\mu)^2=p^2_z+m^2+\mu_5^2+(aB)^2+2(N'+l)eB\pm 2\sqrt{[m^2+2(N'+l)eB](aB)^2+[p^2_z+2(N'+l)eB]\mu_5^2-2mp_z\mu_5 aB},
\end{equation}
which is nothing more than the spectrum of Ref. \cite{Chaudhuri} (see (A29) in this reference), that is, we have the relativistic Landau levels for an electron with AMM in the presence of the CME.

Already for $\theta=\eta=\mu=0$, and doing $\mu_m\to \Delta\mu$, $k_z\to p$, $\sigma=-1$, $m_s=-1/2$ and $m_l\leq 0$ (where implies in $N_{eff}=n$), $B\to H$ and $\mu_5\to -b_0$, \eqref{E2} reduces to the following spectrum
\begin{equation}\label{axialvectorcondensed}
E^2=m^2+p^2+2eHn+(\Delta\mu H)^2+b_0^2\pm 2\sqrt{[m(\Delta\mu H)+b_0p]^2+2eHn[(\Delta\mu H)^2+b_0^2]},
\end{equation}
which is nothing more than the spectrum of Ref. \cite{Bubnov} (see (2.9) in this reference, as well as \cite{Frolov}, which is the original article in which such a spectrum appears), that is, we have the relativistic Landau levels for an electron with AMM that interacts with an axial-vector condensed (or axial-vector background that violates the Lorentz invariance. In particular, the origin this background is found in Refs. \cite{Colladay1,Colladay2}). According to this reference, the axial-vector $b_0$ represents the chiral chemical potential $\mu_5$ of the CME. However, we believe the sign in the first term of the square root is incorrect; that is, we think the correct form should be $[m(\Delta\mu H)-b_0p]^2$. In fact, by doing $b_0\to\mu_5$ in \cite{Bubnov,Frolov} (with $\gamma^\mu\gamma^0=-\gamma^0\gamma^5$), we obtain exactly the Lagrangian of the CME, given by $\mathcal{L}_{\mu_5}=\mu_5\bar{\psi}\gamma^0\gamma^5\psi$. Therefore, we see that the CME can be ``converted'' into the axial-vector condensate (and vice versa) by doing $\mu_5 \leftrightarrow b_0$.

For $\theta=\eta=0$, $\mu=\mu_5=0$ (absence of CME), $\sigma=-1$, doing $k_z \to p_z$, $B\to H$, $n\to N'$ and $m_l \to l\geq 0$, using $s/2$ instead of $m_s$ and $h$, $a\mu_B=ae/2m$ instead of $\mu_m$, and defining $H_c\equiv m^2/e$ (in SI units would be $H_c\equiv m^2c^3/e\hslash^2$), \eqref{E2} reduces to the following spectrum
\begin{equation}\label{L2}
E_{N',l}=\pm m\sqrt{\left(\frac{p_z}{m}\right)^2+\left[\sqrt{1+[2(N'+l)+1+s]\frac{H}{H_c}}+\frac{sa}{2}\frac{H}{H_c}\right]^2},
\end{equation}
or yet (defining $H^*\equiv H/H_c$, $p\equiv p_z/m$ and $n\equiv N'+l$)
\begin{equation}\label{L3}
E_{n}=\pm m\sqrt{p^2+\left[\sqrt{1+[2n+1+s]H^*}+\frac{sa}{2}H^*\right]^2},
\end{equation}
which is nothing more than the spectrum of Ref. \cite{Connell} (see (10) or (A27) in this reference), and of Refs. \cite{Bautista,Bubnov,Rodionov} for $[2n+1+s]\to 2n$ and $saH^\star/2\to \theta a H^\star$, $saH^\star/2\to \zeta a H^\star$, and $saH^\star/2\to \xi(\frac{\mu-\mu_0}{2\mu_0})H^\star$, that is, we have the relativistic Landau levels for an electron with AMM. In particular, in Ref. \cite{Connell}, there are two parameters, $s=\pm 1$ (spin variable) and $\xi=\pm 1$ (no name), where sometimes it is considered $s=\xi$, such as happens in \eqref{L3}. However, according to Refs. \cite{Bautista,Bubnov,Rodionov}, the parameters $\theta$, $\zeta$, and $\xi$ are the spin polarization/projection in the direction of the magnetic field. As we will see below (about the CME), these parameters should be the projection of spin in the direction of linear momentum, i.e., the helicity itself (however, since $p_z$ has the same direction as the uniform external magnetic field, we believe the correct statement would be ``... in the direction of the $z-$linear momentum''). In fact, twice the helicity since $h=\pm 1/2$, i.e., $\theta=\zeta=\xi=2h$. Besides, the spectrum \eqref{L2} (or \eqref{L3}) can also be found (in a way/with another form) in Refs. \cite{Ternov,Broderick,Frolov,Kawaguchi,Paulucci,Ferrer}. However, for $\mu=\mu_5=0$, $\sigma=-1$ and $k_z=0$ (absence of the third spatial dimension, i.e., adopting the polar coordinates system), \eqref{E2} reduces to the following spectrum
\begin{equation}\label{L4}
E^\kappa_{N_{eff}}=\mu_m B+\kappa\sqrt{m^2+2m\tau\lambda\omega_c N_{eff}}, \ \ N_{eff}=\left(n+\frac{1}{2}+\frac{\vert m_j-\frac{s}{2}\vert+\left(m_j+\frac{s}{2}\right)}{2}\right),
\end{equation}
which is nothing more than the spectrum of Ref. \cite{Oliveira5} (see (38) in this reference), that is, we have the relativistic Landau levels for an electron with AMM in a 2D NC phase space or in the $(2+1)$-dimensional NC Minkowski spacetime (in this case, or better, the spacetime with two spatial dimensions, it no longer makes sense to talk about helicity (or carrying $h$) since the helicity operator, given by $\Lambda=\frac{\vec{S}\cdot\vec{p}}{\vert \vec{p}\vert}$ \cite{Greiner} (or $\Lambda_z=\frac{S_z p_z}{\vert \vec{p}_z\vert}$, considering only the direction-$z$), is only well-defined in three spatial dimensions, that is, Dirac fermions in $(2+1)$-dimensions do not have positive and negative helicity (or the same physical sense) such as in $(3+1)$-dimensions. In other words, as in $(2+1)$-dimensions, the linear momentum $\vec{p}$ ``lives'' only in the plane, and as the ``spin'' (quotes) of the fermion is perpendicular to the plane, it implies that the ``spin'' cannot be projected onto $\vec{p}$). Furthermore, Ref. \cite{Abyaneh} also obtained the relativistic Landau levels for an electron/positron in an NC space (but with $\theta\neq 0$ and $\eta=\mu_m=0$). That is, doing $n\to k=0,1,2,\ldots$ and taking $\mu=\mu_5=k_z=\eta=0$ and $\mu_m=0$ (without AMM) in \eqref{E2}, we obtain the Landau levels of this reference for the electron ($\sigma=-1$) with $m_l\leq 0$ (where $N_{eff}=k+(1+s)/2$)) and for the positron ($\sigma=+1$) with $m_l\geq 0$ (where $N_{eff}=k+(1-s)/2$)), respectively.

Now, for $\theta=\eta=0$, $\mu_m=0$, $k_z\to p_3$, $h\to s/2$, $\sigma=-1$, $m_s=-1/2$ and $m_l\leq 0$ (where implies in $N_{eff}=n$), \eqref{E2} reduces to the following spectrum
\begin{equation}\label{L5}
E_n=-\mu\pm\sqrt{m^2+\left[\sqrt{p^2_3+2neB}+s\mu_5\right]^2},
\end{equation}
which is nothing more than the spectrum of Ref. \cite{Fukushima} (see (27) in this reference) for $E\to\omega$ and $\mu\to0$ (see also \cite{Fuk,Fukushima2}), and of Ref. \cite{Sheng} (see (24) in this reference) for $\mu\neq 0$ (however, we find that the term $s\mu_5$ in this reference has the wrong sign), that is, we have the relativistic Landau levels for an electron/positron in the presence of the CME (or simply the relativistic Landau levels for the CME). So, for \cite{Fukushima,Fuk,Fukushima2}, the parameter $s$ is treated/defined as being the spin, while for \cite{Sheng}, as being the helicity. However, $s$ cannot describe the spin since spin is described by the quantum number $m_s$ (as we have already discussed many times in the text). On the other hand, doing/defining $\vert\vec{p}\vert=\vert \vec{k}\vert=\sqrt{p^2_3+2neB}$, we obtain the continuous energy spectrum (or simply the continuous spectrum) of the CME from Refs. \cite{Ghosh1,Ghosh2,Farias,Pasqualotto}, where $s$ represents the helicity (actually, twice the helicity, i.e., $s=2h$). Therefore, the parameter $h$ in \eqref{E2} is really the helicity (while the spin $m_s$ is contained in $N_{eff}$). Furthermore, it is interesting to note that by doing $\mu\to 0$, $\mu_5\to b_0$ and $s\to \zeta$ in \eqref{L5}, we obtain the relativistic Landau levels for an electron/positron that interacts with an axial-vector condensed \cite{Bubnov}.

Finally, for $\theta=\eta=\mu_5=\mu_m=0$, and doing $\mu\to \mu_e$, $m\to m_e$, $k_z\to p_z$, $m\omega_c\to \vert q\vert H$, $m\sigma\omega_c\to qH$, $2m_s\to\lambda=\pm 1$, and $\vert m_l\vert-\sigma m_l=0$ (i.e., electron with $m_j<0$ and positron with $m_j>0$), \eqref{E2} reduces to the following spectrum 
\begin{equation}
E^{\pm}_{n}=-\mu_e\pm\sqrt{m^2_e+p^2_z+\vert q\vert H(2n+1)-qH\lambda},
\end{equation}
or better
\begin{equation}
E_{electron/positron}=\mp \mu_e+\sqrt{m^2_e+p^2_z+\vert q\vert H(2n+1)-qH\lambda},
\end{equation}
which is nothing more than the spectrum of Ref. \cite{Dvornikov3} (see (2.2) in this reference, where the total/complete energy spectrum is in the exponentials. However, only the ground state/lowest Landau level ($n=0$) of the electron ($q=-e$) with $\lambda=-1$ (spin down) and of the positron ($q=+e$) with $\lambda=+1$ (spin up) was considered).

\section{Graphical analysis of the spectrum}\label{sec4}

Here, let us graphically analyze (via 2D graphs) the behavior of the relativistic spectrum (of the electron/positron) as a function of the magnetic field $B$, fermion and chiral chemical potential $\mu$ and $\mu_5$, $z$-momentum $k_z$, and the position and momentum NC parameters $\theta$ and $\eta$ for three different values of $n$ and $m_j$ (for this, we choose the spectra with the highest energies, that is, settings 1 and 4 with $h=+1/2$). Regarding the values of these quantum numbers, they allow us to analyze the behavior of the spectrum in two different cases, which are: while $n$ varies ($n=0,1,2$), $m_j$ remains fixed ($m_j=\pm 1/2$, i.e., electron with $m_j=+1/2$ and positron with $m_j=-1/2$), and while $m_j$ varies ($m_j=\pm 1/2,\pm 3/2,\pm 5/2$, i.e., electron with $m_j=+1/2,+3/2,+5/2$ and positron with $m_j=-1/2,-3/2,-5/2$), $n$ remains fixed ($n=0$), respectively. To be more specific, in the same graph we will have/show the behavior of the spectrum of the electron and positron (settings 1 and 4) for $n=0,1,2$ with $m_j=\pm 1/2$, and later, for $m_j=\pm 1/2,\pm 3/2,\pm 5/2$ with $n=0$ (i.e., the ground state or lowest Landau level), respectively. Besides, for simplicity and without loss of generality, here, $\mu$, $\mu_5$, and $k_z$ are positive quantities ($0\leq\mu<\infty$, $0\leq\mu_5<\infty$ and $0\leq k_z<\infty$), and we adopted $m=e=\mu_m=1$ (``unit constants'' or ``generalized natural units system'' since we already have $\hslash=c=1$). With this, we have Table \ref{tab2}, where $N_\pm \equiv 2n+1\pm 2m_j$. However, before we do this analysis, it is important to calculate the allowed range (or restriction/limitation) for $B$, $\theta$, and $\eta$ in the case of the electron. That is, unlike the positron, where the magnetic field and the NC parameters has the standard range, given by $0\leq B<\infty$, $0\leq \theta<\infty$ and $0\leq \eta <\infty$ (this does not imply in $\tau=\lambda=0$), in the case of the electron, $B$, $\theta$ and $\eta$ cannot have any (arbitrary) value \cite{Oliveira5,Oliveira8,Oliveira9}. In other words, the conditions $\tau>0$ and $\lambda>0$ imply in certain allowed ranges for $\theta$ and $\eta$, given by $0\leq\theta<4/B$ and $0\leq\eta<B$ (i.e., in this case, the allowed values for $\theta$ and $\eta$ are restricted/limited to a given arbitrary value of $B$) \cite{Oliveira5,Oliveira8,Oliveira9}, while the condition $\tau\lambda>0$ implies in a certain allowed range for $B$, given by $\frac{(4+\theta\eta)-\sqrt{(4+\theta\eta)^2-16\theta\eta}}{2\theta}<B<\frac{(4+\theta\eta)+\sqrt{(4+\theta\eta)^2-16\theta\eta}}{2\theta}$ (i.e., in this case, the allowed values for $B$ are restricted/limited to a given arbitrary value of $\theta$ and $\eta$) \cite{Oliveira5,Oliveira8,Oliveira9}, respectively. In fact, taking $\eta\to 0$, and later $\theta\to 0$, we have $0<B<\infty$, or better, $0\leq B<\infty$, i.e., we recover the standard range for $B$ in the case of the electron (as it should be). On the other hand, taking only $\eta\to 0$ (where $\lambda=1$), we have $0\leq B<4/\theta$, that is, we can obtain from this the allowed range for $\theta$ ($0\leq \theta<4/B$), or vice versa.
\begin{table}[h]
\centering
\begin{small}
\caption{Relativistic spectrum with ``unit constants'' (``generalized natural units system'').} \label{tab2}
\begin{tabular}{c}
\hline
Spectrum \\
\hline
$E^{e^-}_{n,m_j>0}=-\mu+\sqrt{k^2_z+1+B^2+\mu_5^2+(1-\frac{B\theta}{4})(B-\eta)N_++2\sqrt{(B-\mu_5 k_z)^2+(1-\frac{B\theta}{4})(B-\eta)(B^2+\mu_5^2)N_+}}$\\
$E^{e^+}_{n,m_j<0}=\mu+\sqrt{k^2_z+1+B^2+\mu_5^2+(1+\frac{B\theta}{4})(B+\eta)N_-+2\sqrt{(B-\mu_5 k_z)^2+(1+\frac{B\theta}{4})(B+\eta)(B^2+\mu_5^2)N_-}}$\\
\hline
\end{tabular}
\end{small}
\end{table}

\begin{figure}[htbp]
    \centering
    \begin{subfigure}[b]{0.497\textwidth}
        \centering
        \includegraphics[width=\textwidth]{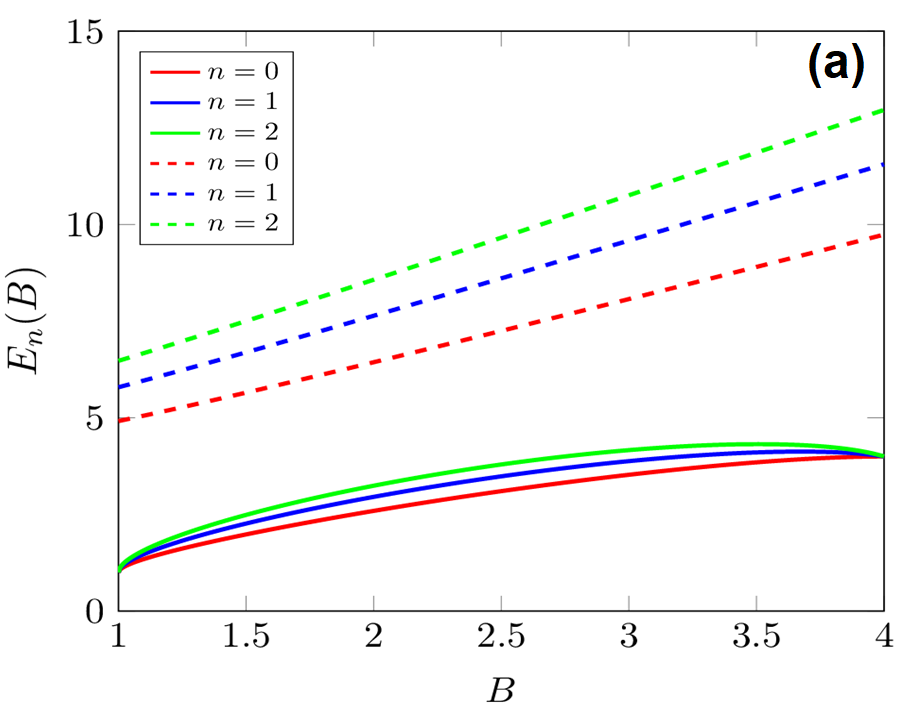}
    \end{subfigure}
    \hfill
    \begin{subfigure}[b]{0.497\textwidth}
        \centering
        \includegraphics[width=\textwidth]{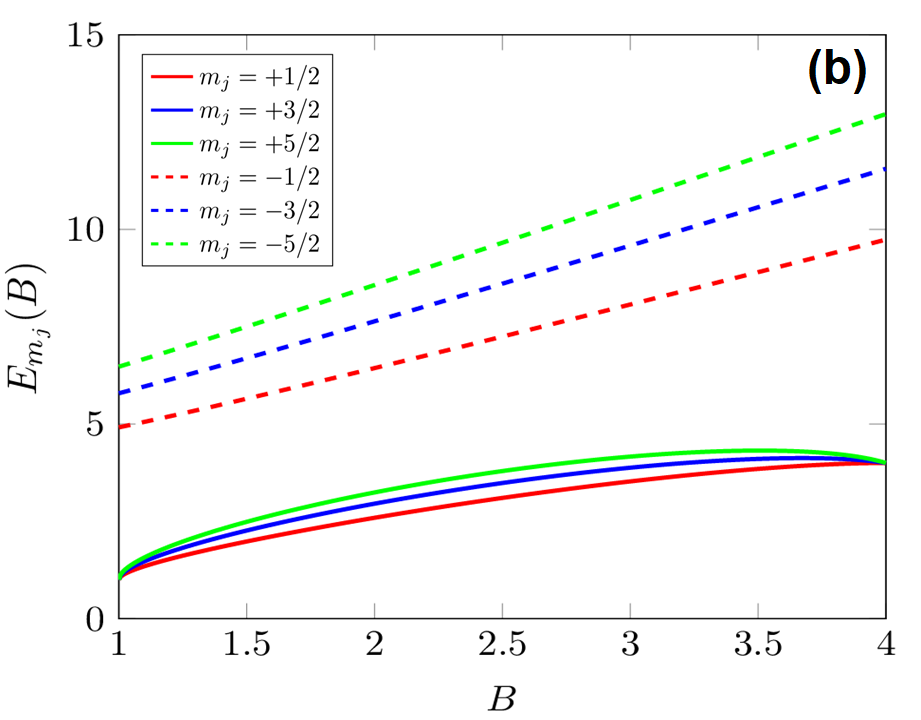}
    \end{subfigure}
    \caption{Behavior of $E_{n}(B)$ vs. $B$ for $n=0,1,2$ with $m_j=\pm 1/2$ (a) and the behavior of $E_{m_j}(B)$ vs. $B$ for $m_j=\pm 1/2,\pm 3/2,\pm 5/2$ with $n=0$ (b).}
    \label{Fig1}
\end{figure}

In that way, in Fig. \ref{Fig1} we have the behavior of $E_{n}(B)$ vs. $B$ for $n=0,1,2$ with $m_j=\pm 1/2$ (a) and the behavior of $E_{m_j}(B)$ vs. $B$ for $m_j=\pm 1/2,\pm 3/2,\pm 5/2$ with $n=0$ (b), where the solid lines are for the electron and the dashed lines for the positron, and we use $\mu=k_z=\mu_5=\theta=\eta=1$, with $1<B<4$. In particular, we clearly see that the behavior of these two cases is exactly the same, that is, the value of the energies with $n$ varying is exactly the same as the energy with $m_j$ varying ($E^{e^\pm}_{n}(B)=E^{e^\pm}_{m_j}(B)$). In fact, this happens because we chose the ground state ($n=0$) for the second case (otherwise, i.e., if $n\geq 1$, the two graphs would be different). So, according to this Figure, we see that the energies of the electron increase with the increase of $n$ and $m_j$ (i.e., $\Delta E^{e^-}_{n}(B)=E^{e^-}_{n+1}(B)-E^{e^-}_{n}(B)>0$ and $\Delta E^{e^-}_{m_j}(B)=E^{e^-}_{m_j+1}(B)-E^{e^-}_{m_j}(B)>0$), while the energies of the positron increase with the increase of $n$ and the decrease of $m_j$ (i.e., $\Delta E^{e^+}_{n}(B)=E^{e^+}_{n+1}(B)-E^{e^+}_{n}(B)>0$ and $\Delta E^{e^+}_{m_j}(B)=E^{e^+}_{m_j-1}(B)-E^{e^+}_{m_j}(B)>0$). Besides, we see that the energies of the electron practically increase as a function of $B$ (or better, most of $B$), that is, the energies increase as the magnetic field increases (however, with one slight exception near $B=4$ for $n=1,2$ or $m_j=+3/2,+5/2$, i.e., there is a small drop in $B>3.5$ for $n=2$ or $m_j=+5/2$ and in $B>3.6$ for $n=1$ or $m_j=+3/2$). Consequently, this implies that the variation of energy as a function of $B$ (or better, most of $B$) is positive (i.e., $\Delta E^{e^-}(B)=E^{e^-}_{final}(B)-E^{e^-}_{initial}(B)>0$). Already in the case of the positron, we see that its energies also increase (actually, they always increase) as a function of $B$; consequently, this implies that the variation of energy as a function of $B$ is positive (i.e., $\Delta E^{e^+}(B)=E^{e^+}_{final}(B)-E^{e^+}_{initial}(B)>0$). Now, comparing the energies of the electron with those of the positron, we see that the energies of the electron are always lower than those of the positron.

\begin{figure}[htbp]
    \centering
    \begin{subfigure}[b]{0.497\textwidth}
        \centering
        \includegraphics[width=\textwidth]{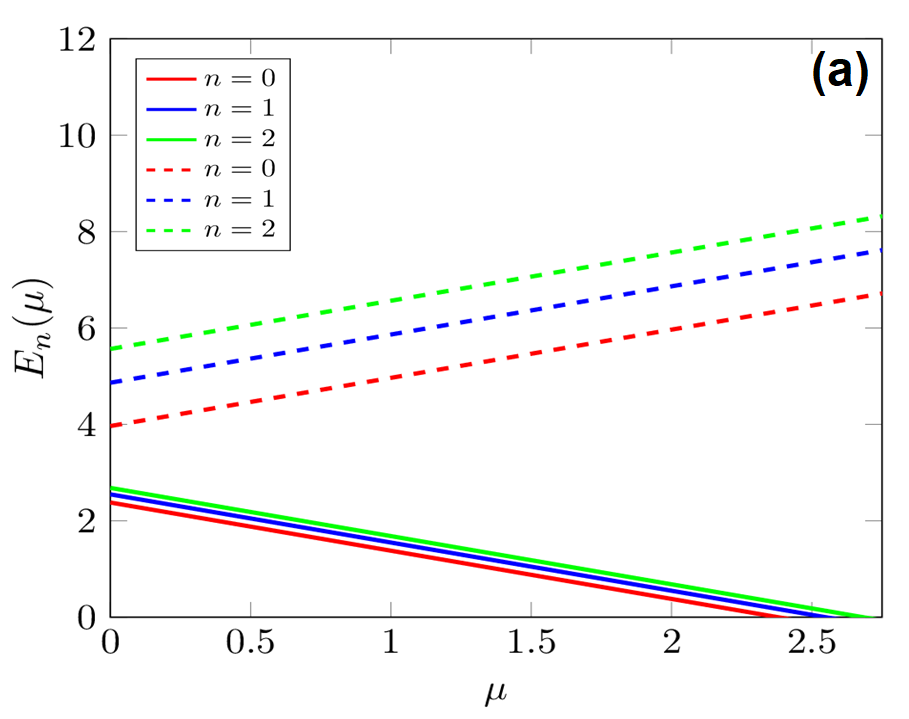}
    \end{subfigure}
    \hfill
    \begin{subfigure}[b]{0.497\textwidth}
        \centering
        \includegraphics[width=\textwidth]{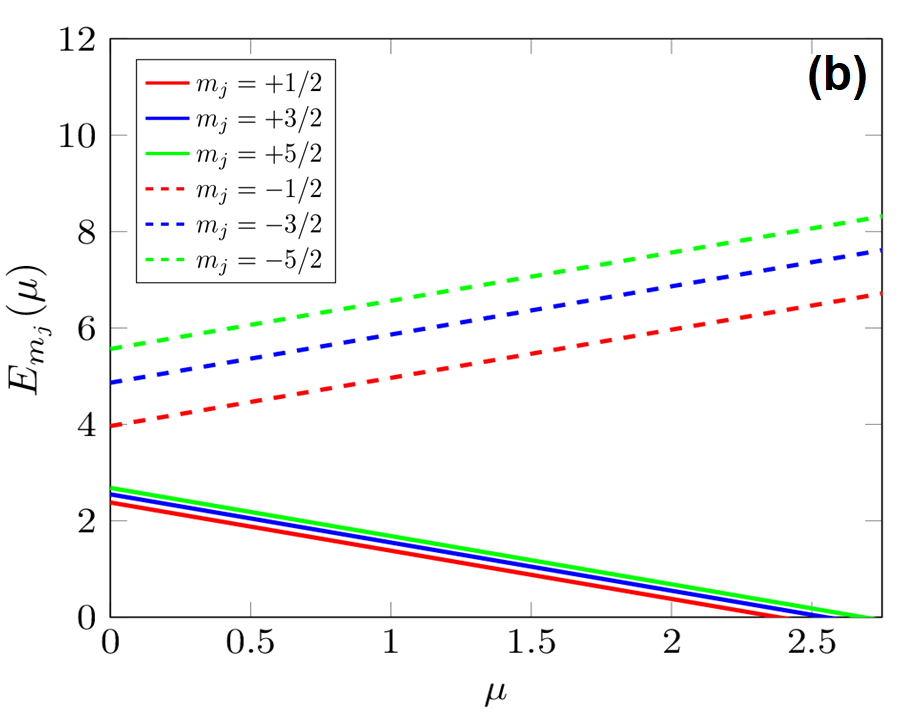}
    \end{subfigure}
    \caption{Behavior of $E_{n}(\mu)$ vs. $\mu$ for $n=0,1,2$ with $m_j=\pm 1/2$ (a) and the behavior of $E_{m_j}(\mu)$ vs. $\mu$ for $m_j=\pm 1/2,\pm 3/2,\pm 5/2$ with $n=0$ (b).}
    \label{Fig2}
\end{figure}

In Fig. \ref{Fig2}, we have the behavior of $E_{n}(\mu)$ vs. $\mu$ for $n=0,1,2$ with $m_j=\pm 1/2$ (a) and the behavior of $E_{m_j}(\mu)$ vs. $\mu$ for $m_j=\pm 1/2,\pm 3/2,\pm 5/2$ with $n=0$ (b), where the solid lines are for the electron and the dashed lines for the positron, and we use $B=k_z=\mu_5=1$, $\theta=3$ (since $\theta<4$), and $\eta=0.5$ (since $\eta<1$). In particular, we clearly see that the behavior of these two cases is exactly the same (such as in the previous Figure), that is, the value of the energies with $n$ varying is exactly the same as the energy with $m_j$ varying ($E^{e^\pm}_{n}(\mu)=E^{e^\pm}_{m_j}(\mu)$). So, according to this Figure, we see that the energies of the electron increase with the increase of $n$ and $m_j$ (i.e., $\Delta E^{e^-}_{n}(\mu)=E^{e^-}_{n+1}(\mu)-E^{e^-}_{n}(\mu)>0$ and $\Delta E^{e^-}_{m_j}(\mu)=E^{e^-}_{m_j+1}(\mu)-E^{e^-}_{m_j}(\mu)>0$), while the energies of the positron increase with the increase of $n$ and the decrease of $m_j$ (i.e., $\Delta E^{e^+}_{n}(\mu)=E^{e^+}_{n+1}(\mu)-E^{e^+}_{n}(\mu)>0$ and $\Delta E^{e^+}_{m_j}(\mu)=E^{e^+}_{m_j-1}(\mu)-E^{e^+}_{m_j}(\mu)>0$). Besides, we see that the energies of the electron decrease and those of the positron increase as a function of $\mu$, that is, the chemical potential has the function of decreasing the energies of the electron and increasing those of the positron. Consequently, this implies that the variation of energy as a function of $\mu$ is negative for the electron (i.e., $\Delta E^{e^-}(\mu)=E^{e^-}_{final}(\mu)-E^{e^-}_{initial}(\mu)<0$) and positive for the positron (i.e., $\Delta E^{e^+}(\mu)=E^{e^+}_{final}(\mu)-E^{e^+}_{initial}(\mu)>0$). It is important to note that in the case of the electron, the energy also cannot decrease as a function of $\mu$ until it reaches zero ($E^{e^-}_{n,m_j}(\mu)\neq 0$), that is, it would also violate the RQM (we have talked about this before). Therefore, as in $\mu\approx 2.4$ (for $n=0$ or $m_j=\pm 1/2$), $\mu\approx 2.5$ (for $n=1$ or $m_j=\pm 3/2$), and $\mu\approx2.7$ (for $n=2$ or $m_j=\pm 5/2$) implies in $E^{e^-}_{n,m_j}(\mu)=0$, we can say that these values are physically forbidden values for the electron. Now, comparing the energies of the electron with those of the positron, we see that the energies of the electron are always lower than those of the positron.

\begin{figure}[htbp]
    \centering
    \begin{subfigure}[b]{0.497\textwidth}
        \centering
        \includegraphics[width=\textwidth]{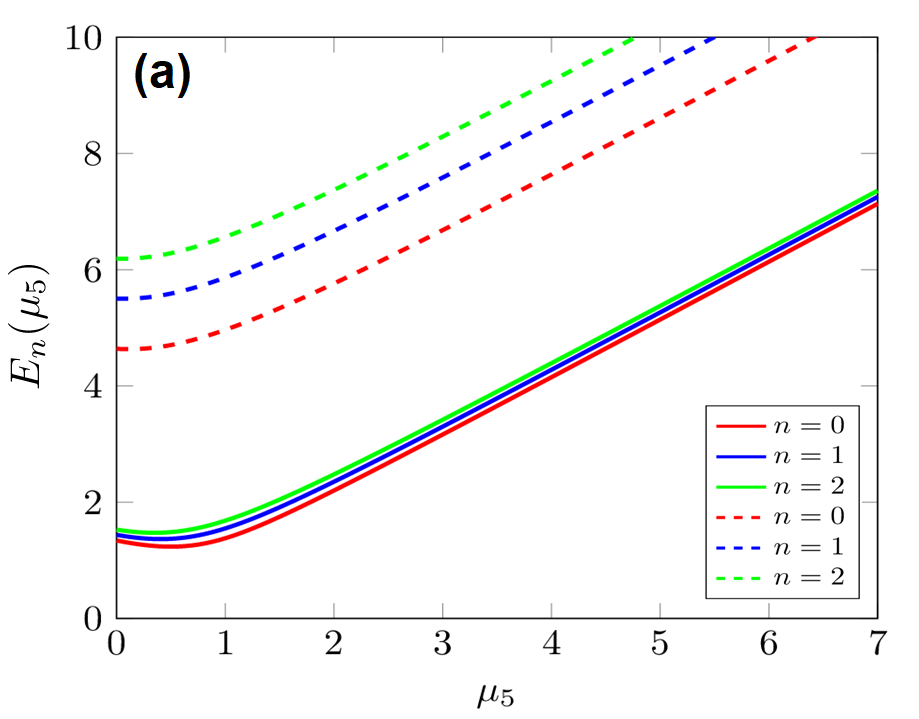}
    \end{subfigure}
    \hfill
    \begin{subfigure}[b]{0.497\textwidth}
        \centering
        \includegraphics[width=\textwidth]{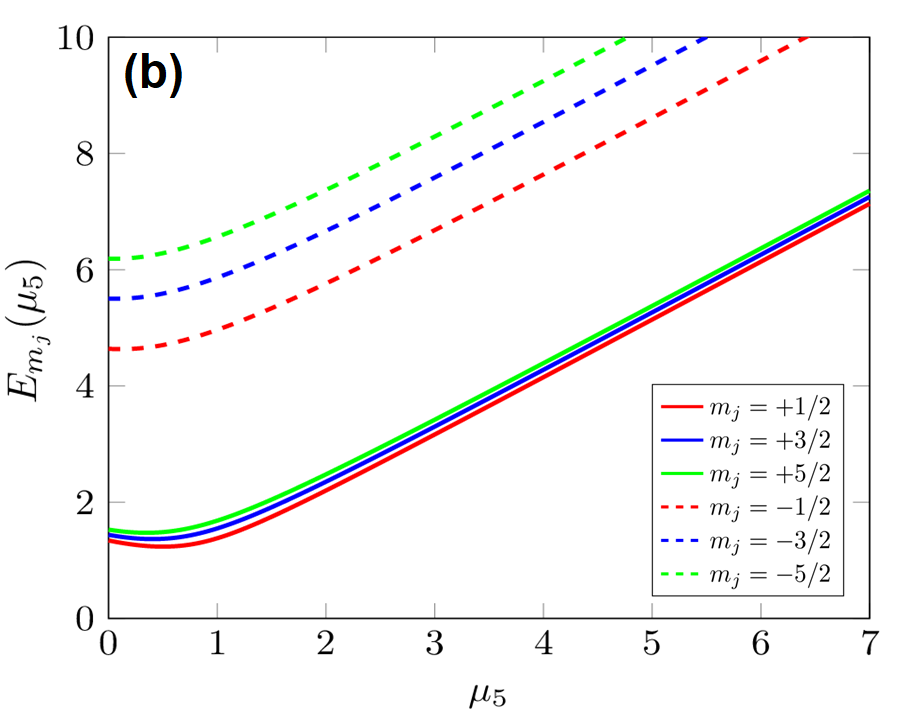}
    \end{subfigure}
    \caption{Behavior of $E_{n}(\mu_5)$ vs. $\mu_5$ for $n=0,1,2$ with $m_j=\pm 1/2$ (a) and the behavior of $E_{m_j}(\mu_5)$ vs. $\mu_5$ for $m_j=\pm 1/2,\pm 3/2,\pm 5/2$ with $n=0$ (b).}
    \label{Fig3}
\end{figure}

In Fig. \ref{Fig3}, we have the behavior of $E_{n}(\mu_5)$ vs. $\mu_5$ for $n=0,1,2$ with $m_j=\pm 1/2$ (a) and the behavior of $E_{m_j}(\mu_5)$ vs. $\mu_5$ for $m_j=\pm 1/2,\pm 3/2,\pm 5/2$ with $n=0$ (b), where the solid lines are for the electron and the dashed lines for the positron, and we use $B=k_z=\mu=1$, $\theta=3$ (since $\theta<4$), and $\eta=0.5$ (since $\eta<1$). In particular, we clearly see that the behavior of these two cases is exactly the same (such as in the previous Figures), that is, the value of the energies with $n$ varying is exactly the same as the energy with $m_j$ varying ($E^{e^\pm}_{n}(\mu_5)=E^{e^\pm}_{m_j}(\mu_5)$). So, according to this Figure, we see that the energies of the electron increase with the increase of $n$ and $m_j$ (i.e., $\Delta E^{e^-}_{n}(\mu_5)=E^{e^-}_{n+1}(\mu_5)-E^{e^-}_{n}(\mu_5)>0$ and $\Delta E^{e^-}_{m_j}(\mu_5)=E^{e^-}_{m_j+1}(\mu_5)-E^{e^-}_{m_j}(\mu_5)>0$), while the energies of the positron increase with the increase of $n$ and the decrease of $m_j$ (i.e., $\Delta E^{e^+}_{n}(\mu_5)=E^{e^+}_{n+1}(\mu_5)-E^{e^+}_{n}(\mu_5)>0$ and $\Delta E^{e^+}_{m_j}(\mu_5)=E^{e^+}_{m_j-1}(\mu_5)-E^{e^+}_{m_j}(\mu_5)>0$). In addition, we see that the energies of the electron can increase or decrease (slightly) as a function of $\mu_5$, that is, they increase for $\mu_5>0.5$ and decrease for $\mu_5<0.5$. Consequently, this implies that the variation of energy as a function of $\mu_5$ can be positive or negative (i.e., $\Delta E^{e^-}(\mu_5>0.5)=E^{e^-}_{final}(\mu_5)-E^{e^-}_{initial}(\mu_5)>0$ and $\Delta E^{e^-}(\mu_5<0.5)=E^{e^-}_{final}(\mu_5)-E^{e^-}_{initial}(\mu_5)<0$). Already in the case of the positron, the energies can be (approximately) constant (for $\mu_5<0.5$) or increase (for $\mu_5>0.5$) as a function of $\mu_5$ (i.e., $\Delta E^{e^+}(\mu_5<0.5)=E^{e^+}_{final}(\mu_5)-E^{e^+}_{initial}(\mu_5)\approx 0$ and $\Delta E^{e^+}(\mu_5>0.5)=E^{e^+}_{final}(\mu_5)-E^{e^+}_{initial}(\mu_5)>0$). Now, comparing the energies of the electron with those of the positron, we see that the energies of the electron are always lower than those of the positron.

\begin{figure}[htbp]
    \centering
    \begin{subfigure}[b]{0.497\textwidth}
        \centering
        \includegraphics[width=\textwidth]{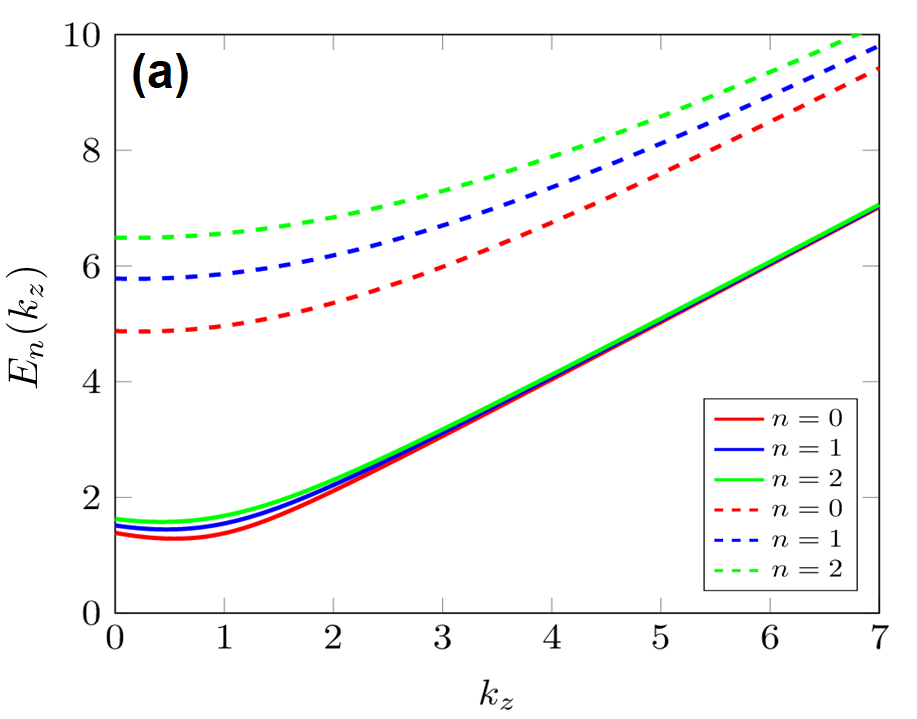}
    \end{subfigure}
    \hfill
    \begin{subfigure}[b]{0.497\textwidth}
        \centering
        \includegraphics[width=\textwidth]{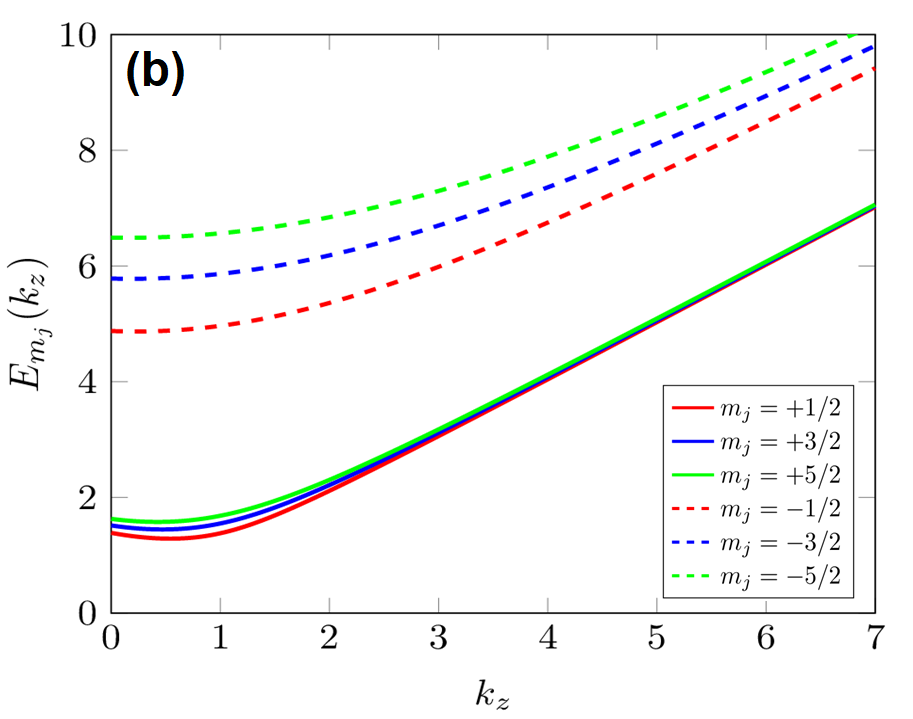}
    \end{subfigure}
    \caption{Behavior of $E_{n}(k_z)$ vs. $k_z$ for $n=0,1,2$ with $m_j=\pm 1/2$ (a) and the behavior of $E_{m_j}(k_z)$ vs. $k_z$ for $m_j=\pm 1/2,\pm 3/2,\pm 5/2$ with $n=0$ (b).}
    \label{Fig4}
\end{figure}

In Fig. \ref{Fig4}, we have the behavior of $E_{n}(k_z)$ vs. $k_z$ for $n=0,1,2$ with $m_j=\pm 1/2$ (a) and the behavior of $E_{m_j}(k_z)$ vs. $k_z$ for $m_j=\pm 1/2,\pm 3/2,\pm 5/2$ with $n=0$ (b), where the solid lines are for the electron and the dashed lines for the positron, and we use $B=\mu=\mu_5=1$, $\theta=3$ (since $\theta<4$), and $\eta=0.5$ (since $\eta<1$). In particular, we clearly see that the behavior of these two cases is exactly the same (such as in the previous Figures), that is, the value of the energies with $n$ varying is exactly the same as the energy with $m_j$ varying ($E^{e^\pm}_{n}(k_z)=E^{e^\pm}_{m_j}(k_z)$). So, according to this Figure (indeed, very similar to Fig. \ref{Fig3}), we see that the energies of the electron increase with the increase of $n$ and $m_j$ (i.e., $\Delta E^{e^-}_{n}(k_z)=E^{e^-}_{n+1}(k_z)-E^{e^-}_{n}(k_z)>0$ and $\Delta E^{e^-}_{m_j}(k_z)=E^{e^-}_{m_j+1}(k_z)-E^{e^-}_{m_j}(k_z)>0$), while the energies of the positron increase with the increase of $n$ and the decrease of $m_j$ (i.e., $\Delta E^{e^+}_{n}(k_z)=E^{e^+}_{n+1}(k_z)-E^{e^+}_{n}(k_z)>0$ and $\Delta E^{e^+}_{m_j}(k_z)=E^{e^+}_{m_j-1}(k_z)-E^{e^+}_{m_j}(k_z)>0$). However, in the case of the electron, for $k_z>2$, the energy levels become sufficiently close. In addition, we see that the energies of the electron can increase or decrease (slightly) as a function of $k_z$, that is, they increase for $k_z>0.5$ and decrease for $k_z<0.5$. Consequently, this implies that the variation of energy as a function of $k_z$ can be positive or negative (i.e., $\Delta E^{e^-}(k_z>0.5)=E^{e^-}_{final}(k_z)-E^{e^-}_{initial}(k_z)>0$ and $\Delta E^{e^-}(k_z<0.5)=E^{e^-}_{final}(k_z)-E^{e^-}_{initial}(k_z)<0$). Already in the case of the positron, the energies can be (approximately) constant (for $k_z<0.5$) or increase (for $k_z>0.5$) as a function of $k_z$ (i.e., $\Delta E^{e^+}(k_z<0.5)=E^{e^+}_{final}(k_z)-E^{e^+}_{initial}(k_z)\approx 0$ and $\Delta E^{e^+}(k_z>0.5)=E^{e^+}_{final}(k_z)-E^{e^+}_{initial}(k_z)>0$). Now, comparing the energies of the electron with those of the positron, we see that the energies of the electron are always lower than those of the positron.

\begin{figure}[htbp]
    \centering
    \begin{subfigure}[b]{0.497\textwidth}
        \centering
        \includegraphics[width=\textwidth]{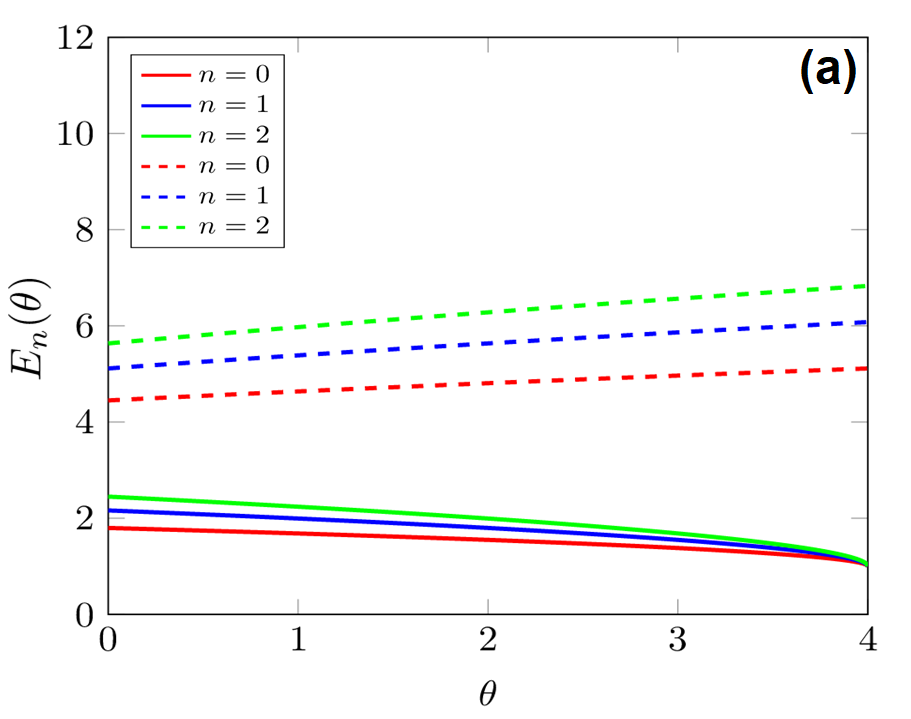}
    \end{subfigure}
    \hfill
    \begin{subfigure}[b]{0.497\textwidth}
        \centering
        \includegraphics[width=\textwidth]{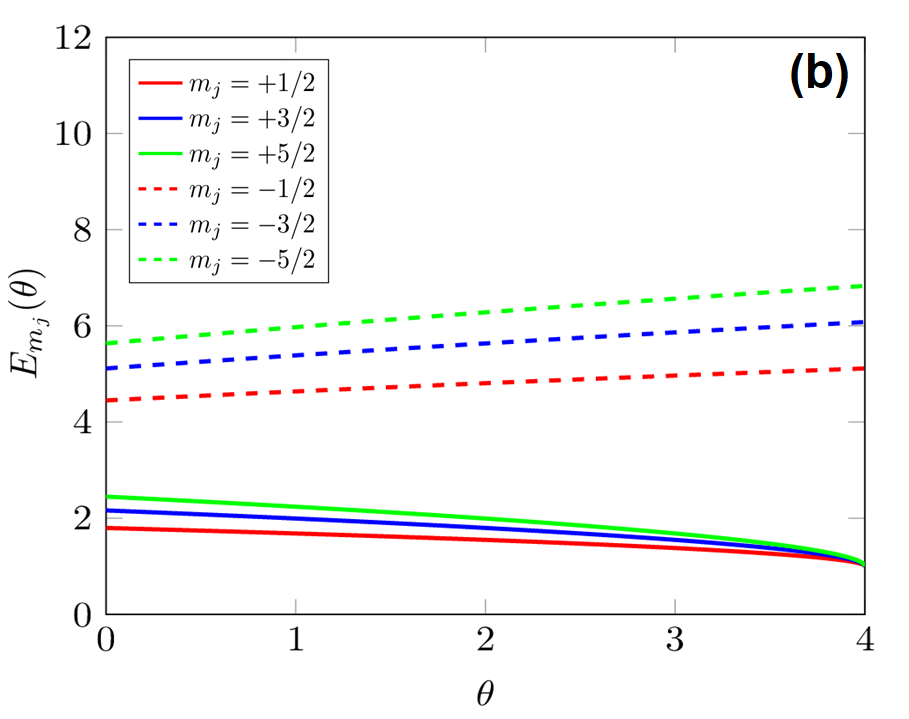}
    \end{subfigure}
    \caption{Behavior of $E_{n}(\theta)$ vs. $\theta$ for $n=0,1,2$ with $m_j=\pm 1/2$ (a) and the behavior of $E_{m_j}(\theta)$ vs. $\theta$ for $m_j=\pm 1/2,\pm 3/2,\pm 5/2$ with $n=0$ (b).}
    \label{Fig5}
\end{figure}

\begin{figure}[htbp]
    \centering
    \begin{subfigure}[b]{0.497\textwidth}
        \centering
        \includegraphics[width=\textwidth]{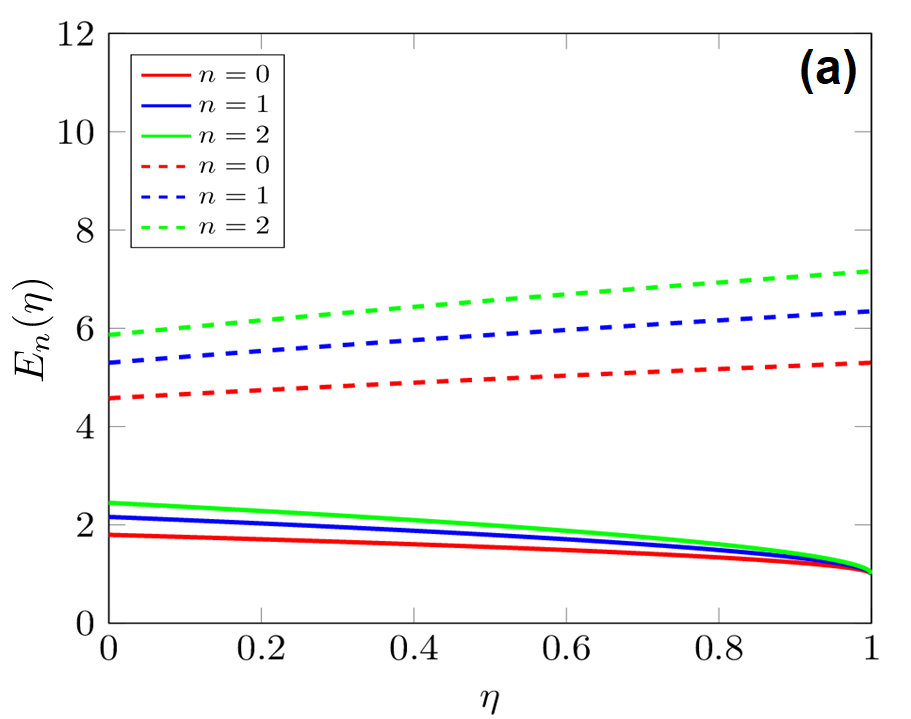}
    \end{subfigure}
    \hfill
    \begin{subfigure}[b]{0.497\textwidth}
        \centering
        \includegraphics[width=\textwidth]{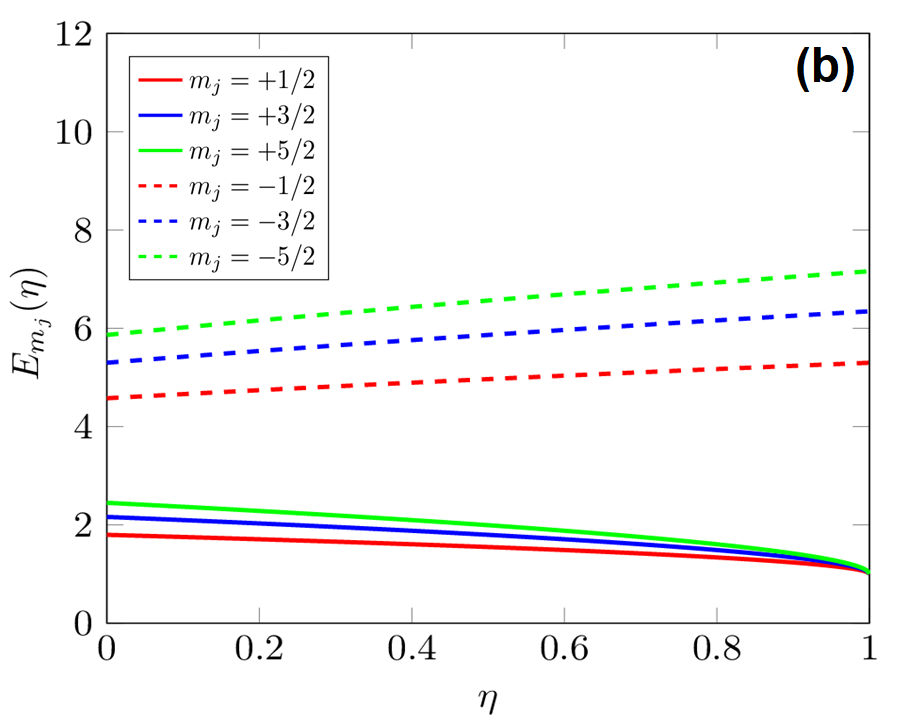}
    \end{subfigure}
    \caption{Behavior of $E_{n}(\eta)$ vs. $\eta$ for $n=0,1,2$ with $m_j=\pm 1/2$ (a) and the behavior of $E_{m_j}(\eta)$ vs. $\eta$ for $m_j=\pm 1/2,\pm 3/2,\pm 5/2$ with $n=0$ (b).}
    \label{Fig6}
\end{figure}

In Fig. \ref{Fig5}, we have the behavior of $E_{n}(\theta)$ vs. $\theta$ for $n=0,1,2$ with $m_j=\pm 1/2$ (a) and the behavior of $E_{m_j}(\theta)$ vs. $\theta$ for $m_j=\pm 1/2,\pm 3/2,\pm 5/2$ with $n=0$ (b), where the solid lines are for the electron and the dashed lines for the positron, and we use $k_z=B=\mu=\mu_5=1$, $\eta=0.5$ (since $\eta<1$) and $0\leq \theta<4$ (since $\theta<4$). In particular, we clearly see that the behavior of these two cases is exactly the same (such as in the previous Figures), that is, the value of the energies with $n$ varying is exactly the same as the energy with $m_j$ varying ($E^{e^\pm}_{n}(\theta)=E^{e^\pm}_{m_j}(\theta)$). So, according to this Figure, we see that the energies of the electron increase with the increase of $n$ and $m_j$ (i.e., $\Delta E^{e^-}_{n}(\theta)=E^{e^-}_{n+1}(\theta)-E^{e^-}_{n}(\theta)>0$ and $\Delta E^{e^-}_{m_j}(\theta)=E^{e^-}_{m_j+1}(\theta)-E^{e^-}_{m_j}(\theta)>0$), while the energies of the positron always increase with the increase of $n$ and the decrease of $m_j$ (i.e., $\Delta E^{e^+}_{n}(\theta)=E^{e^+}_{n+1}(\theta)-E^{e^+}_{n}(\theta)>0$ and $\Delta E^{e^+}_{m_j}(\theta)=E^{e^+}_{m_j-1}(\theta)-E^{e^+}_{m_j}(\theta)>0$). In addition, we see that the energies of the electron decrease as a function of $\theta$, while the energies of the positron always increase as a function of $\theta$. Consequently, this implies that the variation of energy as a function of $\theta$ is negative for the electron (i.e., $\Delta E^{e^-}(\theta)=E^{e^-}_{final}(\theta)-E^{e^-}_{initial}(\theta)<0$) and positive for the positron (i.e., $\Delta E^{e^+}(\theta)=E^{e^+}_{final}(\theta)-E^{e^+}_{initial}(\theta)>0$). Now, comparing the energies of the electron with those of the positron, we see that the energies of the electron are always lower than those of the positron. Already in Fig. \ref{Fig6}, we have the behavior of $E_{n}(\eta)$ vs. $\eta$ for $n=0,1,2$ with $m_j=\pm 1/2$ (a) and the behavior of $E_{m_j}(\eta)$ vs. $\eta$ for $m_j=\pm 1/2,\pm 3/2,\pm 5/2$ with $n=0$ (b), where the solid lines are for the electron and the dashed lines for the positron, and we use $k_z=B=\mu=\mu_5=1$, $\theta=3$ (since $\theta<4$) and $0\leq\eta<1$ (since $\theta<1$). In particular, this Figure is very similar to the one in Fig. \ref{Fig5}. Therefore (or by analogy), the energies of the electron increase with the increase of $n$ and $m_j$, and decrease as a function of $\eta$, while the energies of the positron increase with the increase of $n$ and the decrease of $m_j$, and increase as a function of $\eta$, where the energies of the electron are always lower than those of the positron.

\section{Conclusions}\label{sec5}

In this paper, we analyze the relativistic energy spectrum (or relativistic Landau levels) for charged Dirac fermions with AMM in the presence of the CME and of a NC phase space, where we work with the DE in cylindrical coordinates in the $(3+1)$-dimensional Minkowski spacetime. Using a similarity transformation (given by a unitary operator $U(\phi)$) in order to simplify the problem (i.e., convert the curvilinear gamma matrices into Cartesian gamma matrices), we obtain four coupled first-order differential equations. Subsequently, obtain four non-homogeneous second-order differential equations, where each equation depends on two of the four components of the Dirac spinor. To solve these equations exactly and analytically, we use a change of variable, the asymptotic behavior, and the Frobenius method. Consequently, we obtain the relativistic spectrum for the electron/positron, where such a spectrum is quantized in terms of the radial quantum number $n$ and of the angular quantum number $m_j$, and explicitly depends on the helicity $h$, position and momentum NC parameters $\theta$ and $\eta$, cyclotron frequency $\omega_c$, or NC cyclotron frequency $\omega_c^{NC}=\tau\lambda\omega_c$, anomalous magnetic energy $E_m=\mu_m B$, and on the fermion and chiral chemical potential $\mu$ and $\mu_5$. Thus, we verified that our spectrum generalizes several particular cases found in the literature, i.e., several particular cases can be obtained when we exclude some parameters/quantities and adjust/redefine the quantum numbers in our spectrum.

Regarding the chemical potential $\mu$, it affects/modifies the spectrum of the electron and the positron differently; that is, in the case of the electron, it has the function of decreasing the energies (or shifting all energy levels to lower values), while in the case of the positron, it has the function of increasing the energies (or shifting all energy levels to higher values). Therefore, the energies of the electron are higher in the absence of $\mu$, while in the case of the positron, they are higher in its presence. Now, regarding the parameters $\tau$ and $\lambda$ (dimensionless quantities), we can write them as: $\tau=(1+\sigma\omega_c/\omega_\theta)$ and $\lambda=(1+\sigma\omega_{\eta}/\omega_c)$, where $\omega_\theta\equiv\frac{4}{m\theta}$ and $\omega_{\eta}\equiv\frac{\eta}{m}$ are ``NC angular frequencies'', being $\omega_\theta$ the position NC frequency, and $\omega_{\eta}$ the momentum NC frequency, respectively. Therefore, our spectrum depends on three frequencies, given by $\omega_c$, $\omega_\theta$, and $\omega_\eta$. Furthermore, we also note that even in the absence of the magnetic field ($B=0$), our spectrum still remains quantized/discrete due to the presence of $\eta$ (or $\omega_\eta$), i.e., such NC parameter/frequency acts as a type of ``NC field or potential'' (in a way, that mimics a
uniform magnetic field).

On the other hand, analyzing our spectrum according to the (positive and negative) values of the quantum number $m_j$, or of $m_l$ and $m_s$ (since $m_j=m_l+m_s$), where $m_l=0,\pm 1,\pm 2,\ldots$ and $m_s=\pm 1/2$ are the orbital and spin magnetic quantum numbers, we obtain a Table (see Table \ref{tab1}) which shows four possible settings for the spectrum: two for the electron ($m_j>0$ and $m_j<0$) and two for the positron ($m_j>0$ and $m_j<0$). According to this table, we see that the spectrum of the electron (settings 1 and 2) will always depend on the spin $m_s$ (as well as on $n$) regardless of the value of $m_j$; however, can or not depend on the orbital angular momentum $m_l$ depending on the value of $m_j$ (or even $m_l$), that is, for $m_j>0$ (with $m_l\geq 0$ and $m_s=+1/2$, or $m_l>0$ and $m_s=-1/2$), the spectrum will depend on $m_l$, while for $m_j<0$ (with $m_l<0$ and $m_s=+1/2$, or $m_l\leq 0$ and $m_s=-1/2$), it will not. In other words, if the total/orbital angular momentum of the electron is positive, its spectrum will be affected by $m_l$, and if it is negative, it will not. Consequently, this implies that the spectrum will be larger for $m_j>0$ ($J_z$ is aligned parallel to the magnetic field, i.e., $ \vec{B}\uparrow\uparrow\vec{J}$); therefore, we have $E^{e^-}_{n,m_j>0}>E^{e^-}_{n,m_j<0}$. In particular, even for $m_l=0$, i.e., a null/zero orbital angular momentum (setting 1 for $m_s=+1/2$ and setting 2 for $m_s=-1/2$), the spectra are still not equal (in this case, the spectrum is larger for $m_s=+1/2$, where the ground state ($n=0$) will still depend on the NC parameters). Furthermore, it is important to mention that both $m_j>0$ and $m_j<0$ result in two spectra, where the spectra for each case are the same. That is, for $m_j>0$, we have $E^{e^-}_{n,m_l\geq 0, m_s=+1/2}=E^{e^-}_{n,m_l>0,m_s=-1/2}$ (this means that an electron with orbital angular momentum greater than or equal to zero and spin up, or with orbital angular momentum greater than zero and spin down, it will have the same spectrum), while for $m_j<0$, we have $E^{e^-}_{n,m_l<0, m_s=+1/2}=E^{e^-}_{n,m_l\leq 0,m_s=-1/2}$ (this means that an electron with orbital angular momentum less than zero and spin up, or with orbital angular momentum less than or equal to zero and spin down, it will have the same spectrum).

Already for the case of the positron (settings 3 and 4), something similar and opposite to the electron happens (we can consider this as a consequence of the charge conjugation operation). For example, similar to the case of the electron, the spectrum of the positron will always depend on the spin $m_s$ (as well as on $n$), regardless of the value of $ m_j$; however, can or not depend on the orbital angular momentum $m_l$ depending on the value of $m_j$ (or even $m_l$). However, unlike the case of the electron, the spectrum of the positron for $m_j>0$ (with $m_l\geq 0$ and $m_s=+1/2$, or $m_l>0$ and $m_s=-1/2$), will not depend on $m_l$, while for $m_j<0$ (with $m_l<0$ and $m_s=+1/2$, or $m_l\leq 0$ and $m_s=-1/2$), it will. In other words, if the total/orbital angular momentum of the positron is positive, its spectrum will not be affected by $m_l$, and if it is negative, it will. Consequently, this implies that the spectrum of the positron will be larger for $m_j<0$ ($J_z$ is aligned antiparallel to the magnetic field, i.e., $\vec{B}\uparrow\downarrow\vec{J}$); therefore, we have $E^{e^+}_{n,m_j<0}>E^{e^+}_{n,m_j>0}$. Now, similar to the electron, even for $m_l=0$, i.e., a null/zero orbital angular momentum (setting 3 for $m_s=+1/2$ and setting 4 for $m_s=-1/2$), the spectra of the positron are still not equal (in this case, the spectrum is larger for $m_s=-1/2$, where the ground state ($n=0$) will still depend on the NC parameters). Besides (still similar to the electron), both
$m_j>0$ and $m_j<0$ also result in two spectra, where the spectra for each case are the same, that is, for $m_j>0$, we have $E^{e^+}_{n,m_l\geq 0, m_s=+1/2}=E^{e^+}_{n,m_l>0,m_s=-1/2}$ (this means that a positron with orbital angular momentum greater than or equal to zero and spin up, or with orbital angular momentum greater than zero and spin down, it will have the same spectrum), while for $m_j<0$, we have $E^{e^+}_{n,m_l<0, m_s=+1/2}=E^{e^+}_{n,m_l\leq 0,m_s=-1/2}$ (this means that a positron with orbital angular momentum less than zero and spin up, or with orbital angular momentum less than or equal to zero and spin down, it will have the same spectrum).

Furthermore, we also graphically analyze the behavior of the spectrum (of the electron/positron) as a function of $B$, $\mu$, $\mu_5$, $k_z$, and $\theta$ and $\eta$ for three different values of $n$ and $m_j$. That is, these two quantum numbers allow us to analyze the behavior of the spectrum in two different cases, which are: while $n$ varies ($n=0,1,2$), $m_j$ remains fixed ($m_j=\pm 1/2$), and while $m_j$ varies ($m_j=\pm 1/2,\pm 3/2,\pm 5/2$), $n$ remains fixed ($n=0$). So, in graphs $E_{n}(B)$ vs. $B$ and $E_{m_j}(B)$ vs. $B$, we see that the value of the energies with $n$ varying is exactly the same as the energy with $m_j$ varying, where the energies of the electron increase with the increase of $n$ and $m_j$, while the energies of the positron increase with the increase of $n$ and the decrease of $m_j$. Besides, we see that the energies of the electron practically increase as a function of $B$ (but with one slight exception near $B=4$). Consequently, this implies that the variation of energy as a function of $B$ is positive. For the positron, we see that its energies always increase as a function of $B$. Consequently, this implies that the variation of energy as a function of $B$ is always positive. Now, comparing the energies of the electron with those of the positron, we see that the energies of the electron are always lower than those of the positron (in particular, this happens in all graphs).

In graphs $E_{n}(\mu)$ vs. $\mu$ and $E_{m_j}(\mu)$ vs. $\mu$, we see that the value of the energies with $n$ varying is exactly the same as the energy with $m_j$ varying, where the energies of the electron increase with the increase of $n$ and $m_j$, while the energies of the positron increase with the increase of $n$ and the decrease of $m_j$. Besides, we see that the energies of the electron decrease and those of the positron increase as a function of $\mu$, that is, the chemical potential has the function of decreasing the energies of the electron and increasing those of the positron. Consequently, this implies that the variation of energy as a function of $\mu$ is negative for the electron and positive for the positron. It is important to note that in the case of the electron, the energy also cannot decrease as a function of $\mu$ until it reaches zero, that is, it would also violate the RQM. In graphs $E_{n}(\mu_5)$ vs. $\mu_5$ and $E_{m_j}(\mu_5)$ vs. $\mu_5$, we see that the value of the energies with $n$ varying is exactly the same as the energy with $m_j$ varying, where the energies of the electron increase with the increase of $n$ and $m_j$, while the energies of the positron can increase or decrease with the increase of $n$ and the decrease of $m_j$. In addition, we see that the energies of the electron can increase or decrease (slightly) as a function of $\mu_5$, that is, they increase for $\mu_5>0.5$ and decrease for $\mu_5<0.5$. Consequently, this implies that the variation of energy as a function of $\mu_5$ can be positive or negative. Already in the case of the positron, the energies can be (approximately) constant (for $\mu_5<0.5$) or increase (for $\mu_5>0.5$) as a function of $\mu_5$.

In graphs $E_{n}(k_z)$ vs. $k_z$ and $E_{m_j}(k_z)$ vs. $k_z$ (indeed, very similar to graph of $E$ vs. $\mu_5$), we see that the energies with $n$ varying is exactly the same as the energy with $m_j$ varying, where the energies of the electron increase with the increase of $n$ and $m_j$, while the energies of the positron increase with the increase of $n$ and the decrease of $m_j$. However, in the case of the electron, for $k_z>2$, the energy levels become sufficiently close. In addition, we see that the energies of the electron can increase or decrease (slightly) as a function of $k_z$, that is, they increase for $k_z>0.5$ and decrease for $k_z<0.5$. Consequently, this implies that the variation of energy as a function of $k_z$ can be positive or negative. Already in the case of the positron, the energies can be (approximately) constant (for $k_z<0.5$) or increase (for $k_z>0.5$) as a function of $k_z$. In graphs $E_{n}(\theta)$ vs. $\theta$ and $E_{m_j}(\theta)$ vs. $\theta$, we see that the energies with $n$ varying is exactly the same as the energy with $m_j$ varying, where the energies of the electron increase with the increase of $n$ and $m_j$, while the energies of the positron increase with the increase of $n$ and the decrease of $m_j$. In addition, we see that the energies of the electron decrease as a function of $\theta$, while the energies of the positron increase as a function of $\theta$. Consequently, this implies that the variation of energy as a function of $\theta$ is negative for the electron and positive for the positron. Already in graphs $E_{n}(\eta)$ vs. $\eta$ and $E_{m_j}(\eta)$ vs. $\eta$, they are very similar to the graphs of energies varying with $\theta$; therefore (or by analogy), the energies of the electron increase with the increase of $n$ and $m_j$, and decrease as a function of $\eta$, while the energies of the positron increase with the increase of $n$ and the decrease of $m_j$, and increase as a function of $\eta$, where the energies of the positron are always greater than those of the electron.

\section*{Acknowledgments}

\hspace{0.5cm}The author would like to thank the anonymous referees for the careful reading of the paper as well as for the remarkable suggestions and constructive comments that substantially helped in improving the quality of the paper. In addition, the author would also like to thank the Conselho Nacional de Desenvolvimento Cient\'{\i}fico e Tecnol\'{o}gico (CNPq) for financial support through the postdoc grant No. 175392/2023-4, as well as the Department of Physics at the Universidade Federal da Para\'{i}ba (UFPB) for hospitality and support.

\section*{Data availability statement}

\hspace{0.5cm} This manuscript has no associated data or the data will not be deposited. [Authors’ comment: Data sharing not applicable to this article as no datasets were generated or analysed during the current study].

\section*{Code Availability Statement}

\hspace{0.5cm} This manuscript has no associated code/software. [Author’s comment: Code/Software sharing not applicable to this article as no code/software was generated or analysed during the current study].


\appendix
\section{Detailed calculations on the energy spectrum}\label{sec6}

Here, we provide some more detailed calculations on how the energy spectrum was obtained. So, using the Frobenius method (as well as the asymptotic behavior for large and small $\rho$), we can write the solutions of \eqref{dirac18} as
\begin{equation}\label{A1}
f_1 (\rho)=e^{-\rho/2}\rho^{\gamma}\sum_{N=0}^\infty a_N \rho^N, \ \  f_2 (\rho)=e^{-\rho/2}\rho^{\gamma'}\sum_{N=0}^\infty b_N \rho^N, \ \ f_3 (\rho)=e^{-\rho/2}\rho^{\gamma}\sum_{N=0}^\infty c_N \rho^N, \ \ f_4 (\rho)=e^{-\rho/2}\rho^{\gamma'}\sum_{N=0}^\infty d_N \rho^N,
\end{equation}
or better
\begin{equation}\label{A2}
f_1 (\rho)=e^{-\rho/2}\sum_{N=0}^\infty a_N \rho^{N+\gamma}, \ \ f_2 (\rho)=e^{-\rho/2}\sum_{N=0}^\infty b_N \rho^{N+\gamma'}, \ \ f_3 (\rho)=e^{-\rho/2}\sum_{N=0}^\infty c_N \rho^{N+\gamma}, \ \ f_4 (\rho)=e^{-\rho/2}\sum_{N=0}^\infty d_N \rho^{N+\gamma'},
\end{equation}
where its first and second derivatives are given by
\begin{eqnarray}\label{derivatives}
&& \frac{df_1 (\rho)}{d\rho}=-\frac{1}{2}e^{-\rho/2}\sum_{N=0}^\infty a_N \rho^{N+\gamma}+e^{-\rho/2}\sum_{N=0}^\infty (N+\gamma) a_N\rho^{N+\gamma-1},
\nonumber\\
&& \frac{d^2f_1 (\rho)}{d\rho^2}=\frac{1}{4}e^{-\rho/2}\sum_{N=0}^\infty a_N \rho^{N+\gamma}-e^{-\rho/2}\sum_{N=0}^\infty (N+\gamma)a_N \rho^{N+\gamma-1}+e^{-\rho/2}\sum_{N=0}^\infty (N+\gamma)(N+\gamma-1) a_N\rho^{N+\gamma-2},
\nonumber\\
&& \frac{df_2 (\rho)}{d\rho}=-\frac{1}{2}e^{-\rho/2}\sum_{N=0}^\infty b_N \rho^{N+\gamma'}+e^{-\rho/2}\sum_{N=0}^\infty (N+\gamma') b_N\rho^{N+\gamma'-1},
\nonumber\\
&& \frac{d^2f_2 (\rho)}{d\rho^2}=\frac{1}{4}e^{-\rho/2}\sum_{N=0}^\infty b_N \rho^{N+\gamma'}-e^{-\rho/2}\sum_{N=0}^\infty (N+\gamma')b_N \rho^{N+\gamma'-1}+e^{-\rho/2}\sum_{N=0}^\infty (N+\gamma')(N+\gamma'-1) b_N\rho^{N+\gamma'-2},
\nonumber\\
&& \frac{df_3 (\rho)}{d\rho}=-\frac{1}{2}e^{-\rho/2}\sum_{N=0}^\infty c_N \rho^{N+\gamma}+e^{-\rho/2}\sum_{N=0}^\infty (N+\gamma) c_N\rho^{N+\gamma-1},
\nonumber\\
&& \frac{d^2f_3 (\rho)}{d\rho^2}=\frac{1}{4}e^{-\rho/2}\sum_{N=0}^\infty c_N \rho^{N+\gamma}-e^{-\rho/2}\sum_{N=0}^\infty (N+\gamma)c_N \rho^{N+\gamma-1}+e^{-\rho/2}\sum_{N=0}^\infty (N+\gamma)(N+\gamma-1) c_N\rho^{N+\gamma-2},
\nonumber\\
&& \frac{df_4 (\rho)}{d\rho}=-\frac{1}{2}e^{-\rho/2}\sum_{N=0}^\infty d_N \rho^{N+\gamma'}+e^{-\rho/2}\sum_{N=0}^\infty (N+\gamma') d_N\rho^{N+\gamma'-1},
\nonumber\\
&& \frac{d^2f_4 (\rho)}{d\rho^2}=\frac{1}{4}e^{-\rho/2}\sum_{N=0}^\infty d_N \rho^{N+\gamma'}-e^{-\rho/2}\sum_{N=0}^\infty (N+\gamma')d_N \rho^{N+\gamma'-1}+e^{-\rho/2}\sum_{N=0}^\infty (N+\gamma')(N+\gamma'-1)d_N\rho^{N+\gamma'-2}.
\nonumber\\
\end{eqnarray}

Consequently, this implies that
\begin{eqnarray}\label{A3}
&& \rho\frac{d^2 f_1(\rho)}{d\rho^2}+\frac{df_1 (\rho)}{d\rho}=e^{-\rho/2}\left[\frac{1}{4}\sum_{N=0}^\infty a_N \rho^{N+\gamma+1}-\sum_{N=0}^\infty\left(N+\gamma+\frac{1}{2}\right)a_N \rho^{N+\gamma}+\sum_{N=0}^\infty (N+\gamma)^2 a_N \rho^{N+\gamma-1}\right],
\nonumber\\
&& \rho\frac{d^2 f_2 (\rho)}{d\rho^2}+\frac{df_2 (\rho)}{d\rho}=e^{-\rho/2}\left[\frac{1}{4}\sum_{N=0}^\infty b_N \rho^{N+\gamma'+1}-\sum_{N=0}^\infty\left(N+\gamma'+\frac{1}{2}\right)b_N \rho^{N+\gamma'}+\sum_{N=0}^\infty (N+\gamma')^2 b_N \rho^{N+\gamma'-1}\right],
\nonumber\\
&& \rho\frac{d^2 f_3 (\rho)}{d\rho^2}+\frac{df_3(\rho)}{d\rho}=e^{-\rho/2}\left[\frac{1}{4}\sum_{N=0}^\infty c_N \rho^{N+\gamma+1}-\sum_{N=0}^\infty\left(N+\gamma+\frac{1}{2}\right)c_N \rho^{N+\gamma}+\sum_{N=0}^\infty (N+\gamma)^2 c_N \rho^{N+\gamma-1}\right],
\nonumber\\
&& \rho\frac{d^2 f_4(\rho)}{d\rho^2}+\frac{df_4 (\rho)}{d\rho}=e^{-\rho/2}\left[\frac{1}{4}\sum_{N=0}^\infty d_N \rho^{N+\gamma'+1}-\sum_{N=0}^\infty\left(N+\gamma'+\frac{1}{2}\right)d_N \rho^{N+\gamma'}+\sum_{N=0}^\infty (N+\gamma')^2 d_N \rho^{N+\gamma'-1}\right].
\nonumber\\
\end{eqnarray}

So, substituting \eqref{A3} into \eqref{dirac18}, we obtain (we have already eliminated the exponentials, and after a simplification)
\begin{equation}\label{A4}
\left(
           \begin{array}{c}
            \sum_{N=0}^\infty\{\left[\mathbb{B}_1+\frac{\sigma}{2}\left(m_j-\frac{1}{2}\right)-(N+\gamma'+\frac{1}{2})\right]d_N-\mathbb{B}_2 b_N\} \rho^{N+\gamma'}+\sum_{N=0}^\infty N(N+2\gamma') d_N \rho^{N+\gamma'-1}\\
            \sum_{N=0}^\infty\{\left[\mathbb{D}_1+\frac{\sigma}{2}\left(m_j-\frac{1}{2}\right)-(N+\gamma'+\frac{1}{2})\right]b_N-\mathbb{D}_2d_N\} \rho^{N+\gamma'}+\sum_{N=0}^\infty N(N+2\gamma') b_N \rho^{N+\gamma'-1}\\
            \sum_{N=0}^\infty\{\left[\mathbb{C}_1+\frac{\sigma}{2}\left(m_j+\frac{1}{2}\right)-(N+\gamma+\frac{1}{2})\right]c_N-\mathbb{C}_2 a_N\}\rho^{N+\gamma}+\sum_{N=0}^\infty N(N+2\gamma)c_N \rho^{N+\gamma-1} \\
            \sum_{N=0}^\infty\{\left[\mathbb{A}_1+\frac{\sigma}{2}\left(m_j+\frac{1}{2}\right)-(N+\gamma+\frac{1}{2})\right]a_N-\mathbb{A}_2 c_N\}\rho^{N+\gamma}+\sum_{N=0}^\infty N(N+2\gamma)a_N \rho^{N+\gamma-1} \\
           \end{array}
         \right)=0,
\end{equation}
or yet
\begin{equation}\label{A5}
\left(
           \begin{array}{c}
            \sum_{N=0}^\infty\left\{\left[\mathbb{B}_1-N-\frac{1}{2}-\frac{\vert m_j+\frac{1}{2}\vert-\sigma(m_j-\frac{1}{2})}{2}\right]d_N-\mathbb{B}_2 b_N\right\} \rho^{N+\gamma'}+\sum_{N=1}^\infty N(N+\vert m_j+\frac{1}{2}\vert) d_N \rho^{N+\gamma'-1}\\
            \sum_{N=0}^\infty\left\{\left[\mathbb{D}_1-N-\frac{1}{2}-\frac{\vert m_j+\frac{1}{2}\vert-\sigma(m_j-\frac{1}{2})}{2}\right]b_N-\mathbb{D}_2d_N\right\} \rho^{N+\gamma'}+\sum_{N=1}^\infty N(N+\vert m_j+\frac{1}{2}\vert) b_N \rho^{N+\gamma'-1}\\
            \sum_{N=0}^\infty\left\{\left[\mathbb{C}_1-N-\frac{1}{2}-\frac{\vert m_j-\frac{1}{2}\vert-\sigma(m_j+\frac{1}{2})}{2}\right]c_N-\mathbb{C}_2 a_N\right\}\rho^{N+\gamma}+\sum_{N=1}^\infty N(N+\vert m_j-\frac{1}{2}\vert)c_N \rho^{N+\gamma-1} \\
            \sum_{N=0}^\infty\left\{\left[\mathbb{A}_1-N-\frac{1}{2}-\frac{\vert m_j-\frac{1}{2}\vert-\sigma(m_j+\frac{1}{2})}{2}\right]a_N-\mathbb{A}_2 c_N\right\}\rho^{N+\gamma}+\sum_{N=1}^\infty N(N+\vert m_j-\frac{1}{2}\vert)a_N \rho^{N+\gamma-1} \\
           \end{array}
         \right)=0,
\end{equation}
where we use $\gamma=\frac{\vert m_j-\frac{1}{2}\vert}{2}$ and $\gamma'=\frac{\vert m_j+\frac{1}{2}\vert}{2}$, and the fact that $\sum_{N=0}^{\infty}N(N+\vert m_j\pm \frac{1}{2}\vert)=\sum_{N=1}^{\infty}N(N+\vert m_j\pm \frac{1}{2}\vert)$.

Now, doing $N\to N-1$ in the first terms, we have
\begin{equation}\label{A6}
\left(
           \begin{array}{c}
            \sum_{N=1}^\infty\left\{\left[\mathbb{B}_1-N+\frac{1}{2}-\frac{\vert m_j+\frac{1}{2}\vert-\sigma(m_j-\frac{1}{2})}{2}\right]d_{N-1}-\mathbb{B}_2 b_{N-1}\right\} \rho^{N+\gamma'-1}+\sum_{N=1}^\infty N(N+\vert m_j+\frac{1}{2}\vert) d_N \rho^{N+\gamma'-1}\\
            \sum_{N=1}^\infty\left\{\left[\mathbb{D}_1-N+\frac{1}{2}-\frac{\vert m_j+\frac{1}{2}\vert-\sigma(m_j-\frac{1}{2})}{2}\right]b_{N-1}-\mathbb{D}_2d_{N-1}\right\} \rho^{N+\gamma'-1}+\sum_{N=1}^\infty N(N+\vert m_j+\frac{1}{2}\vert) b_N \rho^{N+\gamma'-1}\\
            \sum_{N=1}^\infty\left\{\left[\mathbb{C}_1-N+\frac{1}{2}-\frac{\vert m_j-\frac{1}{2}\vert-\sigma(m_j+\frac{1}{2})}{2}\right]c_{N-1}-\mathbb{C}_2 a_{N-1}\right\}\rho^{N+\gamma-1}+\sum_{N=1}^\infty N(N+\vert m_j-\frac{1}{2}\vert)c_N \rho^{N+\gamma-1} \\
            \sum_{N=1}^\infty\left\{\left[\mathbb{A}_1-N+\frac{1}{2}-\frac{\vert m_j-\frac{1}{2}\vert-\sigma(m_j+\frac{1}{2})}{2}\right]a_{N-1}-\mathbb{A}_2 c_{N-1}\right\}\rho^{N+\gamma-1}+\sum_{N=1}^\infty N(N+\vert m_j-\frac{1}{2}\vert)a_N \rho^{N+\gamma-1} \\
           \end{array}
         \right)=0,
\end{equation}
or yet
\begin{equation}\label{A7}
\left(
           \begin{array}{c}
            \sum_{N=1}^\infty\left(
           \begin{array}{c}
            \left\{\left[\mathbb{B}_1-N+\frac{1}{2}-\frac{\vert m_j+\frac{1}{2}\vert-\sigma(m_j-\frac{1}{2})}{2}\right]d_{N-1}-\mathbb{B}_2 b_{N-1}+N(N+\vert m_j+\frac{1}{2}\vert) d_N\right\}\\
            \left\{\left[\mathbb{D}_1-N+\frac{1}{2}-\frac{\vert m_j+\frac{1}{2}\vert-\sigma(m_j-\frac{1}{2})}{2}\right]b_{N-1}-\mathbb{D}_2d_{N-1}+N(N+\vert m_j+\frac{1}{2}\vert) b_N\right\}\\
           \end{array}
         \right)\rho^{N+\gamma'-1}\\
           \sum_{N=1}^\infty\left(
           \begin{array}{c}
           \left\{\left[\mathbb{C}_1-N+\frac{1}{2}-\frac{\vert m_j-\frac{1}{2}\vert-\sigma(m_j+\frac{1}{2})}{2}\right]c_{N-1}-\mathbb{C}_2 a_{N-1}+N(N+\vert m_j-\frac{1}{2}\vert)c_N\right\}\\
            \left\{\left[\mathbb{A}_1-N+\frac{1}{2}-\frac{\vert m_j-\frac{1}{2}\vert-\sigma(m_j+\frac{1}{2})}{2}\right]a_{N-1}-\mathbb{A}_2 c_{N-1}+N(N+\vert m_j-\frac{1}{2}\vert)a_N\right\}\\
           \end{array}
         \right)\rho^{N+\gamma-1}\\
           \end{array}
         \right)=0,
\end{equation}
where implies (since $\rho^{N+\gamma-1}\neq 0$ and $\rho^{N+\gamma'-1}\neq 0$)
\begin{equation}\label{A8}
\left(
           \begin{array}{c}
            \left\{\left[\mathbb{B}_1-N+\frac{1}{2}-\frac{\vert m_j+\frac{1}{2}\vert-\sigma(m_j-\frac{1}{2})}{2}\right]d_{N-1}-\mathbb{B}_2 b_{N-1}+N(N+\vert m_j+\frac{1}{2}\vert) d_N\right\}\\
            \left\{\left[\mathbb{D}_1-N+\frac{1}{2}-\frac{\vert m_j+\frac{1}{2}\vert-\sigma(m_j-\frac{1}{2})}{2}\right]b_{N-1}-\mathbb{D}_2d_{N-1}+N(N+\vert m_j+\frac{1}{2}\vert) b_N\right\}\\
           \left\{\left[\mathbb{C}_1-N+\frac{1}{2}-\frac{\vert m_j-\frac{1}{2}\vert-\sigma(m_j+\frac{1}{2})}{2}\right]c_{N-1}-\mathbb{C}_2 a_{N-1}+N(N+\vert m_j-\frac{1}{2}\vert)c_N\right\}\\
            \left\{\left[\mathbb{A}_1-N+\frac{1}{2}-\frac{\vert m_j-\frac{1}{2}\vert-\sigma(m_j+\frac{1}{2})}{2}\right]a_{N-1}-\mathbb{A}_2 c_{N-1}+N(N+\vert m_j-\frac{1}{2}\vert)a_N\right\}\\
           \end{array}
         \right)=0.
\end{equation}

Now, doing $N\to N+1$, we have
\begin{equation}\label{A9}
\left(
           \begin{array}{c}
            \left\{\left[\mathbb{B}_1-N-\frac{1}{2}-\frac{\vert m_j+\frac{1}{2}\vert-\sigma(m_j-\frac{1}{2})}{2}\right]d_{N}-\mathbb{B}_2 b_{N}+(N+1)(N+1+\vert m_j+\frac{1}{2}\vert) d_{N+1}\right\}\\
            \left\{\left[\mathbb{D}_1-N-\frac{1}{2}-\frac{\vert m_j+\frac{1}{2}\vert-\sigma(m_j-\frac{1}{2})}{2}\right]b_{N}-\mathbb{D}_2d_{N}+(N+1)(N+1+\vert m_j+\frac{1}{2}\vert) b_{N+1}\right\}\\
           \left\{\left[\mathbb{C}_1-N-\frac{1}{2}-\frac{\vert m_j-\frac{1}{2}\vert-\sigma(m_j+\frac{1}{2})}{2}\right]c_{N}-\mathbb{C}_2 a_{N}+(N+1)(N+1+\vert m_j-\frac{1}{2}\vert)c_{N+1}\right\}\\
            \left\{\left[\mathbb{A}_1-N-\frac{1}{2}-\frac{\vert m_j-\frac{1}{2}\vert-\sigma(m_j+\frac{1}{2})}{2}\right]a_{N}-\mathbb{A}_2 c_{N}+(N+1)(N+1+\vert m_j-\frac{1}{2}\vert)a_{N+1}\right\}\\
           \end{array}
         \right)=0.
\end{equation}

According to Refs. \cite{Chaudhuri,Connell}, to obtain a well-behaved function (finite or normalizable solution), the power series must terminate at some $N=N'$, such that the recursion relations spits out $a_{N'+1}=b_{N'+1}=c_{N'+1}=d_{N'+1}=0$ (for convenience, here, we will use $n$ instead of $N'$). Therefore, applying this condition in \eqref{A9}, we obtain
\begin{equation}\label{A10}
\left(
\begin{array}{cc}
 \mathbb{B}_1-N_{2,4} & \ \ -\mathbb{B}_2 \\
 -\mathbb{D}_2 & \ \ \mathbb{D}_1-N_{2,4} \\
\end{array}
\right)\left(
           \begin{array}{c}
            d_{n} \\
            b_{n} \\
           \end{array}
         \right)=0, \ \ \left(
\begin{array}{cc}
 \mathbb{C}_1-N_{1,3} & \ \ -\mathbb{C}_2 \\
 -\mathbb{A}_2 & \ \ \mathbb{A}_1-N_{1,3} \\
\end{array}
\right)\left(
           \begin{array}{c}
            c_{n} \\
            a_{n} \\
           \end{array}
         \right)=0,
\end{equation}
where we define
\begin{equation}\label{A11}
N_{2,4}\equiv\left(n+\frac{1}{2}+\frac{\vert m_j+\frac{1}{2}\vert-\sigma(m_j-\frac{1}{2})}{2}\right), \ \ N_{1,3}\equiv\left(n+\frac{1}{2}+\frac{\vert m_j-\frac{1}{2}\vert-\sigma(m_j+\frac{1}{2})}{2}\right).
\end{equation}

According to Refs. \cite{Chaudhuri,Connell}, for the equations in \eqref{A10} to have non-trivial solutions ($f_{1,2,3,4}(\rho)\neq 0$), we must have
\begin{equation}\label{A12}
det\left(
\begin{array}{cc}
 \mathbb{B}_1-N_{2,4} & \ \ -\mathbb{B}_2 \\
 -\mathbb{D}_2 & \ \ \mathbb{D}_1-N_{2,4} \\
\end{array}
\right)=0, \ \ det\left(
\begin{array}{cc}
 \mathbb{C}_1-N_{1,3} & \ \ -\mathbb{C}_2 \\
 -\mathbb{A}_2 & \ \ \mathbb{A}_1-N_{1,3} \\
\end{array}
\right)=0,
\end{equation}
where implies that
\begin{eqnarray}
&& \mathbb{B}_1 \mathbb{D}_1-\mathbb{B}_2 \mathbb{D}_2-N_{2,4}[\mathbb{B}_1+\mathbb{D}_1]+N_{2,4}^2=0,\label{A13}\\
&& \mathbb{C}_1 \mathbb{A}_1-\mathbb{C}_2 \mathbb{A}_2-N_{1,3}[\mathbb{C}_1+\mathbb{A}_1]+N_{1,3}^2=0,\label{A14}
\end{eqnarray}
or yet
\begin{eqnarray}
&& \mathbb{B}'_1 \mathbb{D}'_1-\mathbb{B}'_2 \mathbb{D}'_2-4m\omega N_{2,4}[\mathbb{B}'_1+\mathbb{D}'_1]+16m^2\omega^2N_{2,4}^2=0,\label{A15}\\
&& \mathbb{C}'_1 \mathbb{A}'_1-\mathbb{C}'_2 \mathbb{A}'_2-4mN_{1,3}[\mathbb{C}'_1+\mathbb{A}'_1]+16m^2\omega^2 N_{1,3}^2=0,\label{A16}
\end{eqnarray}
where we use \eqref{BDCA}.

So, to obtain the energy spectra, we must substitute \eqref{define} and \eqref{BD} in \eqref{A15}, as well as substitute \eqref{define} and \eqref{CA} in \eqref{A16}. Starting with \eqref{A15}, the spectrum is obtained as follows. Therefore, through \eqref{define} and \eqref{BD}, we have
\begin{align}\label{A17}
\mathbb{B}'_1 \mathbb{D}'_1-\mathbb{B}'_2 \mathbb{D}'_2 &=(E_-\bar{E}_+ + A_+ A_-)(E_+\bar{E}_- + A_+ A_-)-(A_-\bar{E}_- + E_- A_+)(A_-\bar{E}_+ + E_+ A_+),
\nonumber\\
&= E_-E_+\bar{E}_-\bar{E}_+ +[A_+A_-]^2-A^2_-\bar{E}_-\bar{E}_+-A^2_+ E_- E_+,
\end{align}
where
\begin{align}\label{A18}
E_-E_+\bar{E}_-\bar{E}_+ &=\left[\frac{(E+\mu)-(m+\mu_m B)}{\tau}\right]\left[\frac{(E+\mu)+(m+\mu_m B)}{\tau}\right]\left[\frac{(E+\mu)-(m-\mu_m B)}{\tau}\right]\left[\frac{(E+\mu)+(m-\mu_m B)}{\tau}\right],
\nonumber\\
&=\frac{1}{\tau^4}\left[(E+\mu)^2-(m+\mu_m B)^2\right]\left[(E+\mu)^2-(m-\mu_m B)^2\right],
\nonumber\\
&=\frac{1}{\tau^4}\left\{\left[(E+\mu)^2-(m^2+(\mu_m B)^2)\right]^2-(2m\mu_m B)^2\right\},
\nonumber\\
&=\frac{1}{\tau^4}\left\{(E+\mu)^4-2(E+\mu)^2(m^2+(\mu_m B)^2)+(m^2+(\mu_m B)^2)^2-(2m\mu_m B)^2\right\},
\nonumber\\
[A_+A_-]^2 &=\left[\frac{(\mu_5+k_z)}{\tau}\frac{(\mu_5-k_z)}{\tau}\right]^2=\frac{(\mu_5^2-k_z^2)^2}{\tau^4},
\nonumber\\
A^2_-\bar{E}_-\bar{E}_+ &=\left[\frac{(\mu_5-k_z)}{\tau}\right]^2 \left[\frac{(E+\mu)-(m-\mu_m B)}{\tau}\right]\left[\frac{(E+\mu)+(m-\mu_m B)}{\tau}\right],
\nonumber\\
&=\frac{(\mu_5-k_z)^2}{\tau^4}\left[(E+\mu)^2-(m-\mu_m B)^2\right]
\nonumber\\
A^2_-E_-E_+ &=\left[\frac{(\mu_5+k_z)}{\tau}\right]^2 \left[\frac{(E+\mu)-(m+\mu_m B)}{\tau}\right]\left[\frac{(E+\mu)+(m+\mu_m B)}{\tau}\right],
\nonumber\\
&=\frac{(\mu_5+k_z)^2}{\tau^4}\left[(E+\mu)^2-(m+\mu_m B)^2\right],
\end{align}
being
\begin{align}\label{A19}
A^2_-\bar{E}_-\bar{E}_+ +A^2_-E_-E_+&=\frac{(\mu_5-k_z)^2}{\tau^4}\left[(E+\mu)^2-(m-\mu_m B)^2\right]+\frac{(\mu_5+k_z)^2}{\tau^4}\left[(E+\mu)^2-(m+\mu_m B)^2\right]
\nonumber\\
&=\frac{1}{\tau^4}\left[2(E+\mu)^2(\mu_5^2+k^2_z)-2(\mu_5^2+k_z^2)(m^2+(\mu_m B)^2)-8m\mu_5k_z\mu_m B\right].
\end{align}

Consequently, we have
\begin{align}\label{A20}
\mathbb{B}'_1 \mathbb{D}'_1-\mathbb{B}'_2 \mathbb{D}'_2 &=\frac{1}{\tau^4}\left[(E+\mu)^4-2(E+\mu)^2(m^2+(\mu_m B)^2+\mu_5^2+k^2_z)+m^4+(\mu_m B)^4+(\mu_5^2-k_z^2)^2\right]
\nonumber\\
&+\frac{1}{\tau^4}\left[2(\mu_5^2+k_z^2)(m^2+(\mu_m B)^2)+8m\mu_5k_z\mu_m B\right].
\end{align}

Besides, we have
\begin{equation}\label{A21}
\mathbb{B}'_1+\mathbb{D}'_1=(E_-\bar{E}_+ + A_+ A_-)+(E_+\bar{E}_- + A_+ A_-)=E_-\bar{E}_++E_+\bar{E}_- + 2A_+ A_-,
\end{equation}
where
\begin{align}\label{A22}
A_+A_- &=\frac{(\mu_5+k_z)}{\tau}\frac{(\mu_5-k_z)}{\tau}=\frac{(\mu_5^2-k_z^2)}{\tau^2},
\nonumber\\
E_-\bar{E}_+ &=\left[\frac{(E+\mu)-(m+\mu_m B)}{\tau}\right]\left[\frac{(E+\mu)+(m-\mu_m B)}{\tau}\right],
\nonumber\\
&=\frac{1}{\tau^2}\left[(E+\mu)^2-m^2+(\mu_m B)^2-2(E+\mu)\mu_m B\right],
\nonumber\\
E_+\bar{E}_- &=\left[\frac{(E+\mu)+(m+\mu_m B)}{\tau}\right]\left[\frac{(E+\mu)-(m-\mu_m B)}{\tau}\right],
\nonumber\\
&=\frac{1}{\tau^2}\left[(E+\mu)^2-m^2+(\mu_m B)^2+2(E+\mu)\mu_m B\right].
\end{align}

Consequently, we have
\begin{equation}\label{A23}
\mathbb{B}'_1+\mathbb{D}'_1=\frac{1}{\tau^2}\left[2(E+\mu)^2-2m^2+2(\mu_m B)^2+2\mu_5^2-2k_z^2\right].
\end{equation}

Therefore, substituting \eqref{A20} and \eqref{A23} into \eqref{A15}, we obtain
\begin{align}\label{A24}
(E+\mu)^4 & -2(E+\mu)^2(m^2+(\mu_m B)^2+\mu_5^2+k^2_z) +m^4+(\mu_m B)^4+(\mu_5^2-k_z^2)^2+2(\mu_5^2+k_z^2)(m^2+(\mu_m B)^2)
\nonumber\\
& +8m\mu_5k_z\mu_m B-4\tau^2m\omega N_{2,4}\left[2(E+\mu)^2-2m^2+2(\mu_m B)^2+2\mu_5^2-2k_z^2\right]+16\tau^4 m^2\omega^2N_{2,4}^2=0,
\end{align}
or yet
\begin{align}\label{A25}
(E+\mu)^4 & -2(E+\mu)^2(m^2+(\mu_m B)^2+\mu_5^2+k^2_z+4\tau^2m\omega N_{2,4})+k_z^4+m^4+\mu_5^4+(\mu_m B)^4+2(m k_z)^2+2(\mu_5\mu_m B)^2
\nonumber\\
& -2(\mu_5 k_z)^2-2(m\mu_m B)^2+2(m\mu_5)^2+2(k_z\mu_m B)^2+8m\mu_5k_z\mu_m B+8\tau^2m\omega N_{2,4}\left[k_z^2+m^2-(\mu_m B)^2-\mu_5^2\right]
\nonumber\\
& +16\tau^4 m^2\omega^2N_{2,4}^2=0.
\end{align}

However, using the following relation
\begin{align}\label{A26}
[k_z^2+m^2-\mu_5^2-(\mu_mB)^2]^2+4[m\mu_5+k_z \mu_m B]^2 & =k_z^4+m^4+\mu_5^4+(\mu_m B)^4+2(m k_z)^2+2(\mu_5\mu_m B)^2-2(\mu_5 k_z)^2
\nonumber\\
& -2(m\mu_m B)^2+2(m\mu_5)^2+2(k_z\mu_m B)^2+8m\mu_5k_z\mu_m B,
\end{align}
we can then simplify \eqref{A25} as
\begin{align}\label{A27}
(E+\mu)^4 & -2(E+\mu)^2(m^2+(\mu_m B)^2+\mu_5^2+k^2_z+4\tau^2m\omega N_{2,4})+[k_z^2+m^2-\mu_5^2-(\mu_mB)^2]^2+4[m\mu_5+k_z \mu_m B]^2
\nonumber\\
& +8\tau^2m\omega N_{2,4}\left[k_z^2+m^2-(\mu_m B)^2-\mu_5^2\right]+16\tau^4 m^2\omega^2N_{2,4}^2=0.
\end{align}

In particular, this results in a fourth-degree polynomial equation, given by
\begin{equation}\label{A28}
(E+\mu)^4+\mathcal{B}(E+\mu)^2+\mathcal{C}=0,
\end{equation}
where we define
\begin{eqnarray}
&& \mathcal{B}\equiv -2[k^2_z+m^2+\mu_5^2+(\mu_m B)^2+4\tau^2m\omega N_{2,4}],\label{A29}\\
&& \mathcal{C}\equiv [k_z^2+m^2-\mu_5^2-(\mu_mB)^2]^2+4[m\mu_5+k_z \mu_m B]^2+8\tau^2m\omega N_{2,4}\left[k_z^2+m^2-\mu_5^2-(\mu_m B)^2\right]+16\tau^4 m^2\omega^2N_{2,4}^2.\nonumber
\\ \label{A30}
\end{eqnarray}

Before continuing, it is important to note that comparing our \eqref{A30} with (A27) from Ref. \cite{Chaudhuri}, we see that this reference made a ``small error'', that is, it obtained $4(m \mu_5 + p_z a B)$, while the correct result would be $4(m \mu_5 + p_z a B)^2$. However, we believe it was just a minor typo as it was later corrected. Then, using Bhaskara's formula, the solution of \eqref{A28} is given by the following (four) relativistic dispersion relations (since it has the form $E^2=p^2+m^2$)
\begin{eqnarray}
&& (E^++\mu)^2=\frac{-\mathcal{B}+\sqrt{\mathcal{B}^2-4\mathcal{C}}}{2},\\
&& (E^-+\mu)^2=\frac{-\mathcal{B}-\sqrt{\mathcal{B}^2-4\mathcal{C}}}{2},\\
&& (E^++\mu)^2=\frac{-\mathcal{B}+\sqrt{\mathcal{B}^2-4\mathcal{C}}}{2}\\
&& (E^-+\mu)^2=\frac{-\mathcal{B}-\sqrt{\mathcal{B}^2-4\mathcal{C}}}{2},
\end{eqnarray}
or better (compact form)
\begin{equation}\label{A31}
(E^\pm+\mu)^2=\frac{-\mathcal{B}\pm\sqrt{\mathcal{B}^2-4\mathcal{C}}}{2}.
\end{equation}

Developing the term $\mathcal{B}^2-4\mathcal{C}$, we have
\begin{align}\label{A32}
\mathcal{B}^2-4\mathcal{C}&=4[k^2_z+m^2+\mu_5^2+(\mu_m B)^2+4\tau^2m\omega N_{2,4}]^2-4[k_z^2+m^2-\mu_5^2-(\mu_mB)^2]^2-16[m\mu_5+k_z \mu_m B]^2
\nonumber\\
&-32\tau^2m\omega N_{2,4}\left[k_z^2+m^2-\mu_5^2-(\mu_m B)^2\right]-64\tau^4 m^2\omega^2N_{2,4}^2,
\nonumber\\
&=4\{[k^2_z+m^2+\mu_5^2+(\mu_m B)^2]^2-[k_z^2+m^2-\mu_5^2-(\mu_mB)^2]^2+16\tau^2m\omega N_{2,4}[\mu_5^2+(\mu_m B)^2]-4[m\mu_5+k_z \mu_m B]^2\},
\nonumber\\
&=16\{[k_z^2+m^2][\mu_5^2+(\mu_m B)^2]+4\tau^2m\omega N_{2,4}[\mu_5^2+(\mu_m B)^2]-[m\mu_5+k_z \mu_m B]^2\},
\nonumber\\
&=16\{[m^2+4\tau^2m\omega N_{2,4}](\mu_m B)^2+[k_z^2+4\tau^2m\omega N_{2,4}]\mu_5^2-2m\mu_5k_z\mu_m B]\},
\nonumber\\
&=16\{[m\mu_m B-\mu_5 k_z]^2+4\tau^2m\omega N_{2,4}[(\mu_m B)^2+\mu_5^2]\}.
\end{align}

Therefore, using \eqref{A29} and \eqref{A32}, and knowing that $\omega=\frac{\lambda\omega_c}{2\tau}$, \eqref{A31} becomes
\begin{equation}\label{A33}
(E^\pm+\mu)^2=k^2_z+m^2+\mu_5^2+(\mu_m B)^2+2m\tau\lambda\omega_c N_{2,4}\pm2\sqrt{[m\mu_m B-\mu_5 k_z]^2+2m\tau\lambda\omega_c N_{2,4}[(\mu_m B)^2+\mu_5^2]},
\end{equation}
where generates the following relativistic energy spectrum (since it has the form $E^\pm=\pm\sqrt{p^2+m^2}$)
\begin{equation}\label{A34}
E^\pm_{N_{2,4}}=-\mu\pm\sqrt{k^2_z+m^2+\mu_5^2+(\mu_m B)^2+2m\tau\lambda\omega_c N_{2,4}\pm2\sqrt{[m\mu_m B-\mu_5 k_z]^2+2m\tau\lambda\omega_c N_{2,4}[(\mu_m B)^2+\mu_5^2]}}.
\end{equation}

Similarly, the spectrum generated through \eqref{A16} is given by
\begin{equation}\label{A35}
E^\pm_{N_{1,3}}=-\mu\pm\sqrt{k^2_z+m^2+\mu_5^2+(\mu_m B)^2+2m\tau\lambda\omega_c N_{1,3}\pm2\sqrt{[m\mu_m B-\mu_5 k_z]^2+2m\tau\lambda\omega_c N_{1,3}[(\mu_m B)^2+\mu_5^2]}}.
\end{equation}

That is, the spectrum \eqref{A35} is nothing more than the spectrum \eqref{A34} with $N_{2,4}\to N_{1,3}$.


\begin{thebibliography}{99}
\section*{References}

\bibitem{P1} P. A. M. Dirac, Proc. R. Soc. London A {\bf 117}, 610–624 (1928).

\bibitem{P2} P. A. M. Dirac, Proc. R. Soc. London A {\bf 118}, 351–361 (1928).

\bibitem{Greiner} W. Greiner, {\it Relativistic Quantum Mechanics: Wave Equations}, vol. 3 (Springer, Berlin, 2000).

\bibitem{Bjorken} J. D. Bjorken, and S. D. Drell, {\it Relativistic quantum mechanics} (Mcgraw-Hill College, New York, San Francisco, Toronto, London, 1964).

\bibitem{Grandy} W. T. Grandy, {\it Relativistic quantum mechanics of leptons and fields}, vol. 41 (Springer Science and Business Media, 2012).

\bibitem{Thomson} M. Thomson, {\it Modern particle physics} (Cambridge University Press, 2013).

\bibitem{Griffiths} D. Griffiths, {\it Introduction to elementary particles} (John Wiley \& Sons, 2020).

\bibitem{Halzen} F. Halzen, and A. D. Martin, {\it Quarks and leptons: an introductory course in modern particle
physics} (John Wiley \& Sons, 1984)

\bibitem{Moshinsky} M. A. Moshinsky, and J. Szczepaniak, J. Phys. A: Math. Gen. {\bf 22}, L817 (1989).

\bibitem{Franco} J. A. Franco-Villafane et al, Phys. Rev. Lett. 111, 170405 (2013).

\bibitem{Oliveira1} R. R. S. Oliveira, Eur. Phys. J. C {\bf 79}, 79 (2019).

\bibitem{Oliveira2} R. R. S. Oliveira, R. V. Maluf, and C. A. S. Almeida, Ann. Phys. {\bf 400}, 1-8 (2019).

\bibitem{Cunha} M. M. Cunha, H. S. Dias, and E. O. Silva, Phys. Rev. D {\bf 102}, 105020 (2020).

\bibitem{Aharonov} Y. Aharonov, and A. Casher, Phys. Rev. Lett. {\bf 53}, 319 (1984).

\bibitem{Hagen} C. Hagen, Phys. Rev. Lett. {\bf 64}, 2347 (1990).

\bibitem{Oliveira3} R. R. S. Oliveira et al, Eur. Phys. J. Plus {\bf 134}, 495 (2019).

\bibitem{Oliveira4} R. R. S. Oliveira, Gen. Relativ. Gravit. {\bf 51}, 120 (2019).

\bibitem{Boada} O. Boada et al, New J. Phys. {\bf 13}, 035002 (2011).

\bibitem{Lamata} L. Lamata et al, Phys. Rev. Lett. {\bf 98}, 253005 (2007).

\bibitem{Schakel} A. M. Schakel, Phys. Rev. D {\bf 43}, 1428 (1991).

\bibitem{Miransky} V. A. Miransky, and I. A. Shovkovy, Phys. Rep. {\bf 576}, 1-209 (2015).

\bibitem{Oliveira5} R. R. S. Oliveira, G. Alencar, and R. R. Landim, Gen. Relativ. Gravit. {\bf 55}, 15 (2023).

\bibitem{Huerta} C. H. Alderete et al, Nat. Commun. {\bf 11}, 3720 (2020).

\bibitem{Bermudez1} A. Bermudez, M. A. M. Delgado, and A. Luis, Phys. Rev. A {\bf 77}, 063815 (2008).

\bibitem{Bermudez2} A. Bermudez, M. A. M. Delgado, and E. Solano, Phys. Rev. Lett. {\bf 99}, 123602 (2007).

\bibitem{Oliveiraneutrinos} R. R. S. Oliveira, JHEP {\bf 2025}, 1-11 (2025).

\bibitem{Novoselov} K. S. Novoselov et al, Nature {\bf 438}, 197-200 (2005).

\bibitem{Gonzalez} J. Gonzalez, F. Guinea, and M. A. Vozmediano, Nucl. Phys. B {\bf 406}, 771-794 (1993).

\bibitem{McCann} E. McCann, and V. I. Fal’ko, J. Phys. Condens. Matter {\bf 16}, 2371 (2004).

\bibitem{Ahrens} S. Ahrens et al, New J. Phys. {\bf 17}, 113021 (2015).

\bibitem{Armitage} N. P. Armitage, E. J. Mele, and A. Vishwanath, Rev. Mod. Phys. {\bf 90}, 015001 (2018).

\bibitem{Hasan} M. Z. Hasan, and C. L. Kane, Rev. Mod. Phys. {\bf 82}, 3045 (2010).

\bibitem{Qi} X. L. Qi, and S. C. Zhang, Rev. Mod. Phys. {\bf 83}, 1057-1110 (2011).

\bibitem{Guvendi} A. Guvendi, O. Mustafa, and S. G. Dogan, Phys. Lett. B {\bf 862}, 139313 (2025).

\bibitem{Oliveira6} R. R. S. Oliveira, Gen. Relativ. Gravit. {\bf 52}, 88 (2020).

\bibitem{Bakke} K. Bakke, and C. Furtado, Ann. Phys. (N.Y.) {\bf 336}, 489-504 (2013).

\bibitem{O} R. R. S. Oliveira, Int. J. Mod. Phys. D {\bf 34}, 2450065 (2025).

\bibitem{Oliveira7} R. R. S. Oliveira, Class. Quantum Grav. {\bf 41}, 175017 (2024).

\bibitem{Ahmed} F. Ahmed, Eur. Phys. J. C {\bf 79}, 534 (2019).

\bibitem{Bragança} E. A. F. Bragança, R. L. L. Vitória, H. Belich, and E. B. de Mello, Eur. Phys. J. C 80, 1-11 (2020).

\bibitem{Cho} H. T. Cho, Phys. Rev. D {\bf 68}, 024003 (2003).

\bibitem{Dolan} S. R. Dolan, and D. Dempsey, Class. Quantum Gravity {\bf 32}, 184001 (2015).

\bibitem{Hendi} S. H. Hendi, JHEP {\bf 2012}, 1-20 (2012).

\bibitem{Finster} F. Finster, J. Smoller, and S. T. Yau, J. Math. Phys. {\bf 41}, 2173-2194 (2000).

\bibitem{Badawi} A. Al-Badawi, I. Sakallı, and S. Kanzi, Ann. Phys. {\bf 412}, 168026 (2020).

\bibitem{Broderick}  A. Broderick, M. Prakash, and J. M. Lattimer, Astrophys. J. {\bf 537}, 351 (2000).

\bibitem{Bautista} E. Bautista, Phys. Rev. D {\bf 48}, 783 (1993).

\bibitem{Behncke} H. Behncke, Math. Z. {\bf 174}, 213-225 (1980).

\bibitem{Dyck} S. R. V. Dyck Jr., P. B. Schwinberg, and H. G. Dehmelt, Phys. Rev. Lett. {\bf 59}, 26 (1987).

\bibitem{Bubnov} A. F. Bubnov, N. V. Gubina, V. C. Zhukovsky, Phys. Rev. D {\bf 96}, 016011 (2017).

\bibitem{Ferrer} E. J. Ferrer, and A. Hackebill, Phys. Rev. C {\bf 99}, 065803 (2019).

\bibitem{Frolov} I. E. Frolov, and V. C. Zhukovsky, J. Phys. A: Math. Theor. {\bf 40}, 10625 (2007).

\bibitem{Kawaguchi} M. Kawaguchi, I. Siddique, and M. Huang, Eur. Phys. J. C {\bf 85}, 246 (2025).

\bibitem{Paulucci} L. Paulucci, E. J. Ferrer, V. de La Incera, and J. E. Horvath, Phys. Rev. D {\bf 83}, 043009 (2011).

\bibitem{Rodionov} V. N. Rodionov, J. Exp. Theor. Phys. {\bf 98}, 395–402 (2004).

\bibitem{Dvornikov} M. Dvornikov, EPJ Web Conf. {\bf 191}, 05008 (2018).

\bibitem{Struckmeier} J. Struckmeier, D. Vasak, A. Redelbach, and H. Stoecker, Class. Quantum Gravity {\bf 41}, 175014 (2024).

\bibitem{Ternov} I. M. Ternov, V. G. Bagrov, A. M. Khapaev, JETP {\bf 21}, 613 (1965).

\bibitem{Connell} R. F. O'Connell, Phys. Rev. {\bf 176}, 1433 (1968).

\bibitem{Fukushima} K. Fukushima, D. E. Kharzeev, and H. J. Warringa, Phys. Rev. D {\bf 78}, 074033 (2008).

\bibitem{Fuk} K. Fukushima, Lect. Notes Phys. {\bf 871}, 241 (2013).

\bibitem{Kharzeev} D. E. Kharzeev, L. D. McLerran, and H. J. Warringa, Nucl. Phys. A {\bf 803}, 227 (2008).

\bibitem{Warringa} H. J. Warringa, J. Phys. G {\bf 35}, 104012 (2008).

\bibitem{Li} Q. Li et al. Nat. Phys. {\bf 12}, 550-554 (2016).

\bibitem{Yamamoto} A. Yamamoto, Phys. Rev. Lett. {\bf 107}, 031601 (2011).

\bibitem{Dvornikov2} M. Dvornikov, and V. B. Semikoz, Phys. Rev. D {\bf 95}, 043538 (2017).

\bibitem{Isachenkov} M. V. Isachenkov, and A. V. Sadofyev, Phys. Lett. B {\bf 697}, 404-406 (2011).

\bibitem{Kharzeev2} D. E. Kharzeev, J. Liao, S. A. Voloshin, G. Wang, Prog. Part. Nucl. Phys. {\bf 88}, 1–28 (2016).

\bibitem{Li2} W. Li, and G. Wang, Ann. Rev. Nucl. Part. Sci. {\bf 70}, 293–321 (2020).

\bibitem{Buividovich} P. V. Buividovich, M. N. Chernodub, E. V. Luschevskaya, and M. I. Polikarpov, Phys. Rev. D {\bf 80}, 054503  (2009).

\bibitem{Kalaydzhyan} T. Kalaydzhyan, and I. Kirsch, Phys. Rev. Lett. {\bf 106}, 211601 (2011).

\bibitem{Sigl} G. Sigl, and N. Leite, JCAP {\bf 2016}, 025 (2016).

\bibitem{Fukushima2} K. Fukushima, M. Ruggieri, and R. Gatto, Phys. Rev. D {\bf 81}, 114031 (2010).

\bibitem{Ghosh1} S. Ghosh, N. Chaudhuri, P. Roy, and S. Sarkar, Phys. Rev. D {\bf 109}, 016021 (2024).

\bibitem{Ghosh2} S. Ghosh, N. Chaudhuri, S. Sarkar, and P. Roy, Phys. Rev. D {\bf 111}, 076012 (2025).

\bibitem{Pasqualotto} A. E. B. Pasqualotto, R. L. S. Farias, W. R. Tavares, S. S. Avancini, and G. Krein, Phys. Rev. D {\bf 107}, 096017 (2023).

\bibitem{Farias} R. L. S. Farias, D. C. Duarte, G. Krein, and R. O. Ramos, Phys. Rev. D {\bf 94}, 074011 (2016).

\bibitem{Sheng} X. Sheng, D. H. Rischke, David Vasak, and Q. Wang, Eur. Phys. J. A {\bf 54}, 21 (2018).

\bibitem{Gusynin} V. P. Gusynin, V. A. Miransky, and I. A. Shovkovy, Phys. Rev. D {\bf 52}, 4747 (1995).

\bibitem{Shushpanov} I. A. Shushpanov, and A. V. Smilga, Phys. Lett. B {\bf 402}, 351 (1997).

\bibitem{Cohen} T. D. Cohen, D. A. McGady, and E. S. Werbos, Phys. Rev. C {\bf 76}, 055201 (2007).

\bibitem{Fraga} E. S. Fraga, and A. J. Mizher, Phys. Rev. D {\bf 78}, 025016 (2008).

\bibitem{Agasian} N. O. Agasian, and S. M. Fedorov, Phys. Lett. B {\bf 663}, 445 (2008).

\bibitem{Alford} M. G. Alford, J. Berges, and K. Rajagopal, Nucl. Phys. B {\bf 571}, 269 (2000).

\bibitem{Ferrer2} E. J. Ferrer, V. Incera, and C. Manuel, Phys. Rev. Lett. {\bf 95}, 152002 (2005).

\bibitem{Fukushima3} K. Fukushima, and H. J. Warringa, Phys. Rev. Lett. {\bf 100}, 032007 (2008).

\bibitem{Metlitski} M. A. Metlitski, and A. R. Zhitnitsky, Phys. Rev. D {\bf 72}, 045011 (2005).

\bibitem{Son} D. T. Son, and M. A. Stephanov, Phys. Rev. D {\bf 77}, 014021 (2008).

\bibitem{Chaudhuri} N. Chaudhuri, A. Mukherjee, S. Ghosh, S. Sarkar, P. Roy, Eur. Phys. J. A {\bf 58}, 82 (2022).

\bibitem{Snyder1} H. S. Snyder, Phys. Rev. {\bf 71}, 38 (1947).

\bibitem{Snyder2} H. S. Snyder, Phys. Rev. {\bf 72}, 68 (1947).

\bibitem{Moffat} J. W. Moffat, Phys. Lett. B {\bf 491}, 345-352 (2000).

\bibitem{Majid} S. Majid, Lect. Notes Phys. {\bf 541}, 227 (2000).

\bibitem{Szabo1} R. J. Szabo, Gen. Relativ. Gravit. {\bf 42}, 1 (2010).

\bibitem{Addazi} A. Addazi et al, Prog. Part. Nucl. Phys. {\bf 125}, 103948, (2022).

\bibitem{Ho} P. M. Ho, H. C. Kao, Phys. Rev. Lett. {\bf 88}, 151602 (2002).

\bibitem{Bertolami1} O. Bertolami et al, Phys. Rev. D {\bf 72}, 025010 (2005).

\bibitem{Seiberg} N. Seiberg, and E. Witten, JHEP {\bf 09}, 032 (1999).

\bibitem{Gomis} J. Gomis, and T. Mehen, Nucl. Phys. B. {\bf 591}, 265 (2000).

\bibitem{Douglas} M. R. Douglas, and N. A. Nekrasov, Rev. Mod. Phys. {\bf 73}, 977 (2001).

\bibitem{Szabo} R. J. Szabo, Phys. Rep. {\bf 378}, 207 (2003).

\bibitem{Djemai} A. E. F. Djemai, and H. Smail,  Commun. Theor. Phys. {\bf 41}, 837 (2004).

\bibitem{Melic} B. Melic et al, Phys. Rev. D {\bf 72}, 057502 (2005).

\bibitem{Buric} M. Buric et al, Phys. Rev. D {\bf 75}, 097701 (2007).

\bibitem{Schupp} P. Schupp et al, Eur. Phys. J. C {\bf 36}, 405 (2004).

\bibitem{Abel} S. A. Abel et al, JHEP {\bf 09}, 074 (2006).

\bibitem{Pikovski} I. Pikovski et al, Nature Phys. {\bf 8}, 393 (2012).

\bibitem{Carlson} C. E. Carlson, C. D. Carone, and R. F. Lebed, Phys. Lett. B {\bf 518}, 201 (2001).

\bibitem{Riad} I. F. Riad, and M. M. Sheikh-Jabbari, J. High Energy Phys. {\bf 2000}, 045 (2000).

\bibitem{Nicolini} P. Nicolini, A. Smailagic, and E. Spallucci, Phys. Lett. B {\bf 632}, 547 (2006).

\bibitem{Bastos1} C. Bastos et al, Phys. Rev. D {\bf 78}, 023516 (2008).

\bibitem{Calmet} X. Calmet et al, Eur. Phys. J. C {\bf 23}, 363 (2002).

\bibitem{Berto} O. Bertolami et al, Phys. Rev. D {\bf 72}, 025010 (2005).

\bibitem{Dayi} O. F. Dayi, and A. Jellal, J. Math. Phys. {\bf 43}, 4592 (2002).

\bibitem{Gangopadhyay} S. Gangopadhyay, A. Saha, and A. Halder, Phys. Lett. A {\bf 379}, 2956-2961 (2015).

\bibitem{Gamboa} J. Gamboa, M. Loewe, and J. C. Rojas, Phys. Rev. D {\bf 64}, 067901 (2001).

\bibitem{Nascimento} J. P. Nascimento, V. Aguiar, and I. Guedes, Physica A {\bf 477}, 65 (2017).

\bibitem{Santos} E. S. Santos, and G. R. de Melo, Int. J. Theor. Phys. {\bf 50}, 332-338 (2011).

\bibitem{Chaichian} M. Chaichian, M. M. S. Jabbari, and A. Tureanu, Eur. Phys. J. C {\bf 36}, 251-252 (2004).

\bibitem{Haghighat} M. Haghighat, and M. Khorsandi, Eur. Phys. J. C {\bf 75}, 1-7 (2015).

\bibitem{Ribeiro} L. R. Ribeiro, E. Passos, C. Furtado, and J. R. Nascimento, Eur. Phys. J. C {\bf 56}, 597-606 (2008).

\bibitem{Yu} X. Yu, and K. Li, Phys. Rev. A {\bf 84}, 033617 (2011).

\bibitem{Maireche} A. Maireche, Mod. Phys. Lett. A {\bf 35}, 2050015 (2020).

\bibitem{Hassanabadi1} H. Hassanabadi et al, J. Math. Phys. {\bf 55}, 033502 (2014).

\bibitem{Maluf} R. V. Maluf, Int. J. Mod. Phys. A {\bf 26}, 4991-5003 (2011).

\bibitem{Yang} Z. H. Yang et al, Int. J. Theor. Phys. {\bf 49}, 644-651 (2010).

\bibitem{Oliveira8} R. R. S. Oliveira,  Gen. Relativ. Gravit. {\bf 56}, 30 (2024).

\bibitem{Oliveira9} R. R. S. Oliveira, and R. R. Landim, Phys. Scr. {\bf 99} 045917 (2024).

\bibitem{Oliveira10} R. R. S. Oliveira, G. Alencar, and R. R. Landim, Phys. Scr. {\bf 99}, 035226 (2024).

\bibitem{Oliveira11} R. R. S. Oliveira, R. R. Landim, Acta Phys. Pol. B {\bf 55}, 10-A1 (2024).

\bibitem{Adorno} T. C. Adorno et al, Phys. Lett. B {\bf 682}, 235-239 (2009).

\bibitem{Alavi} S. A. Alavi, and N. Rezaei, Pramana - J. Phys. {\bf 88}, 77 (2017).

\bibitem{Nath} D. Nath et al, EPL {\bf 123}, 20008 (2018).

\bibitem{Boumali} A. Boumali, and H. Hassanabadi, Z. Naturforsch. {\bf 70}, 619–627 (2015).

\bibitem{Abyaneh} M. Z. Abyaneh, and M. Farhoudi, Eur. Phys. J. Plus {\bf 136}, 863 (2021).

\bibitem{Mirza} B. Mirza, and M. Zarei, Eur. Phys. J. C {\bf 32}, 583-586 (2004).

\bibitem{Jing1} J. Jing et al, Phys. Lett. B {\bf 808}, 135660 (2020).

\bibitem{Jing2} J. Jing et al, Int. J. Mod. Phys. A {\bf 37}, 2250172 (2022).

\bibitem{Ma1} K. Ma J. H. Wang, and H. X. Yang, Phys. Lett. B {\bf 756}, 221-227 (2016).

\bibitem{Ma2} K. Ma, Y. J. Ren, and Y. H. Wang, Phys. Rev. D {\bf 97}, 115011 (2018).

\bibitem{Ma3} L. B. Ma et al, Int. J. Theor. Phys. {\bf 62}, 178 (2023).

\bibitem{Maireche2} A. Maireche, Indian J. Phys. {\bf 98}, 803–820 (2024).

\bibitem{Bastos} C. Bastos et al, Int. J. Mod. Phys. A {\bf 28}, 1350064 (2013).

\bibitem{Qolizadeh} M. Qolizadeh, S. M. Motevalli, and S. S. Hosseini, Int. J. Theor. Phys. {\bf 64}, 1-13 (2025).

\bibitem{Halder} A. Halder, S. Gangopadhyay, and A. Saha, EPL {\bf 149}, 60001 (2025).

\bibitem{Schluter} P. Schluter, K. H. Wietschorke, and W. Greiner, J. Phys. A: Math. Gen. {\bf 16}, 1999 (1983).

\bibitem{Villalba} V. M. Villalba, and A. R. Maggiolo, Eur. Phys. J. B {\bf 22}, 31 (2001).

\bibitem{Thomas} L. C. Thomas, T. Dezen, E. B. Grohs, and C. T. Kishimoto, Phys. Rev. D {\bf 101}, 063507 (2020).

\bibitem{Leary} C. C. Leary, D. Reeb, and M. G. Raymer, New J. Phys. {\bf 10}, 103022 (2008).

\bibitem{Pal} P. B. Pal, Int. J. Mod. Phys. A {\bf 7}, 5387–5459 (1992).

\bibitem{Studenikin} A. I. Studenikin, and A. Ternov, Phys. Lett. B {\bf 608}, 107-114 (2005).

\bibitem{Akhmedov} E. K. Akhmedov, and A. Wilhelm, JHEP {\bf 2013}, 1–41 (2013).

\bibitem{Dvornikov3} M. Dvornikov, and V. B. Semikoz, JCAP {\bf 08}, 021 (2018).

\bibitem{Colladay1} D. Colladay, and V. Alan Kostelecký, Phys. Rev. D {\bf 55}, 6760 (1997).

\bibitem{Colladay2} D. Colladay, and V. Alan Kostelecký, Phys. Rev. D {\bf 58}, 116002 (1998).

\end{thebibliography}
\end{document}